\documentclass{jfm}
\usepackage{graphicx}
\usepackage{epstopdf, epsfig}
\usepackage{bm}
\usepackage{amsmath}
\usepackage{tensor}
\usepackage{siunitx}
\usepackage{natbib}
\usepackage{color}
\usepackage[]{hyperref}

\shorttitle{Modelling fluid deformable surfaces}
\shortauthor{A. Torres-S\'anchez, D. Mill\'an and M. Arroyo}

\title{Modelling fluid deformable surfaces with an emphasis on biological interfaces}

\author{Alejandro Torres-S\'anchez\aff{1},
  Daniel Mill\'an\aff{1,2}
 \and Marino Arroyo\aff{1,3}\corresp{\email{marino.arroyo@upc.edu}}}

\affiliation{\aff{1}LaC\`aN, Universitat Polit\`ecnica de 
Catalunya-BarcelonaTech, 08034 Barcelona, Spain
\aff{2}CONICET \& Facultad de Ciencias Aplicadas a la Industria, Universidad Nacional de Cuyo, San Rafael, Argentina
\aff{3} Institute for Bioengineering of Catalonia, The Barcelona Institute of Science and Technology, 08028 Barcelona, Spain
}

\begin{document}

\maketitle

\begin{abstract}
Fluid deformable surfaces are ubiquitous in cell and tissue biology, including lipid bilayers, the actomyosin cortex, or epithelial cell sheets. These interfaces exhibit a complex interplay between elasticity, low Reynolds number interfacial hydrodynamics, chemistry, and geometry, and govern important biological processes such as cellular traffic, division, migration, or tissue morphogenesis. To address the modelling challenges posed by this class of problems, in which interfacial phenomena tightly interact with the shape and dynamics of the surface, we develop a general continuum mechanics and computational framework for fluid deformable surfaces. The dual solid-fluid nature of fluid deformable surfaces challenges classical Lagrangian or Eulerian descriptions of deforming bodies. Here, we extend the notion of Arbitrarily Lagrangian-Eulerian (ALE) formulations, well-established for bulk media, to deforming surfaces. To systematically develop models for fluid deformable surfaces, which consistently treat all couplings between fields and geometry, we follow a nonlinear Onsager formalism according to which the dynamics minimize a Rayleighian functional where dissipation, power input and energy release rate compete. Finally, we propose new computational methods, which build on Onsager's formalism and our ALE formulation, to deal with the resulting stiff system of higher-order of partial differential equations. We apply our theoretical and computational methodology to classical models for lipid bilayers and the cell cortex. The methods developed here allow us to formulate/simulate these models for the first time in their full three-dimensional generality, accounting for finite curvatures and finite shape changes.
\end{abstract}


\section{Introduction\label{sec:intro}}

Fluid deformable surfaces are a common motif in cell and tissue biology. For instance, lipid bilayers are fluid thin sheets that define the boundary of cells and compartmentalize them. They are the base material for the plasma membrane, the endoplasmic reticulum, mitochondria, or the Golgi apparatus. From a mechanical viewpoint, lipid bilayers are remarkable soft materials exhibiting a solid-fluid duality: while they store elastic energy when stretched or bent, as solid shells \citep{Lipowsky1991-pp}, they cannot store elastic energy under in-plane shear, a situation under which they flow as viscous two-dimensional fluids \citep{Dimova2006-tu}. This solid-fluid duality is tightly intertwined with membrane geometry: shape changes induce lipid flows that bring material from one part of the membrane to another \citep{Evans1994-yg}, whereas flows in the presence of curvature generate out-of-plane forces, which further curve the membrane \citep{Rahimi2013-ns}. The solid-fluid duality of membranes is essential for cell function; it is required during cell motility and migration \citep{Arroyo2012-nb,Lieber2015-fs}, membrane trafficking \citep{Sprong2001-wb,Rustom2004-ar}, or to enable the mechano-adaptation of cells to stress \citep{Staykova2013-fu,Kosmalska2015-ow}. Furthermore, the in-plane fluidity of the membrane allows membrane inclusions, such as proteins, to diffuse \citep{Sens2008-ne}. On the other hand, lipid bilayers are chemically responsive. Chemo-mechanical couplings can trigger tubulation \citep{Roux2002-en}, phase separation \citep{Bacia2005-nv}, budding and fission \citep{Staneva2004-ng,Zhou2005-uc}, or pearling \citep{Khalifat2014-pm}. 

Another important instance of biological fluid surface is the actomyosin cortex, a thin network of cross-linked actin filaments lying immediately beneath the plasma membrane of animal cells \citep{Bray1988-ak}.  Within this network, myosin motors exert active forces by consuming chemical energy in the form of adenosine triphosphate (ATP), that generate active tension \citep{Salbreux2012-fa}. Furthermore, the cell cortex undergoes dynamic remodelling, or turnover, in less than one minute, as a result of the polymerization and depolymerization of actin and the binding and unbinding of cross-linkers \citep{Howard2001-nl}. The cell cortex behaves as an elastic network at short time-scales and as a quasi-two-dimensional viscous fluid at longer time-scales due to  turnover. The interplay between remodelling, elasticity, and active forces in this thin cortical layer plays a critical role in different cellular processes such as cytokinesis \citep{Levayer2012-na}, or migration \citep{Bergert2015-my, Ruprecht2015-hz, Callan-Jones2016-tp}, where the coupling between shape and actin flows becomes apparent. 


In summary, fluid deformable surfaces are ubiquitous interfaces in biology, adopting three dimensional dynamical shapes, involving chemo-mechanical couplings, and exhibiting a dual solid-fluid behaviour. The mechanics of these biological interfaces plays an essential role in processes from the subcelullar to the tissue scale (not discussed here). However and despite recent efforts \citep{Salbreux2017-ln,Sahu2017-nz,Sauer2017-oc}, a general theoretical and computational framework to describe the multiphysics and geometry-dependent mechanics of these systems has been lacking. Towards filling this gap, here we develop a three-dimensional non-linear modelling and simulation framework for fluid deformable surfaces. Even though such interfacial fluids are embedded in a bulk fluid or confined to substrates, here we focus only on the mechanics of the surface. The coupling of interfaces with a bulk fluid \citep{Salac2011-hy,Woodhouse2012-rm,Farutin2012-ca,Shen2018-cx,Laadhari2017-yf} or a substrate \citep{Staykova2013-fu} has been examined extensively in other works. 


Different mathematical representations of the kinematics of fluid deformable surfaces have been proposed. In a common approach \citep{Secomb1982-cf,Barthes-Biesel1985-hp}, the velocity field of the interface is defined as the restriction to the surface of the velocity field of the bulk fluid assuming a no-slip condition.  The governing equations for the interface are then obtained in terms of time-dependent projection operators. Even if the interaction with a bulk fluid is not considered, the interfacial velocity can be extended to a tubular neighbourhood in 3D around the interface to find the governing equations, which can be shown to be independent of the extension \citep{Dziuk2007-gp, Dziuk2013-md}. This approach has been applied to the numerical simulations of lipid membranes \citep{Rodrigues2015-za,Barrett2015-eg, Barrett2016-pp} and to the numerical solution of vector PDEs, such as Navier-Stokes, on surfaces of fixed shape \citep{Hansbo2016-gp,Reuther2018-ky,Fries2018-lv}.  This framework can be adapted to parametrization-free descriptions of surfaces using level-sets \citep{Dziuk2013-md,Burman2015-jp}. However, by extending the problem to Euclidean space, this approach hides much of the geometric structure of the governing equations. Furthermore, it is not obvious how to extend it to bilayer interfaces such as lipid membranes, in which individual monolayers are bound to the mid-surface but can slip relative to each other. An alternative approach, pioneered by  \citet{Scriven1960-ij}, distinguishes between the intrinsic (tangential) velocity of particles as seen from within the surface, and the extrinsic (normal) surface velocity, which changes its shape and thus its metric tensor \citep{Aris1962-ye}. This approach, revisited in different theoretical and computational works \citep{Hu2007-ow,Arroyo2009-xz,Rahimi2012-sa,Sahu2017-nz}, requires the language and computational tools of differential geometry, provides a clear geometric picture of the governing equations, and eloquently shows the tight interplay between shape changes and interfacial flows. Here, we show that, by decoupling shape changes and tangential flows, this approach can naturally generalize Arbitrarily Lagrangian-Eulerian (ALE) methods, well established for bulk media \citep{Hirt1974-hr,Donea2003-oa}, to fluid deformable surfaces.  Thus, this formalism (1) alleviates the large distortions of a pure Lagrangian framework, which usually requires intensive remeshing \citep{Rodrigues2015-za}, and (2) allows us to deal with multilayer systems by considering independent tangential velocities for each monolayer.


To deal with the multiphysics aspects of fluid surfaces, we base our approach on a nonlinear Onsager's formalism \citep{Arroyo2018-vd,Doi2011-ni,Mielke2012-oo,Peletier2014-ku}, which provides a unified variational framework for the dissipative dynamics of soft-matter systems. In this formalism, the dynamics minimize a Rayleighian functional and result from the interplay between energetic driving forces, dissipative drag forces and external forces, each of them deriving from potentials that are the sum of individual contributions for each physical mechanism.  Complex models coupling different physics can be assembled by just adding more terms to the energy and dissipation potentials, and encoding in them the interactions between the different physical mechanisms. Thus, this framework provides a transparent and thermodynamically consistent method to generate complex models. Onsager's formalism is applicable to capillarity, elasticity, low Reynolds number hydrodynamics, reaction-diffusion systems, and provides a natural framework to model biological activity. In different contexts, similar ideas have been referred to by different names, such as extremal principles in non-equilibrium thermodynamics studied in physics \citep{Martyushev2006-un,Lebon2008-sz}, in materials modelling \citep{Ziegler1958-ou,Ziegler1987-wt,Ortiz1999-fl,Fischer2014-ik} or in atmospheric transport processes \citep{Paltridge1975-qr}. The Onsager formalism used here generalizes previous minimum principles identified in low Reynolds number hydrodynamics coupled to capillary \citep{Skalak1970-sy} or viscoelastic interfaces \citep{Secomb1982-cf,Dorries1996-ht}.

In addition to the geometric and multiphysics aspects of the theory, the three-dimensional simulation of fluid surfaces requires specialized numerical methods since the resulting equations (1) usually involve higher-order derivatives of the parametrization, (2) lead to a mixed system of elliptic and hyperbolic partial differential equations and (3) are stiff and difficult to integrate in time \citep{Rahimi2013-ns}. Indeed, surface shape enters into the energy and dissipation expressions through curvature, which involves second-order derivatives of the parametrization. From a finite element method (FEM) perspective, this requires the basis functions parametrizing the surface to be in $H^2$ (square-integrable functions whose  first- and second-order derivatives are also square-integrable). Here, we resort to subdivision surfaces, which have already been used to study the equilibrium shapes of lipid bilayers \citep{Feng2006-gi,Ma2008-mf} and to analyze thin shells \citep{Cirak2000-tq, Cirak2001-bz, Cirak2011-wf,Zhang2014-eg,Li2018-xu}. Based on a time-incremental version of Onsager's formalism, we develop variational time-integrators \citep{Ortiz1999-fl,Peco2013-ws}, which are nonlinearly and unconditionally stable and allow us to adapt the time-step spanning orders of magnitude during the dynamics of fluid deformable surfaces. 

The paper is structured as follows. In section \ref{mathe}, we develop a theoretical description of fluid surfaces, including Lagrangian, Eulerian and ALE formulations. We introduce the rate-of-deformation tensor and the Reynolds transport theorem. We also describe a useful set of tools to represent the kinematics of fluid deformable surfaces. In section \ref{models}, we describe several classical models of fluid surfaces to describe the dynamics of lipid bilayers and the cell cortex. We show how Onsager's variational formalism provides a direct and transparent tool to derive complex governing equations. In section \ref{computational}, we describe the discretization, both in time and space, of the equations governing the dynamics of general fluid surfaces. We introduce a variational time-integrator based on Onsager's formalism, and show how to discretize the different fields defined on the surface. In section \ref{examples}, we exercise the models in section \ref{models} through several examples simulated using the techniques described in section \ref{computational}. Finally, we conclude in section \ref{conclusions} with a summary and discussion of the manuscript, along with suggestions for future work.

\section{Mathematical description of fluid deformable surfaces\label{mathe}}

In this section, we mathematically describe fluid surfaces as a two-dimensional continua moving and deforming in Euclidean space. One way to represent this kind of systems is through a Lagrangian parametrization of the surface, $\bm{\phi}(\bm{\xi},t)$, in which a material particle is identified with a point $\bm{\xi}^*$ in parametric domain and $\bm{\phi}(\bm{\xi}^*,t)$ follows its trajectory in time. However, Lagrangian parametrizations present two major drawbacks for the description of fluid surfaces. 
First, due to the fluid nature of the interface, Lagrangian parametrizations suffer from very large distortions that are difficult to accommodate with conventional discretization schemes. Second, a single Lagrangian parametrization cannot track simultaneously all material particles in a multilayer interface. For example, in a lipid bilayer, two material particles representing lipid molecules from each monolayer occupy the same position on the surface. A single Lagrangian parametrization cannot track the time-evolution of both simultaneously because they can slip relative to each other.

In this section, we examine the definition of Lagrangian, Eulerian and ALE parametrizations of material surfaces and establish their relations. Associated with the flow generated by these parametrizations, we define the Lagrangian, Eulerian and ALE time-derivatives of fields on the surface. We then introduce the right Cauchy deformation tensor and the rate-of-deformation tensor, which characterizes the rate at which lengths, angles and areas transform on the time-evolving surface. We examine time-derivatives of integrals on time-evolving surfaces, and derive the form of Reynolds transport theorem and conservation of mass for the Lagrangian, Eulerian and ALE descriptions. Finally, we introduce some mathematical tools to represent the kinematics of fluid surfaces.

Throughout the manuscript, we make extensive use of the differential geometry of surfaces, including the definition of the metric tensor or first fundamental form $\bm{g}$,  second fundamental form or shape operator $\bm{k}$, covariant differentiation $\bnabla$, and Lie derivation $L_{\bm{v}}$, along with push-forwards and pull-backs by maps. 
Contravariant components of a tensor are denoted by superscripts, whereas covariant components are denoted by subscripts; for instance, the components of the metric tensor are denoted by $g_{ab}$, whereas the components of a tangent vector are denoted by $v^a$. We use Latin letters to denote indices running from $1$ to $2$, representing tensors on the tangent space of the surface, and Greek letters to denote indices running from $1$ to $3$, used for tensors in Euclidean space.
We follow Einstein's notation: contravariant and covariant indices with the same label are implicitly summed $T_{a\cdots} T^{a\cdots} = \sum_{a=1}^2 T_{a\cdots} T^{a\cdots}$. 
We refer to \cite{Do_Carmo2016-kq, Do_Carmo1992-bx,Willmore1996-ct} for background texts about the differential geometry of surfaces and manifolds. 

\subsection{Lagrangian, Eulerian and ALE parametrizations}
\begin{figure}
	\begin{center}
		\includegraphics[width=3in]{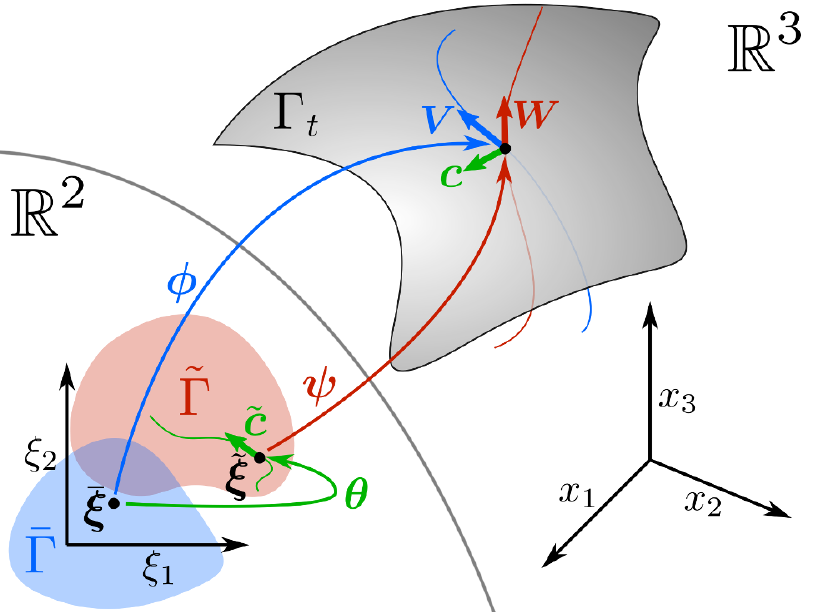}
			\caption{\label{params} A Lagrangian parametrization $\bm{\phi}(\bm{\xi},t)$ maps a domain $\bar{\Gamma}\subset\mathbb{R}^2$ onto a time-evolving surface $\Gamma_t$. Fixing a point $\bar{\bm{\xi}}$ in $\bar{\Gamma}$, the curve in $\mathbb{R}^3$ generated by $\bm{\phi}$ follows the time evolution of a material particle (blue). The velocity of this particle at time $t$ is given by $\bm{V}$. An alternative parametrization $\bm{\psi}(\bm{\xi},t)$ maps the parametric domain $\tilde{\Gamma}$ onto $\Gamma_t$. The composition $\bm{\theta}=\bm{\psi}^{-1}\circ\bm{\phi}$ characterizes the motion of material particles in $\tilde{\Gamma}$. The curve in $\tilde{\Gamma}$ generated by the mapping $\bm{\theta}$ for $\bar{\bm{\xi}}$ fixed (green) indicates how the parametric position of a material particle evolves with time in $\tilde{\Gamma}$. At time $t$ this curve has a velocity $\tilde{\bm{c}}$. The curve constructed from the map $\bm{\psi}$ by fixing $\tilde{\bm{\xi}}=\bm{\theta}(\bar{\bm{\xi}},t)$ (red) does not follow the time-evolution of any material particle in general. At time $t$ this curve has a velocity $\bm{W}$. The velocities $\bm{V}$ and $\bm{W}$ are related by $\bm{V}=\bm{W}+\bm{c}$, where $\bm{c}$ is the push-forward of $\tilde{\bm{c}}$ by $\bm{\psi}_t$.}
	\end{center}
\end{figure}

We consider the parametrization of a two-dimensional continuum $\Gamma_t$ moving and deforming in $\mathbb{R}^3$. In a Lagrangian parametrization of $\Gamma_t$, $\bm{\phi}:\bar{\Gamma}\times\mathcal{I}\ni\left(\bm{\xi},t\right)\mapsto\bm{x}\in\Gamma_t$, where $\bar{\Gamma}\subset\mathbb{R}^2$  and $\mathcal{I}\subset\mathbb{R}$,  a point $\bm{\xi}=(\xi_1,\xi_2)\in\bar{\Gamma}$ identifies a material particle and the curve obtained by fixing $\bm{\xi}$, $\bm{\phi}_{\bm{\xi}}(t)=\bm{\phi}\left(\bm{\xi},t\right)$, is its trajectory in $\mathbb{R}^3$ (see figure \ref{params}). We focus on a specific chart, although the arguments presented in this section can be trivially extended to surfaces covered by an atlas of charts. For  systems with multiple components, e.g.~multilayer systems, where material particles of different components coexist at the same point $\bm{x}\in\Gamma_t$, a single Lagrangian parametrization of $\Gamma_t$ does not exist. This is the case of a lipid bilayer, where a point $\bm{x}\in\Gamma_t$ has simultaneously attached two material particles belonging to each monolayer. Nevertheless, we can always define a Lagrangian parametrization for each of the components of the system independently so that the results in this and following sections can be applied to each component (monolayer) separately. The time derivative of the Lagrangian parametrization is  the material velocity 
\begin{equation}
\bar{\bm{V}}\left(\bm{\xi},t\right) = \frac{d}{dt} \bm{\phi}_{\bm{\xi}}(t) = \partial_t \bm{\phi}\left(\bm{\xi},t\right).
\end{equation}
The spatial velocity $\bm{V}$ on $\Gamma_t$ is obtained by composition with $\bm{\phi}^{-1}_t$
\begin{equation}
\bm{V}(\bm{x},t) = \bar{\bm{V}} \circ \bm\phi^{-1}_t(\bm{x}),
\end{equation}
where $\bm\phi_t(\bm{\xi})=\bm\phi(\bm{\xi},t)$ is obtained by fixing time $t$. In general, $\bm{V}$ has a tangential and a normal component to $\Gamma_t$
\begin{equation}
\label{matvel}
\bm{V} = \bm{v} + v_n\bm{N},
\end{equation}
where $\bm{N}$ is the unit normal to the surface. The normal velocity  $v_n$ characterizes shape changes of $\Gamma_t$ while $\bm{v}$ represents the flow of material tangent to $\Gamma_t$. In the remainder of the paper we denote by upper-case letters vectors with tangential and normal components to $\Gamma_t$ and by lower-case letters vectors that are tangent to $\Gamma_t$. We now introduce an alternative parametrization of the surface $\bm{\psi}:\tilde{\Gamma}\times\mathcal{I}\ni(\bm{\xi},t)\mapsto\bm{x}\in\Gamma_t$, where $\tilde{\Gamma}\subset\mathbb{R}^2$. The curves of constant $\bm{\xi}$, $\bm{\psi}_{\bm{\xi}}(t)=\bm{\psi}(\bm{\xi},t)$ do not follow trajectories of material particles in general. 
The velocity fields associated with this parametrization are
\begin{equation}
\begin{aligned}
\label{alevel}
\tilde{\bm{W}}(\bm{\xi},t) &= \frac{d}{dt} \bm{\psi}_{\bm{\xi}}(t) = \partial_t \bm{\psi}(\bm{\xi},t),\\
\bm{W}(\bm{x},t) &= \tilde{\bm{W}} \circ \bm\psi^{-1}_t(\bm{x})= \bm{w} + w_n\bm{N}.
\end{aligned}
\end{equation}
We can construct a map  relating both parametrizations $\bm{\theta}=\bm{\psi}^{-1}_t\circ\bm{\phi}:\bar{\Gamma}\times\mathcal{I}\rightarrow \tilde{\Gamma}$, a diffeomorphism from $\mathbb{R}^2$ to $\mathbb{R}^2$ at each $t$. The curves of constant $\bm{\xi}$, $\bm{\theta}_{\bm{\xi}}(t) = \bm{\theta}(\bm{\xi},t)$, track the parametric positions of material particles evolving in $\tilde{\Gamma}$, and have a velocity
\begin{equation}
\begin{aligned}
\bar{\bm{c}}(\bm{\xi},t)   &=  \frac{d}{dt}\bm{\theta}_{\bm{\xi}}(t) = \partial_t \bm{\theta}(\bm{\xi},t),\\
\tilde{\bm{c}}(\bm{\xi},t) &=  \bar{\bm{c}} \circ \bm\theta_t^{-1}(\bm{\xi}).
\end{aligned}
\end{equation}
To physically interpret $\tilde{\bm{c}}$, we define its push-forward by $\bm{\psi}_t$ as
\begin{equation}
\bm{c} = \bm{\psi}_{t*} \tilde{\bm{c}} =  \left[D\bm{\psi}_t\tilde{\bm{c}}\right]\circ\bm\psi^{-1}_t,
\end{equation}
where $\bm{\psi}_{t*}$ denotes the push-forward, and $D\bm{\psi}_t$ stands for the differential of $\bm{\psi}_t$, a linear mapping from the tangent space of $\tilde{\Gamma}$ at $\bm{\xi}$, $T_{\bm{\xi}}\tilde{\Gamma}$,  to the tangent space of $\Gamma_t$ at $\bm{x} = \bm{\psi}_t(\bm{\xi})$, $T_{\bm{x}}\Gamma_t$. The components of this vector in the global basis of Euclidean space are
\begin{equation}
c^\alpha = \left(\partial_b\psi^\alpha_t \tilde{c}^b\right)\circ\bm{\psi}^{-1}_t,
\end{equation}
where we have used the notation $\partial_a =\partial_{\xi_a}$. This expression shows that that in the basis $\bm{e}_a=\partial_a\bm{\psi}_t\circ\bm{\psi}^{-1}_t(\bm{x})$ of the tangent space $T_{\bm{x}}\Gamma_t$, the convected basis by $\bm{\psi}$, the components of $\bm{c}$ are simply
\begin{equation}
\label{equalcomponents}
c^a = \tilde{c}^a\circ\bm{\psi}_t^{-1}.
\end{equation}
Thus, in the convected basis, the components in $T_{\bm{\psi}_t^{-1}(\bm{x)}}\tilde{\Gamma}$ coincide with those in $T_{\bm{x}}\Gamma_t$.
Using the chain rule and previous definitions (see appendix \ref{LagALEvel}), we recover the classical relation between Lagrangian and ALE parametrizations in the bulk \citep{Donea2003-oa}, 
\begin{equation}
\bm{V} = \bm{W} + \bm{c},
\end{equation}
and thus $\bm{c}$ represents the relative velocity of material particles with respect to the parametrization given by $\bm{\psi}$. Since $\bm{c}$ is the push-forward of a vector field with respect to $\bm{\psi}$, then it is tangent to $\Gamma_t$. Comparing Eqs.~\eqref{matvel} and \eqref{alevel}, we conclude that
\begin{equation}
v_n = w_n,
\end{equation}
and 
\begin{equation}
\bm{v}=\bm{w}+\bm{c}.
\end{equation} 
This reflects that, since both parametrizations describe the same shape, their normal velocities, characterizing shape changes, must coincide. With this in mind, we can now introduce the notion of Eulerian parametrization in the context of a time-evolving surface. We say that a parametrization  $\bm{\chi}$ is Eulerian if its velocity field is always perpendicular to the surface
\begin{equation}
\partial_t \bm{\chi} \circ \bm{\chi}_t^{-1} = v_n\bm{N}.
\end{equation}
In summary, the parametrization $\bm{\phi}$ is a Lagrangian parametrization that tracks the evolution of material particles as they move with and along $\Gamma_t$. On the other hand, $\bm{\chi}$ is an Eulerian parametrization whose velocity is always perpendicular to $\Gamma_t$ regardless of the tangential flows of material. These parametrizations are special cases of a general parametrization $\bm{\psi}$, which may present tangential movements not consistent with the velocity of material particles.  This kind of parametrization is referred to as an arbitrarily Lagrangian-Eulerian (ALE) parametrization. 

We introduce here some notation. We denote the pull-backs of a tensor $\bm{t}$ on $\Gamma_t$ by the Lagrangian, Eulerian and ALE maps by
\begin{equation}
\label{pullbacks}
\bar{\bm{t}} = \bm{\phi}_t^*\bm{t},\qquad \hat{\bm{t}} = \bm{\chi}_t^*\bm{t},\qquad \tilde{\bm{t}} = \bm{\psi}_t^*\bm{t},
\end{equation}
where $\bm{\phi}^*_t$ denotes the pull-back through $\bm{\phi}_t$. 

\subsection{Material, Eulerian and ALE time derivatives}

We introduce next the concept of time-derivative of fields on $\Gamma_t$. Let us focus for simplicity on a scalar field  over $\Gamma_t$, $f(\bm{x},t)$. We first note that the operator $\partial_t$ acting on $ f(\bm{x},t)$, with the usual meaning of taking the time-derivative at fixed $\bm{x}$, is not well defined since $\bm{x}$ cannot be held fixed on a time-evolving surface in general \citep{Cermelli2005-go}. The idea of time-derivative can be more easily rationalized resorting to a parametrization. Let us first consider the Lagrangian parametrization $\bm{\phi}$. Fixing a point $\bm{\xi}\in\bar{\Gamma}$, we can compute how $f(\bm{x},t)$ changes along the curve $\bm{\phi}_{\bm{\xi}}(t)$. We define the {material time derivative} of the scalar $f$ as
\begin{equation}
D_tf(\bm{x},t) \equiv \left.\frac{d}{dt}f\left(\bm{\phi}_{\bm{\xi}}(t),t\right)\right|_{\bm{\xi}=\bm{\phi}_t^{-1}(\bm{x})}.
\end{equation}
We note that $f\left(\bm{\phi}_{\bm{\xi}}(t),t\right)$ is a function of $t$ only and therefore the right-hand side of the previous expression is the usual derivative of a function of one variable.
By noting that the pull-back of $f$ onto $\bar{\Gamma}$ is  $\bar{f}= \bm{\phi}^{*}_tf=f\circ\bm{\phi}_t$, we can rewrite the previous expression as
\begin{equation}
\begin{aligned}
D_tf(\bm{x},t) &= \left.\frac{d}{dt}\bar{f}\left(\bm{\xi},t\right)\right|_{\bm{\xi}=\bm{\phi}^{-1}_t(\bm{x})} = \partial_t \bar{f} \circ \bm{\phi}_t^{-1}(\bm{x}) \\
&= \partial_t \left( f\circ\bm{\phi}_t \right) \circ \bm{\phi}^{-1}_t(\bm{x})= \bm{\phi}_{t*} \partial_t \left(\bm{\phi}^*_t f\right)(\bm{x}),
\end{aligned}
\end{equation}
where $\partial_t\bar{f}$ has the usual meaning of taking the partial derivative of $\bar{f}$ at fixed $\bm{\xi}$. The last expression in this equation can be worded as the push-forward of the time-derivative of the pull-back of $f$ by the Lagrangian parametrization $\bm{\phi}$. This is nothing but the {Lie-derivative} of $f$ along the flow generated by $\bm{V}$, usually denoted by $L_{\bm{V}}f$, which is an extension to non-tangential vector fields of the usual definition of Lie-derivative \citep{Do_Carmo1992-bx,Arroyo2009-xz}. We note that the Lie-derivative depends on $\bm{\phi}$ only through $\bm{V}$. Thus, we can write the material time-derivative as
\begin{equation}
D_t f = L_{\bm{V}} f = \bm{\phi}_{t*} \partial_t \left(\bm{\phi}^*_t f\right).
\end{equation}
We can equivalently define the {ALE time-derivative} of $f$
\begin{equation}
\tilde\partial_t f \equiv L_{\bm{W}} f = \bm{\psi}_{t*} \partial_t \left(\bm{\psi}^*_t f\right),
\end{equation}
and the {Eulerian time derivative}
\begin{equation}
\label{Eultimeder}
\partial_t f \equiv L_{v_n\bm{N}} \bm{t} = \bm{\chi}_{t*} \partial_t \left(\bm{\chi}^*_t f\right).
\end{equation}
In the left-hand-side of this equation. the meaning of the symbol $\partial_t$ is clear: it measures the rate of change of $f$ along the flow  normal to $\Gamma_t$. If the shape of $\Gamma_t$ remains stationary, then $\partial_t$ recovers the usual meaning of taking the derivative with respect to time at fixed $\bm{x}$. We note that the symbol $\partial_t$ retains the usual meaning when applied to fields on parametric domains, e.g.~$\partial_t\bar{f}=\lim_{\Delta t\rightarrow0}\left(\bar{f}(\bm{\xi},t+\Delta t)-\bar{f}(\bm{\xi},t)\right)/\Delta t$, and should not be confused with the definition  equation \eqref{Eultimeder} for fields on $\Gamma_t$.
The operators $D_t$, $\tilde{\partial}_t$ and $\partial_t$ are related. For instance, using previous definitions in equation \eqref{pullbacks} and the chain rule (see appendix \ref{LagALEder}), we find that
\begin{equation}\label{scalar1}
D_t f= \tilde{\partial}_t f + \bnabla f \bcdot\bm{c},
\end{equation}
where $\bnabla$ denotes the covariant derivative, or here the surface gradient, of $f$.
\subsection{Rate-of-deformation tensor}
\begin{figure}
	\begin{center}
		\includegraphics[width=3in]{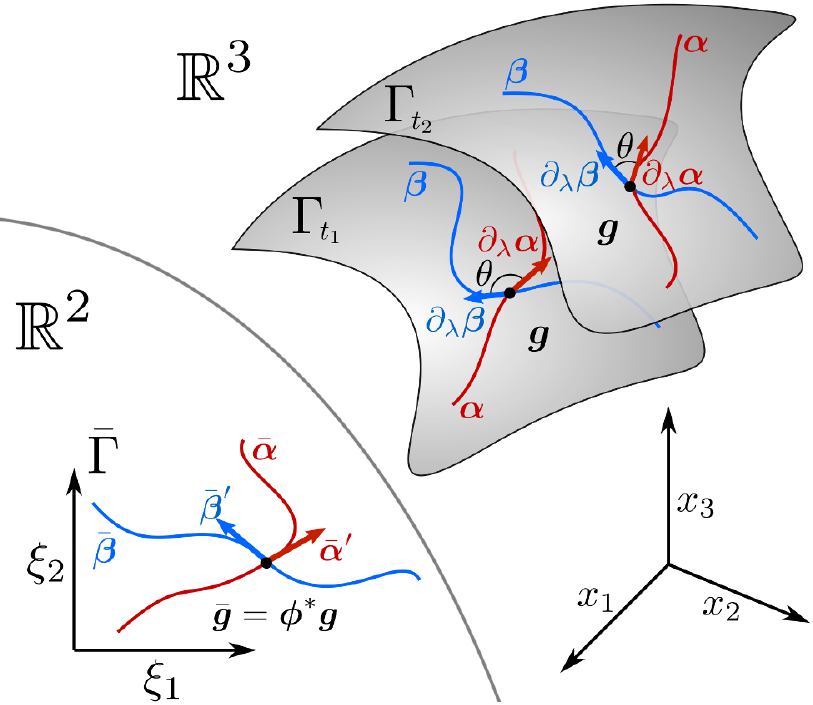}
		\caption{\label{rateofdef} The material curves $\bar{\bm{\alpha}}$ and $\bar{\bm{\beta}}$ are mapped onto $\Gamma_t$ to curves $\bm{\alpha}$ and $\bm{\beta}$ through the Lagrangian parametrization $\bm{\phi}$. As $\Gamma_t$ deforms, the length of material curves and the angle between them change. Through a pull-back, we can induce a metric on $\bar{\Gamma}$, $\bar{\bm{g}}=\bm{\phi}^*\bm{g}$, the right Cauchy-Green deformation tensor, which allows us to compute scalar products such as $\partial_\lambda \bm{\alpha}\bcdot\partial_\lambda \bm{\beta}$ from $\bar{\bm{g}}(\bar{\bm{\alpha}}',\bar{\bm{\beta}}')$. Thus, the deformation of $\Gamma_t$ is encoded on $\bar{\Gamma}$ by $\bar{\bm{g}}$.}
	\end{center}
\end{figure}

An important tensor on $\Gamma_t$ is the first fundamental form or metric tensor $\bm{g}$. The metric tensor induces a scalar product on the tangent space of $\Gamma_t$ that allows us to measure lengths, angles and areas on $\Gamma_t$. Given two tangent vectors to $\Gamma_t$, $\bm{v}$ and $\bm{w}$, the scalar product is defined by
\begin{equation}
\bm{v}\bcdot\bm{w}= \bm{g}\left(\bm{v},\bm{w}\right) = g_{ab}v^aw^b,
\end{equation}
where the notation $\bm{g}(\bcdot,\bcdot)$ views $\bm{g}$ as a bilinear form.
For surfaces in $\mathbb{R}^3$, the metric tensor is defined so that the scalar product on $\Gamma_t$ coincides with the scalar product in $\mathbb{R}^3$. Then, given a basis $\{\bm{e}_1,\bm{e}_2\}$ of $T\Gamma_t$, the tangent bundle of $\Gamma_t$, the components of the metric tensor in this basis are
\begin{equation}
g_{ab}= \bm{e}_a \bcdot \bm{e}_b.
\end{equation}
Let us consider two curves in the parametric domain $\bar{\Gamma}$, given by $\bar{\bm{\alpha}}(\lambda):[-1,1]\rightarrow\bar{\Gamma}$ and $\bar{\bm{\beta}}(\lambda):[-1,1]\rightarrow\bar{\Gamma}$, that cross at $\lambda=0$, and the image of these curves by $\bm{\phi}$, $\bm{\alpha}(\lambda,t)=\bm{\phi}(\bar{\bm{\alpha}}(\lambda),t)$ and $\bm{\beta}(\lambda,t)=\bm{\phi}(\bar{\bm{\beta}}(\lambda),t)$ (see figure \ref{rateofdef}). 
The length of $\bm{\alpha}$ (and equivalently of $\bm{\beta}$) is given by the functional
\begin{equation}
\ell[\bm{\alpha}]=\int_{-1}^1 |\partial_\lambda\bm{\alpha}| d\lambda,
\end{equation}
where $|\bm{v}|=\sqrt{\bm{v}\bcdot\bm{v}}$ is the norm of $\bm{v}$. The angle $\theta$ between curves $\bm{\alpha}$ and $\bm{\beta}$ at their point of intersection is given by
\begin{equation}
\cos \theta = \left[\frac{\partial_\lambda\bm{\alpha}\bcdot \partial_\lambda\bm{\beta}}{|\partial_\lambda\bm{\alpha}||\partial_\lambda\bm{\beta}|}\right]_{\lambda=0}.
\end{equation}
The time-evolution of the lengths of material curves and the angles between them measures how the material deforms. It is interesting to note that the pull-back of $\bm{g}$, $\bar{\bm{g}}=\bm{\phi}^*\bm{g}$, induces a time-dependent scalar product on $\bar{\Gamma}$ that allows us to compute products of deformed vectors from their time-independent description on $\bar{\Gamma}$. For instance, one can easily see that
\begin{equation}
\label{curvesdot}
\left(\partial_\lambda\bm{\alpha} \bcdot \partial_\lambda\bm{\beta}\right)_{\lambda=0}= \left[\left(\bm{g}\circ\bm{\alpha}\right)\left(\partial_\lambda\bm{\alpha}, \partial_\lambda\bm{\beta}\right)\right]_{\lambda=0}=\left[\left(\bar{\bm{g}}\circ\bar{\bm{\alpha}}\right)\left(\bar{\bm{\alpha}}', \bar{\bm{\beta}}'\right)\right]_{\lambda=0}.
\end{equation}
Equivalently,
\begin{equation}
\label{curvesnorm}
\left|\partial_\lambda\bm{\alpha}\right| = \sqrt{\left(\bar{\bm{g}}\circ\bar{\bm{\alpha}}\right)\left(\bar{\bm{\alpha}}', \bar{\bm{\alpha}}'\right)}.
\end{equation}
Thus, scalar products, lengths and angles of material curves on $\Gamma_t$, such as $\bm{\alpha}$ and $\bm{\beta}$, can be measured on $\bar{\Gamma}$, from the time-independent $\bar{\bm{\alpha}}$ and $\bar{\bm{\beta}}$, with the time-dependent scalar product induced by $\bar{\bm{g}}$. It is clear from  Eqs.~\eqref{curvesdot} and \eqref{curvesnorm} that the time-dependence of these measures of local deformation is completely encoded in $\bar{\bm{g}}$. We conclude that the tensor $\bar{\bm{g}}$ characterizes the deformation of $\Gamma_t$. In continuum mechanics, this tensor is referred to as the (right Cauchy-Green) {deformation tensor} and is generally denoted by $\bm{C}$.  The time-derivative of  this tensor defines a new tensor over $\bar{\Gamma}$
\begin{equation}
\bar{\bm{d}} = \frac{1}{2}\partial_t\bar{\bm{g}},
\end{equation}
where the $1/2$ is introduced here to follow the usual convention. The push-forward of this tensor to $\Gamma_t$ by $\bm{\phi}$ defines the so-called {rate-of-deformation tensor},
\begin{equation}
\bm{d}=\frac{1}{2} \bm{\phi}_{t*} \partial_t\bar{\bm{g}} = \frac{1}{2} \bm{\phi}_{t*} \partial_t\left(\bm{\phi}^* \bm{g}\right) = \frac{1}{2}L_{\bm{V}}\bm{g},
\end{equation}
where we recognize again the structure of a Lie derivative, this time applied to the metric tensor. The rate of change of the scalar product can then be written as 
\begin{equation}
\left.\frac{d}{dt}\left(\partial_\lambda\bm{\alpha} \bcdot \partial_\lambda\bm{\beta}\right)\right|_{\lambda=0}=2\left[\left(\bm{d}\circ\bm{\alpha}\right)\left(\partial_\lambda\bm{\alpha}, \partial_\lambda\bm{\beta}\right)\right]_{\lambda=0},
\end{equation}
and the rate of change of the norm as
\begin{equation}
\left[\frac{d}{dt}\left|\partial_\lambda\bm{\alpha}\right|\right] =\frac{1}{\displaystyle \left|\partial_\lambda\bm{\alpha}\right|}\left(\bm{d}\circ\bm{\alpha}\right)\left(\partial_\lambda\bm{\alpha}, \partial_\lambda\bm{\alpha}\right).
\end{equation}
Thus, the rate of change of local deformation of $\Gamma_t$ is encoded in $\bm{d}$. As shown in appendix \ref{RateofDef}, see also \cite{Marsden1994-lj,-Z_Wu2005-re}, the rate-of-deformation tensor for a surface moving in Euclidean space can be written as
\begin{equation}
\label{rateofdeformation}
\bm{d} = \frac{1}{2}\left[ \bnabla \bm{v}  + \left(\bnabla \bm{v}\right)^T\right] - v_n\bm{k},
\end{equation}
where  $\bnabla$ is the covariant derivative, and $\bm{k}$ is the shape operator characterizing the local curvature of the surface and defined as 
\begin{equation}
k_{ab} =- \partial_b\left( \bm{N}\circ\bm{\phi}_t\right)\circ\bm{\phi}_t^{-1}\bcdot\bm{e}_a.
\end{equation} 
From this expression, it is clear that the surface $\Gamma_t$ deforms through tangential flows, which contribute to the rate-of-deformation tensor with the usual term $\left[ \bnabla \bm{v}  + \left(\bnabla \bm{v}\right)^T\right]/2$, but also through the change in shape of $\Gamma_t$, which contributes with the term $- v_n\bm{k}$. This relation illustrates the coupling between tangential flows and shape changes  in the presence of curvature.

\subsection{Reynolds transport theorem and conservation of mass\label{secReynolds}}
\begin{figure}
	\begin{center}
		\includegraphics[width=3in]{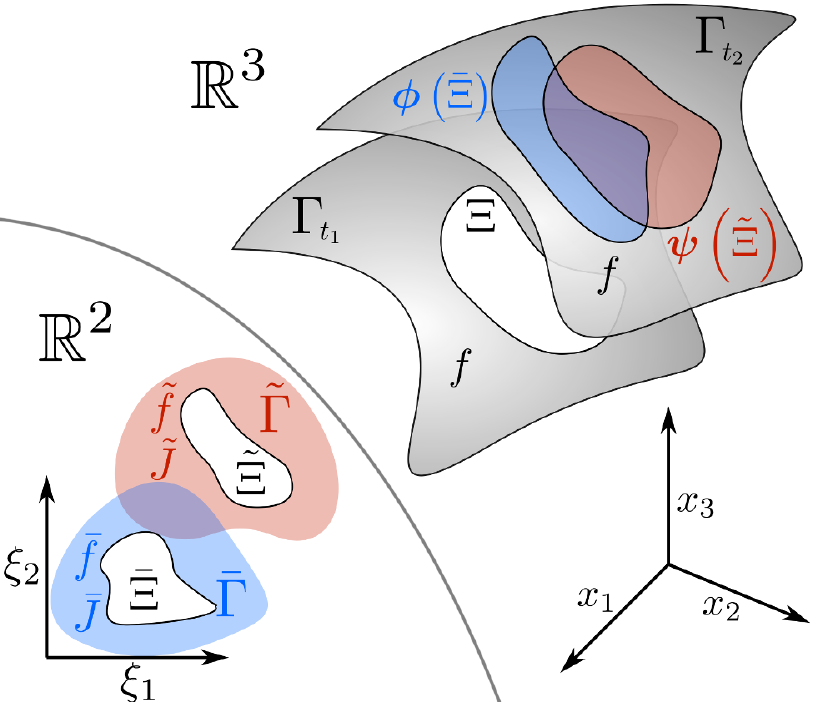}
		\caption{\label{reynolds} Given a domain $\Xi$ on $\Gamma_{t_1}$ and a scalar function $f$, we can compute the integral of $f$ on $\Xi$, $I=\int_{\Xi} f dS$, on $\bar{\Gamma}$ by pulling back the domain onto $\bar{\Gamma}$, $\bar{\Xi}=\bm{\phi}_t^{-1}(\Xi)$, the function $\bar{f}=\bm{\phi}^*f$ and using the Jacobian $\bar{J}=\sqrt{\bar{g}}$, $I=\int_{\bar{\Xi}} \bar{f} \bar{J} d\bm{\xi}$ (blue). The same can be done for the ALE parametrization (red). As $t$ evolves, the domain $\Xi$ evolves differently following the Lagrangian parametrization, $\bm{\phi}(\tilde{\Xi})$, or the ALE parametrization, $\bm{\psi}(\tilde{\Xi})$, and therefore the rate of change of $I$ on $\bar{\Gamma}$, $D_tI$, and on $\tilde{\Gamma}$, $\tilde{\partial}_tI$, are different. These are the material and ALE time-derivatives of $I$.}
	\end{center}
\end{figure}

In this section we extend the concept of Lagrangian, Eulerian and ALE time-derivatives of integrals on $\Gamma_t$.  Consider a subset $\Xi\subset\Gamma_t$, a scalar field $f:\Gamma_t\rightarrow\mathbb{R}$, and define
\begin{equation}
I = \int_{\Xi} f dS.
\end{equation}
To compute this integral, we can pull-back $fdS$ onto $\bar{\Gamma}$
\begin{equation}
I = \int_{\bar{\Xi}} \bar{f} \bar{J} d\bm{\xi}.
\end{equation}
where $\bar{\Xi} = \bm{\phi}_t^{-1}(\Xi)$, $\bar{J}=\sqrt{\bar{g}}$, $\bar{g}=\det(\bar{\bm{g}})=\det\left(D\bm{\phi}^T D\bm{\phi}\right)$ and $d\bm{\xi} = d\xi_1d\xi_2$. We define the material time derivative of $I$ as
\begin{equation}
D_t I = \frac{d}{dt} \int_{\bar{\Xi}} \bar{f} \bar{J} d\bm{\xi}.
\end{equation}
This expression characterizes the rate of change of the integral $I$ when the domain $\Xi$ is a material subset of $\Gamma_t$, i.e.~it evolves following the flow generated by $\bm{\phi}$ (see figure \ref{reynolds}). 
Developing the definition, we have
\begin{equation}
D_t I =  \frac{d}{dt} \int_{\,\bar{\Xi}} \bar{f} \bar{J} d\bm{\xi}
=\int_{\,\bar{\Xi}} \partial_t\left(\bar{f} \bar{J}\right) d\bm{\xi} = \int_{\,\bar{\Xi}} \left(\partial_t\bar{f} \bar{J} + \bar{f} \partial_t\bar{J} \right)d\bm{\xi}.
\end{equation}
The rate of change of $\bar{J}=\sqrt{\bar{g}}$ can be written in terms of $\bm{d}$ by noting that  $\partial_t\bar{J}=\partial_t\bar{g}/(2\bar{J})$  and using Jacobi's formula $\partial_t \bar{g} = \bar{g} \bar{\bm{g}}^{-1}\bm{:}\left(\partial_t \bar{\bm{g}}\right) = \bar{g} \left[\bm{g}^{-1}\bm{:}\left(L_{\bm{V}} \bm{g}\right)\circ\bm{\phi}_t\right] = 2 \bar{g} \left[\text{tr}\bm{d}\circ\bm{\phi}_t\right]$, where $\text{tr}\bm{d} = d^a_{~a} = g^{ab} d_{ab}$ is the trace $\bm{d}$. Thus, we have
\begin{equation}
\partial_t \bar{J} = \bar{J} \left(\text{tr}\bm{d}\circ\bm{\phi}_t\right) = \bar{J}\left[\left(\bnabla\bcdot\bm{v} - v_n H\right)\circ\bm{\phi}_t\right],
\end{equation}
where we have used equation \eqref{rateofdeformation},  $\bnabla\bcdot\bm{v}=\bnabla_a v^a$ is the surface divergence of the tangential vector field $\bm{v}$, and we define the mean curvature as $H=g^{ab}k_{ab}$. Then,
\begin{equation}
\begin{aligned}
D_t I&=\int_{\Xi} D_t f dS + \int_{\,\bar{\Xi}} \bar{f} \bar{J} \left[\left(\bnabla\bcdot\bm{v} - v_n H\right)\circ\bm{\phi}_t\right] d\bm{\xi}\\
&=\int_{\Xi} \left[D_t f  + f \left( \bnabla\bcdot\bm{v}-v_nH \right)\right] dS. \label{reynolds_mat}
\end{aligned}
\end{equation}
Using Eqs.~\eqref{scalar1} and the divergence theorem for surfaces, we can rewrite the previous equation in different ways
\begin{align}
D_t I &=\int_{\Xi} \left[\partial_t f  +  \bnabla\bcdot\left(f\bm{v}\right)-fv_nH\right] dS, \label{reynolds_eul}\\
&=\int_{\Xi} \left[\partial_t f -fv_nH\right] dS + \int_{\partial \Xi}f\bm{v}\bcdot\bm{m} dl,\label{reynolds_eul_border}\\
&=\int_{\Xi} \left[\tilde{\partial}_t f  +  \bnabla\bcdot\left(f\bm{c}\right)+f\left(\bnabla\bcdot\bm{w}-v_nH\right)\right] dS, \label{reynolds_ale}\\
&=\int_{\Xi} \left[\tilde{\partial}_t f + f\left(\bnabla\bcdot\bm{w}-v_nH\right)\right] dS + \int_{\partial \Xi}f\bm{c}\bcdot\bm{m} dl\label{reynolds_ale_border},
\end{align}
where $\partial \Xi$ indicates the boundary curve of $\Xi$ and $\bm{m}$ the outer normal to $\partial \Xi$ and tangent to $\Xi$. Eqs.~\eqref{reynolds_mat}-\eqref{reynolds_ale_border} are the equivalent to Reynold's transport theorem  for material domains in terms of the material, Eulerian and ALE time-derivative of $f$. As for scalar fields, we can extend the notion of material time-derivative of an integral relative to other parametrizations. In particular, we can consider the parametric domain $\tilde{\Xi}=\bm{\psi}_t^{-1}(\Xi)$, and the time-derivative
\begin{equation}
\tilde{\partial}_t I = \frac{d}{dt} \int_{\tilde{\Xi}} \tilde{f} \tilde{J} d\bm{\xi}, 
\end{equation}
where $\tilde{J}=\det\tilde{\bm{g}}$. This time-derivative characterizes the rate of change of $I$ when it follows the flow generated by the ALE parametrization. One can easily prove that
\begin{equation}
D_t I = \tilde{\partial}_t I + \int_{\partial \Xi}f\bm{c}\bcdot\bm{m} dl.
\end{equation}
For an Eulerian parametrization, one equivalently finds
\begin{equation}
D_t I = \partial_t I + \int_{\partial \Xi}f\bm{v}\bcdot\bm{m} dl.
\end{equation}
From these expressions, it is clear that for a closed surface $D_t I=\tilde{\partial}_t I = \partial_t I$. 

The previous expression can be used to derive the statement of conservation of mass on fluid surfaces. Indeed, in the special case of $f = \rho$, the mass density per unit area, conservation of mass for every material sub-domain $\Gamma_t$ requires that
\begin{equation}
D_t \int_{\Xi} \rho dS  = \int_{\Xi} r dS,
\end{equation}
where $r$ is the rate of creation of mass per unit area, which may for instance result from the exchange of material with the bulk. Since this must hold for every subdomain $\Xi$, we can localize the statement to obtain Lagrangian, Eulerian and ALE forms of local conservation of mass
\begin{equation}
\label{masscons}
\begin{aligned}
0 &= D_t\rho + \rho \left( \bnabla\bcdot\bm{v}-v_nH \right) - r,\\
& = \partial_t \rho +  \bnabla\bcdot(\rho\bm{ v})-\rho v_nH- r,\\
& = \tilde{\partial}_t \rho +  \bnabla\bcdot(\rho\bm{ c})+\rho\left(\bnabla\bcdot\bm{w} -  v_nH\right)- r.
\end{aligned}
\end{equation}
For inextensible fluid surfaces in the absence of mass exchange, balance of mass reduces to $D_t\rho=0$, leading to the condition
\begin{equation}
\label{inext}
\bnabla\bcdot\bm{v} - v_nH =\text{tr}\,\bm{d}= 0.
\end{equation} 
Thus, for an inextensible surface with curvature, any shape change must be accompanied by a tangent flow to fulfill the inextensibility constraint, further highlighting  the tight coupling between tangent flows and shape changes in the presence of curvature.

\subsection{Representation of kinematics for fluid deformable surfaces}

In previous sections, we have seen that Lagrangian parametrizations are natural tools to  define the deformation tensor $\bar{\bm{g}}$, the rate-of-deformation tensor $\bm{d}$, and to establish the transport theorem on a time-deforming surface. A time-dependent Lagrangian parametrization contains information about shape changes ($v_n$) and about interfacial flows ($\bm{v}$). In practical computations, however, Lagrangian parametrizations are not well-suited for fluid surfaces because they exhibit large distortions, requiring intensive remeshing \citep{Rodrigues2015-za}, and because a single Lagrangian parametrization cannot describe a multicomponent system like a lipid bilayer, where monolayers can slip relative to each other. In this case, one could consider  a Lagrangian parametrization for each component, which, however, increases the number of degrees of freedom since each parametrization describes both tangential motion and shape, whereas only tangential motions are independent of each other. In the present section, we provide a set of modelling tools, which are useful for a clean formulation of physical models of fluid surfaces and particularly for their numerical discretization.

In the previous section, we have introduced the notion of a time-dependent ALE parametrization $\bm{\psi}$ to describe the time-evolution of a material surface $\Gamma_t$, which can alleviate mesh distortion when dealing with fluid surfaces since it does not follow material particles. We note, however, that $\bm{\psi}$ does not contain information about the tangential motion of material particles (the interfacial flows) given by $\bm{v}$, since  $\bm{v}$ and the tangential velocity of $\bm{\psi}$ differ by a relative velocity $\bm{c}$. This fact confronts us with two issues. First, how to select the tangential velocity of $\bm{\psi}$, which is {arbitrary} in the sense of not being prescribed by any physical law. Second, since $\bm{v}$ needs to be considered as an object independent of $\bm{\psi}$, how to parametrize tangential vector fields? The first issue has been addressed by introducing  a numerical drag, which limits the tangential motion of $\bm{\psi}$  \citep{Rahimi2012-sa,Ma2008-mf}. One could also use the physically unconstrained tangential degrees of freedom of $\bm{\psi}$ to perform dynamical mesh adaptation \citep{Veerapaneni2011-yv}. These approaches, however, increase the number of essential degrees of freedom required to describe shape changes (from one to three) and require parameter tuning. Instead, in section \ref{aleparam} we develop a special kind of ALE parametrization based on an offset \citep{Rangarajan2015-vv}, which parametrizes $\bm{\psi}$ using a scalar field over $\Gamma_t$. Regarding the second issue, we note that interpolating tangent vector fields in a system with multiple charts is delicate, see section \ref{computational}. In section \ref{hodgesection}, we introduce the Hodge decomposition of vector fields in terms of scalar fields, whose interpolation is straightforward.

\subsubsection{An ALE parametrization based on an offset\label{aleparam}}

\begin{figure}
	\begin{center}
		\includegraphics[width=3.5in]{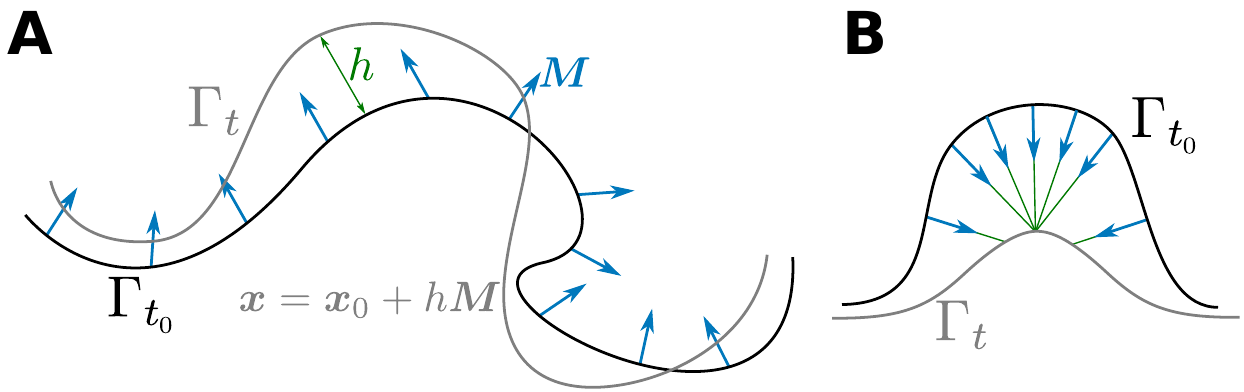}
		\caption{\label{fig_parametrization} Surface parametrization in terms of an offset. (A) The field of directors $\bm{M}$ represents the direction in which the point $\bm{x}_0\in\Gamma_{t_0}$ can evolve. The height function $h$, which may be negative, represents the distance between the point $\bm{x}$ on $\Gamma_t$ and $\Gamma_{t_0}$ in the direction of $\bm{M}$. (B) In this example $\Gamma_t$ lies at the limit of the tubular neighbourhood to $\Gamma_{t_0}$ for the given director field.}
	\end{center}
\end{figure}

We define next a restricted ALE parametrization, which by construction is devoid of the arbitrary freedom associated with tangential motions. Let us consider the state of the surface at a given time $t_0$, $\Gamma_{t_0}$, and a parametrization of this surface $\bm{\psi}_0(\bm{\xi})$. We consider a vector field $\bm{M}(\bm{\xi})$, representing a field of directors over $\Gamma_{t_0}$, with non-zero normal component but not necessarily coinciding with the normal field of $\Gamma_{t_0}$. We define a family of parametrizations of $\Gamma_t$ at time $t>t_0$ in terms of the offset of a point $\bm{x}=\bm{\psi}_0(\bm{\xi})$ along $\bm{M}(\bm{\xi})$,
\begin{equation}
\label{paramgeneral}
\bm{\psi}(\bm{\xi},t)=\bm{\psi}_0(\bm{\xi}) +  h(\bm{\xi},t)\bm{M}(\bm{\xi}),
\end{equation}
see figure \ref{fig_parametrization}A. The field that characterizes the time-evolution of the parametrization is $h$, a simple scalar field on $\tilde{\Gamma}$.
In principle, this approach is not completely general, since it only allows us to parametrize the surfaces lying in the so-called  {tubular neighbourhood}  of $\Gamma_{t_0}$ \citep{Do_Carmo2016-kq}. For some interval $I=(t_0,t_0+\delta t)$, the deformed surface $\Gamma_t$ will lie in the tubular neighbourhood of $\Gamma_{t_0}$ if the time-evolution is smooth. However, after some time, $\Gamma_t$ may leave the tubular neighbourhood of $\Gamma_{t_0}$ (see figure \ref{fig_parametrization}B for an example). A simple solution to this issue is  to then update the reference configuration $\Gamma_{t_0}$. This kind of parametrization,  proposed by \cite{Rangarajan2015-vv}, generalizes the classical Monge parametrization, which is recovered by setting $\Gamma_{t_0}$ to a plane, $\bm{M}$ to its constant normal and $h$ to the height of the surface $\Gamma_t$ with respect to the plane \citep{Do_Carmo2016-kq}. We finally note that for this kind of surface parametrization, we have
\begin{equation}
\bm{W}(\bm{x},t) = \left(\partial_t h\,\bm{M}\right)\circ\bm{\psi}^{-1}(\bm{x},t).
\end{equation}
Since $h$ is a scalar field on $\tilde{\Gamma}$, $\partial_t h$ in this equation has the usual meaning of  time differentiation at fixed $\bm{\xi}$. In practice, $\bm{M}$ can be chosen to be $\bm{N}_0$, the field of normals of the reference surface as in \citep{Rangarajan2015-vv}. This leads to an Eulerian parametrization at $t=t_0$, and close to it at later times. Thus, $\bm{W}$ will have in general non-zero  tangential components, and therefore this parametrization is neither Eulerian nor Lagrangian. Instead, it is an ALE parametrization depending on a generalized height field $h$, in which the arbitrariness is removed by following equation (\ref{paramgeneral}) and choosing the field of directors $\bm{M}$.

\subsubsection{Velocity potentials: Hodge decomposition\label{hodgesection}}

\begin{figure}
	\begin{center}
		\includegraphics[width=3in]{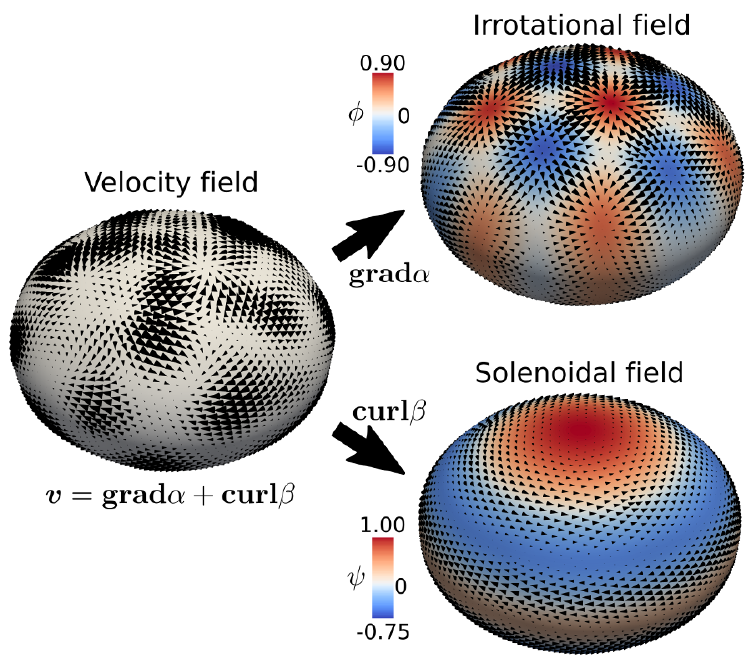}
		\caption{\label{hodge} A vector field on a surface can be decomposed in a solenoidal and a irrotational fields.}
	\end{center}
\end{figure}

Given a vector field $\bm{V}\in\mathbb{R}^3$, it is well-known that $\bm{V}$ admits a decomposition in terms of the gradient of a function $\Phi$ and the curl of a vector potential $\bm{A}$ in what is called the Helmholtz decomposition,
\begin{equation}
\bm{V} = \bnabla\,\Phi + \bnabla\times\,\bm{A},
\end{equation}
where here $\bnabla$ and $\bnabla\times$ stand for the gradient and curl in $\mathbb{R}^3$. For a vector field tangent to a plane embedded in $\mathbb{R}^3$, this can be simplified to 
\begin{equation}
\bm{V} =\bnabla\,\Phi + \bnabla\times\left(\Psi\bm{ \bm{N}}\right),
\end{equation}
where $\bm{N}$ is the normal to the plane and $\Psi$ is a scalar function. Therefore, for a plane embedded in $\mathbb{R}^3$, a tangent vector field can be represented in terms of two scalar fields, $\Phi$ and $\Psi$. This property can be generalized to arbitrary surfaces in terms of their intrinsic differential geometry, i.e.~not relying on their embedding in $\mathbb{R}^3$, as a special case of the Hodge decomposition for $n$-forms \citep{Do_Carmo1992-bx}. A vector field $\bm{v}$ tangent to a surface $\Gamma$ can be decomposed as
\begin{equation}
\label{velhodge}
\bm{v} = \bnabla \alpha + \bnabla\times \beta + \bm{h},
\end{equation}
where $\alpha$ and $\beta$ are scalar fields on $\Gamma$ and $\bm{h}$ is a harmonic vector field, satisfying $\bnabla\bcdot\bm{h}=0$ and $\bnabla\times\bm{h}=0$.  We note that the curl operator $\bnabla\times$ on a surface, an instance of exterior derivative, is defined differently to its counterpart in Euclidean space. For instance, applied on a scalar function $\bnabla\times\beta$ is a vector with components $(\bnabla\times \beta)^a = \epsilon^{ab} \bnabla_b\beta$, where $\bm{\epsilon}$ is the antisymmetric tensor  
\begin{equation}
\bm{\epsilon}^{ab} = J^{-1} \varepsilon^{ab}, 
\end{equation}
with $\varepsilon^{ab}$ the Levi-Civita symbols defined by the matrix
\begin{equation}
[\varepsilon] = \begin{pmatrix}
0 & 1\\
-1 & 0
\end{pmatrix}.
\end{equation}
For simply connected surfaces, i.e.~closed surfaces with genus equal to 0, there is only a trivial harmonic vector field, $\bm{h}=\bm{0}$, and  $\bm{v}$ can be described in terms of the two scalar fields $\alpha$ and $\beta$ (see figure \ref{hodge} for an example on an ellipsoid).
In the absence of shape changes, from equation \eqref{inext} it is clear that an inextensible flow satisfies
$\bnabla\bcdot\bm{v} = 0$. In this case, $\bm{v}$ can be represented in terms of a stream function as $\bm{v}=\bnabla\times\beta$. This approach was introduced by \citet{Secomb1982-cf} to describe flows in fluid surfaces with fixed shape and used more recently by various authors \citep{Morris2015-up,Sigurdsson2016-jt,Reuther2016-yo,Gross2018-nt,Mickelin2018-ug}. However, we note that for inextensible surfaces that change shape, both $\alpha$ and $\beta$ need be considered. In this case, using the fact that $\bnabla\bcdot\bm{v} = \bnabla\bcdot\bnabla\alpha + \bnabla\bcdot\bnabla\times\beta =  \bnabla\bcdot\bnabla\alpha  = \Delta \alpha$, where  $\Delta\alpha =  \bnabla\bcdot\bnabla\alpha$, it follows from equation \eqref{inext} that in an inextensible flow $\alpha$ and $v_n$ satisfy the constraint
\begin{equation}
\Delta \alpha = v_nH.
\end{equation}

\section{Physical models of fluid surfaces\label{models}}

In this section we examine classical models for fluid deformable surfaces, two used to model lipid bilayers and one applicable to the cell cortex. Thanks to the tools introduced above and Onsager's formalism, we derive the corresponding governing equations in their full three-dimensional and nonlinear generality.

\subsection{Lipid bilayers: An inextensible viscous layer with bending energy\label{inextmodel}}

Lipid membranes are interfacial viscous fluids with bending elasticity. The interplay between viscosity and elasticity determines their relaxation dynamics after they are brought out-of-equilibrium by external forces or biological activity. 
These two essential mechanical features of lipid membranes, their out-of-plane elasticity and interfacial viscosity, have often been examined separately. The mechanical equilibrium of lipid bilayers can be understood to a large extent with the classical bending model of Helfrich \citep{Helfrich1973-tn,Lipowsky1991-pp,Julicher1993-xt}. For that reason, studies of lipid bilayers at scales beyond tens of nanometers have mainly focused on this model, e.g.~in investigations of equilibrium configurations of closed vesicles under geometric constraints, such as fixed surface area or fixed enclosed volume \citep{Steigmann1999-vi,Capovilla2002-uj,Tu2004-mn,Feng2006-gi,Rangarajan2015-vv,Sauer2017-oc}. Beyond the Helfrich model, and subsequent refinements such as the Area Difference Elasticity model \citep{Seifert1997-tl}, more general models are required to describe the dynamical transformations that bilayers undergo, which should capture the interfacial dissipative mechanisms that dominate at sub-cellular scales. The interfacial hydrodynamics of bilayer membranes was first examined separately from membrane deformation, i.e.~assuming fixed membrane shape. These studies focused on the mobility of membrane inclusions, such as proteins, starting with the seminal work of \cite{Saffman1975-xx} on planar lipid bilayers. Subsequent studies have considered the effect of fluid boundaries \citep{Stone1998-ea} or the (fixed) shape of the fluid membrane \citep{Levine2004-nu,Henle2010-nl,Sigurdsson2016-jt}. Interfacial flows of vesicles induced by shear bulk flows were also considered at fixed vesicle shape \citep{Secomb1982-cf}. Following the seminal works of \citet{Scriven1960-ij} and \citet{Aris1962-ye} on the hydrodynamics of insoluble fluid films, \citet{Barthes-Biesel1985-hp} examined the interfacial flow of vesicles in a shear flow allowing for infinitesimal shape deformations. More recently, a geometrically non-linear model for an inextensible viscous interfacial fluid with bending rigidity was examined, formulated geometrically, and exercised under the assumption of axisymmetry \citep{Arroyo2009-xz,Arroyo2010-mm}. Along these lines, there is an increasing interest in the community of applied and computational mathematics to develop numerical methods to solve the three-dimensional equations governing inextensible viscous interfaces with curvature elasticity \citep{Nitschke2012-nh,Rodrigues2015-za,Reuther2016-yo,Barrett2016-yi}. This model provides a first approximation to the dynamical behaviour of lipid membranes. 
\begin{figure}
	\begin{center}
		\includegraphics[width=3.5in]{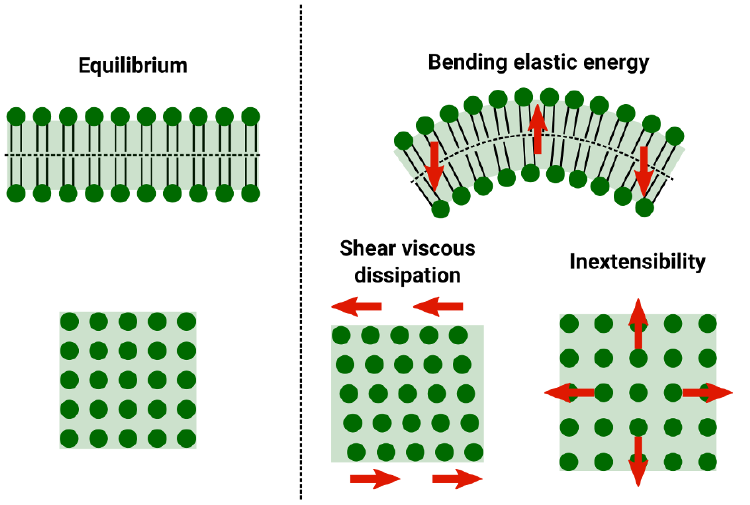}
		\caption{\label{eners_mon} A simple model of a lipid bilayer as an inextensible viscous fluid with bending elasticity.}
	\end{center}
\end{figure}

Here, we formulate this model based on Onsager's variational formalism \citep{Arroyo2009-xz}, and derive the Euler-Lagrange equations. We first introduce the bending energy of the bilayer, or Helfrich energy,
\begin{equation}
\mathcal{F}_H\left[\bm{\phi}\right] = \int_{\Gamma_t} \left[\frac{\kappa}{2}(H-H_0)^2 + \bar{\kappa} K\right]dS,
\end{equation}
where $\kappa$ and $\bar{\kappa}$ are the bending and Gaussian bending modulus respectively, which we assume to be homogeneous on $\Gamma_t$, $H_0$ is the spontaneous curvature of the membrane, and $K$ is the Gaussian curvature $K=\det(k\indices{^{a}_b})$. $\kappa$ is a positive number and $\bar{\kappa}\approx-\kappa$. Thus, this energy penalizes deviations of the mean curvature away from the spontaneous curvature $H_0$ and disfavors regions with negative Gaussian curvature (see figure \ref{eners_mon}). Recalling the Gauss-Bonnet theorem \citep{Do_Carmo2016-kq}, according to which $\int_{\Gamma_t} KdS$ is a topological invariant, we can ignore the second term in the Helfrich energy for closed surfaces of fixed topology. For simplicity, we restrict our attention to symmetric bilayers, with the same composition in both monolayers, for which $H_0=0$. Thus, we can rewrite the Helfrich energy as
\begin{equation}
\label{helfrich2}
\mathcal{F}_H\left[\bm{\phi}\right] = \int_{\Gamma_t} \frac{\kappa}{2}H^2dS.
\end{equation}
The free energy depends on the set of  state variables of the system, usually denoted by $X$. In this case, the material parametrization is the state variable of the system, $X=\{\bm{\phi}\}$ . We note, however, that since the energy only depends on the shape of $\Gamma_t$, $\bm{\phi}$ can be replaced by any other ALE parametrization $\bm{\psi}$. In Onsager's variational principle, dissipative mechanisms are introduced through dissipation potentials. The Newtonian shear rheology of lipid membranes \citep{Dimova2006-tu} is encoded in the following dissipation potential
\begin{equation}
\label{sheardiss}
\mathcal{D}_S\left[\bm{\phi};\bm{V}\right] = \int_{\Gamma_t}\mu |\bm{d}|^2 dS,
\end{equation}
where $\mu$ is the (in-plane) shear viscosity of the monolayer and  $|\bm{d}|^2 = d_{ab} g^{ac} g^{bd}d_{cd}$. Since we assume that the deformation of the membrane is inextensible, only the deviatoric part of $\bm{d}$ matters in the above definition. The dissipation potential depends on the state variable of the system, $\bm{\phi}$, through shape in $dS$ and $\bm{g}$, but primarily on the variable representing the rate of change of the state, $\bm{V}$. The variables that represent the processes that change the state of the system and produce dissipation are called process variables and denoted by $V$; here $V=\{\bm{V}\}$. Here the process variable is simply the time-derivative of the state variable $\bm{V}=\partial_t\bm{\phi}$, but this is not necessarily the case. For instance, if we had used $\bm{\psi}$ rather than $\bm{\phi}$ as state variable, then $\bm{W}=\partial_t\bm{\psi}$ would not be a meaningful variable to encode dissipation since it does not represent a physical velocity. {For thermodynamical consistency, the dissipation potential $\mathcal{D}(X;V)$ must be a convex function of $V$ with minimum at $V=0$ \citep{Arroyo2018-vd}. We further assume that $\mathcal{D}(X;0)=0$, so that $\mathcal{D}(X;V)\ge0$.} If external forces $\bm{F}$ are applied, these introduce a power input
\begin{equation}
\mathcal{P}\left[\bm{\phi};\bm{V}\right] = -\int_{\Gamma_t} \bm{F}\bcdot\bm{V}dS.
\end{equation}
One can also include the dissipation potential associated to the bulk viscous fluid where the membrane is embedded. Here, we ignore bulk hydrodynamical forces to focus on the fluid membrane, an assumption which is physically justified for phenomena below the Saffman-Delbr\"uck lengthscale $l_\text{SD}=\mu/\mu_b$, where $\mu_b$ is the bulk viscosity \citep{Saffman1975-xx,Arroyo2009-xz}. For a lipid membrane, $l_\text{SD}\approx \SI{5}{\micro\meter}$.

Onsager's variational principle establishes a competition between energy release rate, power and dissipation through the Rayleighian, which takes the form
\begin{equation}
\mathcal{R}\left[\bm{\phi};\bm{V}\right] = D_t\mathcal{F}_H[\bm{\phi};\bm{V}]  + \mathcal{D}_S[\bm{\phi};\bm{V}] + \mathcal{P}\left[\bm{\phi};\bm{V}\right].
\end{equation}
Here, the rate of change of the energy $D_t\mathcal{F}_H[\bm{\phi};\bm{V}]$ is 
\begin{equation}
\label{matderhelf}
\begin{aligned}
D_t\mathcal{F}_H[\bm{\phi};\bm{V}] &= \int_{\Gamma_t}\kappa \left\{H\Delta v_n- \left(\frac{1}{2}H^2-|\bm{k}|^2 \right)Hv_n  + \bnabla_a\left(\frac{1}{2} \kappa H^2 v^a\right)\right\} dS,
\end{aligned}
\end{equation}
where we have used that
\citep{Capovilla2002-uj,Arroyo2009-xz}
\begin{equation}
\partial_t H = \Delta v_n + |\bm{k}|^2 v_n.
\end{equation}
Then, Onsager's principle states that process variables minimize the Rayleighian
\begin{equation}
\label{OnsagerGeneral}
\bm{V} = \underset{\bm{U}}{\arg\min}\,\mathcal{R}\left[\bm{\phi};\bm{U}\right],
\end{equation}
subject to constraints $Q[\bm{\phi};\bm{V}]$. Here, we consider that the surface is inextensible
\begin{equation}
Q_1[\bm{\phi};\bm{V}] = \text{tr}\,\bm{d} = \bnabla\bcdot\bm{v}-v_nH = 0.
\end{equation}
Furthermore, due to osmotic effects, it can often be assumed that cells and vesicles maintain their  volume constant, and hence 
\begin{equation}
\label{volumeconstraint}
Q_2[\bm{\phi};\bm{V}] = \int_{\Gamma_t}\bm{V}\bcdot\bm{N} dS = 0.
\end{equation}
To enforce these constraints, we can introduce the Lagrangian
\begin{equation}
\mathcal{L}\left[\bm{\phi};\bm{V},\gamma,P\right] = \mathcal{R}\left[\bm{\phi};\bm{V}\right]  +  \int_{\Gamma_t}\gamma\, Q_1[\bm{\phi};\bm{V}]\,dS  -PQ_2[\bm{\phi};\bm{V}].
\end{equation}
$P$ is the pressure in the vesicle and $\gamma$ is a component of the surface tension. Then, Onsager's principle subject to constraints can be written as a saddle-point problem
\begin{equation}
\label{Onsager_inext}
\{\bm{V},\gamma,P\} =\underset{\bm{U}}{\arg\min}~ \underset{\lambda,S}{\arg\max}\,\mathcal{L}\left[\bm{\phi};\bm{U},\lambda,S\right].
\end{equation}
From the stationarity conditions of equation \eqref{Onsager_inext}, one finds the weak and the strong form of the governing equations (see appendix \ref{weakFormInext}), the latter of which take the form
\begin{align}
\label{linmom}\bnabla_a \bm{\Sigma}^a + \bm{B} &= \bm{0},\\
\text{tr}\,\bm{d} &= 0,\\
\label{strvol}\int_{\Gamma_t} \bm{V}\bcdot\bm{N} dS &= 0.
\end{align}
Here, $\bm{\Sigma}^a$ is the so-called surface stress vector,
\begin{equation}
\bm{\Sigma}^a = \sigma^{ab} \bm{e}_b + \sigma_n^a\bm{N},
\end{equation}
where $\bm{\sigma}$ is the in-plane stress
\begin{equation}
\label{stress1}
\sigma^{ab} = \kappa H\left(\frac{1}{2}Hg^{ab}-k^{ab}\right)+2\mu d^{ab}+\gamma g^{ab},
\end{equation}
and
$\bm{\sigma}_n$ is a vector of normal stresses
\begin{equation}
\label{nstress}
\sigma_n^a = \kappa g^{ab} \bnabla_bH.
\end{equation}
When multiplied by a unit vector $\bm{l}$ in $T_{\bm{x}}\Gamma_t$, $\bm{\Sigma}^bl_b$ is the three-dimensional force per unit length across a curve passing through $\bm{x}$ and perpendicular to $\bm{l}$. Note that $\sigma_n^a$ represents bending moments caused by curvature imbalances. Finally, $\bm{B}$ is the field of (external) body forces
\begin{equation}
\label{bodyf}
\bm{B} = \bm{F} +P\bm{N}.
\end{equation}

Eqs.~\eqref{linmom}-\eqref{strvol} express balance of linear momentum and conservation of mass (inextensibility) and enclosed volume in a fully nonlinear regime. Alternatively, one could have derived equation \eqref{linmom} from local force balance on the membrane as in \citep{Salbreux2017-ln}, and postulated the constitutive laws Eqs.~\eqref{stress1} and \eqref{nstress}. Thus, by starting from different ingredients (a Rayleighian expressing energy release-rate, dissipation and power input) and invoking a variational principle subject to constraints, Onsager's formalism recovers these equations in a systematic and transparent way. As a direct corollary of Onsager's principle, it is easy to see that, in the absence of external power inputs, the free energy is a Lyapunov functional of the dynamics, i.e.~$\mathcal{F}$ is a decreasing functional \citep{Arroyo2018-vd} 
\begin{equation}
\label{lyapunov1}
D_t\mathcal{F} \leq 0,
\end{equation}
which provides a  nonlinear notion of stability for the dynamics.
From a computational point of view, only the weak form of the stationarity conditions issuing from Onsager's formalism is required for a space discretization based on finite elements. Onsager's formalism also provides a framework to formulate nonlinearly stable variational time-integrators, as described in section \ref{computational}. 

We finally note that the choice of state and process variables is not unique. For instance, as mentioned earlier, we can choose an ALE parametrization $\bm{\psi}$ instead of $\bm{\phi}$ as state variable, since the free energy  depends only on shape. The velocity $\bm{V}$, our process variable, then needs to be split into a normal and a tangential components $\bm{V}=\bm{v}+v_n\bm{N}$, where $v_n=\bm{W}\bcdot\bm{N}$. More specifically and for the ALE parametrization in equation \eqref{paramgeneral}, we have $v_n=\partial_t h \bm{M}\bcdot\bm{N}$. We can then rewrite the Lagrangian in terms of $\partial_t h$ and $\bm{v}$. We can further decompose $\bm{v} = \bnabla \alpha + \bnabla\times \beta$.  The governing equations issued from any of these choices look very different but describe the same dynamics. For instance, the dynamics obtained from 
\begin{equation}
\label{Onsager_inext2}
\{\partial_t h,\alpha,\beta,\gamma,P\} = \underset{\{w_h,a,b\}}{\arg\min}~\underset{\lambda,S}{\arg\max}\,\mathcal{L}\left[h;w_h,a,b,\lambda,S\right],
\end{equation}
are equivalent to those resulting from equation \eqref{Onsager_inext}. While the choice of variables $X=\{\bm{\phi}\}$ and $V=\{\bm{V}\}$ is natural from a modelling viewpoint, the choice $X=\{h\}$ and $V=\{v_n,\alpha,\beta\}$ is better suited from a computational viewpoint, as will become clear in section \ref{computational}.

\subsection{Lipid bilayers: The Seifert-Langer model}

\begin{figure}
	\begin{center}
		\includegraphics[width=4.25in]{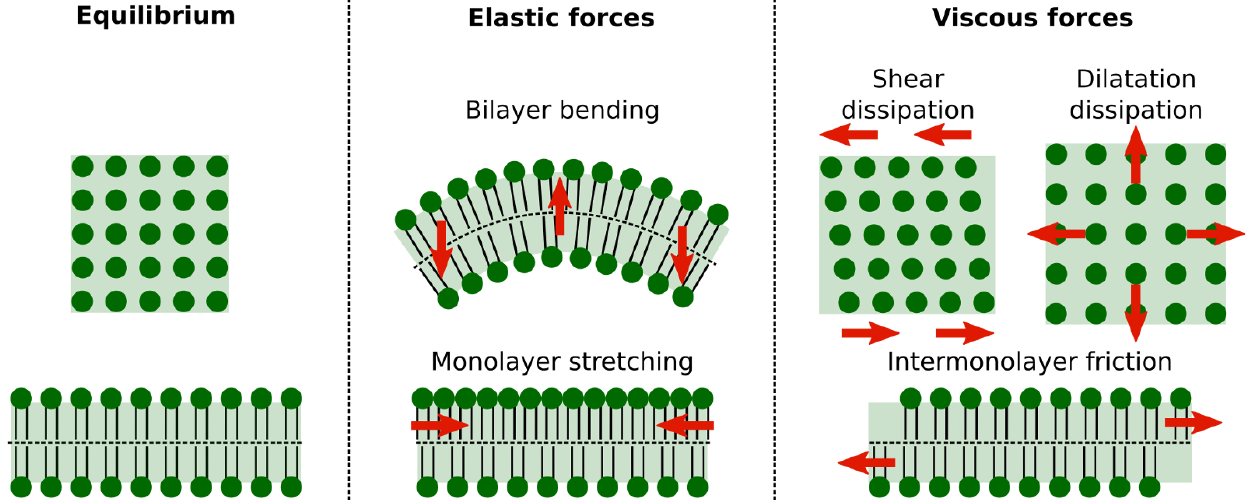}
		\caption{\label{eners} In a basic model incorporating elasticity and hydrodynamics \citep{Seifert1993-ld}, a lipid bilayer stores energy due to bending and monolayer stretching and dissipates energy through shear, dilatation and inter-monolayer friction.}
	\end{center}
\end{figure}
The previous model provides a first approach to the mechanics of lipid bilayers.
It is often overlooked, however, that by ignoring the bilayer architecture it fails to capture many important phenomena.
Seifert and Langer developed a continuum model explicitly accounting for the bilayer architecture and capturing the major energetic driving forces and dissipative drag forces involved in the dynamics of lipid membranes \citep{Seifert1993-ld}. The elastic forces in this theory appear in response to bending  of the membrane, as in the previous model, but also to monolayer stretching (see figure \ref{eners}). As viscous effects, the in-plane Newtonian rheology of the lipid bilayer \citep{Dimova2006-tu} is included through shear and dilatation dissipations, and the frictional coupling between the two monolayers opposing inter-monolayer slippage is also included. This model provided predictions about the relaxation dynamics of membrane fluctuations. Importantly, its material parameters can be experimentally measured \citep{Dimova2006-tu}. The work of \cite{Seifert1993-ld}, along with \citep{Evans1994-yg}, highlighted the role of inter-monolayer friction as a ``hidden'' but significant dissipative effect. This physical model was originally introduced and has been predominantly exercised under the restricted assumptions of linearized disturbances around a planar state \citep{Seifert1993-ld,Fournier2015-ov} or of simplified and fixed membrane shape \citep{Evans1994-yg}. These approximations, however, hide much of the interaction between shape dynamics and interfacial hydrodynamics, which is mediated by membrane curvature. This was demonstrated by the linearization of the theory about spherical or cylindrical configurations \citep{Rahimi2013-za} and by simulations based on a fully non-linear version of this theory, albeit axisymmetric \citep{Rahimi2012-sa}, which further demonstrated the geometry-dependent subtle interplay between all the ingredients in  figure \ref{eners} at multiple scales. Seifert and Langer's (SL) model is conceptually simple, captures sufficient physics to describe a plethora of dynamical phenomena, and can be the basis for more sophisticated dynamical models including for instance lipid tilt near molecular inclusions \citep{Hamm2000-tk, Hamm1998-up} or the physicochemical interaction of lipids with scaffolding or integral proteins \citep{Brochard-Wyart2002-tm,Arroyo2018-vd}.
Here we formulate and develop numerical calculations with this model in a three-dimensional and fully non-linear setup which, to our best knowledge, has not been examined before.
\begin{figure}
	\begin{center}
		\includegraphics[width=2.5in]{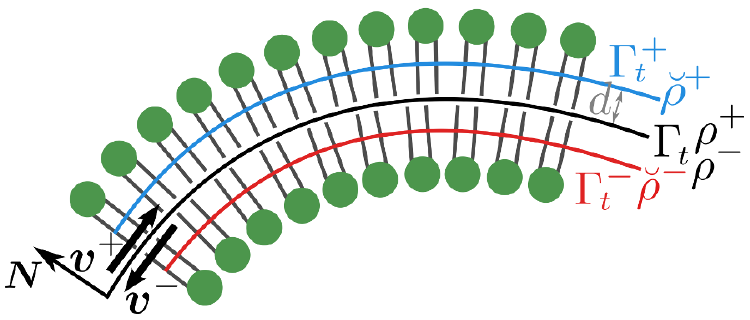}
		\caption{\label{bilayer} Sketch of relevant fields in the SL model. The densities at the monolayer midsurfaces $\breve{\rho}^\pm$ are projected onto the bilayer midsurface leading to the scalar fields $\rho^{\pm}:\Gamma_t\rightarrow \mathbb{R}$. The velocity fields $\bm{v}^\pm$ identify the velocity of the material particles at $\Gamma_t$.}
	\end{center}
\end{figure}
In this model, $\Gamma_t$ characterizes the bilayer mid-surface (see figure \ref{bilayer}).
In addition to the Helfrich energy of the form of equation \eqref{helfrich2}, the Seiftert-Langer model accounts for the stretching elasticity of each of the monolayers through the functional
\begin{equation}
\mathcal{F}_S\left[\bm{\chi},\rho^\pm\right] = \int_{\Gamma_t} \frac{k_S}{2}\left(\breve{\rho}^\pm-1\right)^2 dS = \int_{\Gamma_t} \frac{k_S}{2}\left[\rho^\pm(1\pm dH)-1\right]^2 dS,
\end{equation}
where the fields $\breve{\rho}^\pm$ represent the lipid density at the neutral surface of the upper (+) and lower (-) monolayers, measured in units of the equilibrium density, which differ from the lipid density at the bilayer mid-surface according to
\begin{equation}
\breve{\rho} = \rho^\pm(1\pm dH) + O(Kd^2),
\end{equation}
with $d\approx\SI{1}{\nano\meter}$ the distance between the neutral surfaces and the bilayer mid-surface. Unless otherwise noted, a functional containing $\pm$ implies a summation on the $+$ and $-$ monolayers. For convenience, in this section we use an Eulerian parametrization $\bm{\chi}$ to derive the governing equations. The free energy depends on $\bm{\chi}$ and the two density fields $\rho^+$ and $\rho^-$, representing the density of lipids on the mid-surface, and thus $X=\{\bm{\chi},\rho^+,\rho^-\}$. We take into account three main dissipation mechanisms in the bilayer. First we consider the dissipation due to in-plane shear in each monolayer, which takes the form
\begin{equation}
\mathcal{D}_S\left[\bm{\chi};v_n,\bm{v}^\pm\right] = \int_{\Gamma_t}\mu |\bm{d}^\pm|^2 dS,
\end{equation}
where $\mu$ is the shear viscosity and
\begin{equation}
\bm{d}^\pm = \frac{1}{2}\left\{\bnabla\left(\bm{v}^\pm\right) + \left[\bnabla\left(\bm{v}^\pm\right)\right]^T\right\} - v_n \bm{k},
\end{equation}
is the rate-of-deformation tensor for each monolayer (see equation \eqref{rateofdeformation}).  We consider three process variables, $v_n=\partial_t\bm{\chi}\bcdot\bm{N}$, which determines shape changes, and $\bm{v}^+$ and $\bm{v}^-$, which determine the tangential flow of lipids in each monolayer. Thus, $V=\{v_n,\bm{v}^+,\bm{v}^-\}$.
 Additionally, we consider a dilatational dissipation
\begin{equation}
\mathcal{D}_D\left[\bm{\chi};v_n,\bm{v}^\pm\right] = \frac{1}{2}\int_{\Gamma_t}\lambda \left(\text{tr}\bm{d}^\pm\right)^2 dS,
\end{equation}
where $\lambda$ is the dilatational viscosity. Finally, we consider the inter-monolayer friction caused by the relative slippage of one monolayer with respect to the other
\begin{equation}
\mathcal{D}_I\left[\bm{\chi};v_n,\bm{v}^\pm\right] = \int_{\Gamma_t}b_I |\bm{v}^+-\bm{v}^-|^2 dS,
\end{equation}
where $b_I$ is the inter-monolayer friction coefficient. Thus, the total dissipation is
\begin{equation}
\mathcal{D}\left[\bm{\chi};v_n,\bm{v}^\pm\right] = \mathcal{D}_S\left[\bm{\chi};v_n,\bm{v}^\pm\right] + \mathcal{D}_D\left[\bm{\chi};v_n,\bm{v}^\pm\right]+ \mathcal{D}_I\left[\bm{\chi};v_n,\bm{v}^\pm\right].
\end{equation}
The rate of change of the free energy is
\begin{equation}
\begin{aligned}
D_t\mathcal{F}_S\left[\bm{\chi};\rho^\pm;v_n,\partial_t\rho^\pm\right] = \int_{\Gamma_t} &k_S\left\{\left[ \rho^\pm(1\pm dH)-1\right] \times\vphantom{\frac{1}{2}}\right.\\
&\left[\partial_t\rho^\pm(1\pm dH) \pm d \rho^\pm \left(\Delta v_n+|\bm{k}|^2 v_n\right)\vphantom{\frac{1}{2}}\right. \\
&\left.- \frac{1}{2}\left[\rho^\pm(1\pm dH) - 1\right]v_nH\right]\\
&\left.+\frac{1}{2}\bnabla\bcdot\left( \left[ \rho^\pm(1\pm dH)-1\right]^2 \bm{v}^{\pm}\right) \right\}dS.
\end{aligned}
\end{equation}
Note carefully that $D_t\mathcal{F}_S\left[\bm{\chi};\rho^\pm;v_n,\partial_t\rho^\pm\right]$ depends on $\partial_t \rho^{\pm}$ rather than on the process variables $\bm{v}^\pm$. We invoke the equations encoding conservation of mass
\begin{equation}
\label{procrho}
\partial_t \rho^\pm = \Pi\left(\bm{\chi},\rho^\pm;v_n,\bm{v}^\pm\right) =-\bnabla\bcdot(\rho^\pm\bm{v}^\pm)-\rho^\pm v_n H,
\end{equation}
where $\Pi$ is referred to as a process operator, to express $D_t\mathcal{F}_S$ in terms of the process variables, in equal footing with $\mathcal{D}$, towards applying Onsager's formalism \citep{Rahimi2012-sa}. Process operators, usually linear operators, relate the rate of change of state variables, in this case ${\partial}_t \rho^\pm$, with process variables, $v_n$ and $\bm{v}^\pm$. In general, we write 
\begin{equation}
\label{procgen}
\dot{X} = \Pi(X) V. 
\end{equation}
In the previous model, the process operator was trivial $\dot{X}=V$ ($\partial_t\bm{\phi}=\bm{V}$). 
Using equation \eqref{procrho}, the rate of change of the energy can be written as
\begin{equation}
\begin{aligned}
D_t\mathcal{F}_S\left[\bm{\chi};\rho^\pm;v_n,\partial_t\rho^\pm\right] = \int_{\Gamma_t} &k_S\left\{\left[ \rho^\pm(1\pm dH)-1\right] \times\vphantom{\frac{1}{2}}\right.\\
&\left[\left[-\bnabla\left(\rho^\pm \bm{v}^\pm\right) +\rho^\pm v_n H\right](1\pm dH)\vphantom{\frac{1}{2}}\right. \\
&\left. \pm d \rho^\pm \left(\Delta v_n+|\bm{k}|^2 v_n\right)- \frac{1}{2}\left[\rho^\pm(1\pm dH) - 1\right]v_nH\right]\\
&\left.+\frac{1}{2}\bnabla\bcdot\left( \left[ \rho^\pm(1\pm dH)-1\right]^2 \bm{v}^{\pm}\right) \right\}dS,
\end{aligned}
\end{equation}
and the Lagrangian
\begin{equation}
\mathcal{L}\left[\bm{\chi},\rho^\pm;v_n,\bm{v}^\pm,P\right] = \mathcal{R}\left[\bm{\chi},\rho^\pm;v_n,\bm{v}^\pm\right] - P Q\left[\bm{\chi};v_n\right],
\end{equation}
where here
\begin{equation}
Q\left[\bm{\chi};v_n\right] = \int_{\Gamma_t}v_n dS.
\end{equation}
Then, Onsager's variational principle states that
\begin{equation}
\label{Onsager_bila}
\{v_n,\bm{v}^{\pm},P\} = \underset{\{u_n,\bm{u}^\pm\}}{\arg\min}~\underset{S}{\arg\max}\,\mathcal{L}\left[\bm{\chi},\rho^{\pm};u_n,\bm{u}^{\pm},S\right].
\end{equation}
The stationarity conditions issued from Onsager's principle provide equations for $P$ and the fields $v_n$ and $\bm{v}^\pm$. To find the time-evolution of the density fields $\rho^\pm$, the process operator (equation \eqref{procrho}) needs to be integrated in time. We stress that Onsager's variational principle provides directly the weak form of the problem, which can be directly discretized with finite elements. For completeness, we derive using Onsager's formalism the stress tensor and strong form of the governing equations of SL model, which to the best of our knowledge have not been presented before in the fully nonlinear case. The tangential and normal components of the stress of each monolayer can be identified as (see appendix \ref{weakFormSL})
\begin{equation}
\label{bilayerstress1}
\begin{aligned}
\bm{\sigma}^\pm =&~ k_S\left(\rho^\pm (1\pm dH)-1\right)\left(\frac{1}{2}\left[\rho^\pm (1\pm dH)+1\right]  \bm{g} \mp d\rho^\pm\bm{k}\right) \\
&+\frac{1}{2}\kappa H \left(\frac{1}{2}H \bm{g}-\bm{k}\right)  + 2\mu\bm{d}^\pm+\lambda \text{tr}\bm{d}^\pm \bm{g},
\end{aligned}
\end{equation}
and
\begin{equation}
\label{bilayerstress2}
\bm{\sigma}_n^\pm = \left[\frac{\kappa}{2}+ k_S\left(\rho^{\pm}d\right)^2\right] \bnabla H \pm  dk_S\left(2\rho^\pm(1\pm dH)-1\right)\bnabla\rho^\pm.
\end{equation}
Note that, aside from the terms involving $k_S$, the expressions are similar to those of previous model. Density imbalances generate a source of in-plane stress, but also lead to bending moments. Funthermore, the bending rigidity of the bilayer is $\kappa + 2 k_S(\rho^\pm d)^2$, which includes the effect of Helfrich and stretching energies. Balance of linear momentum tangent to the surface on the upper and lower monolayers reads
\begin{equation}
\begin{aligned}
\bnabla\bcdot\bm{\sigma}^+ + \bm{k}\bm{\sigma}^+_n&=b_I\left(\bm{v}^+-\bm{v}^-\right),\\
\bnabla\bcdot\bm{\sigma}^- + \bm{k}\bm{\sigma}^-_n&= b_I\left(\bm{v}^--\bm{v}^+\right),
\end{aligned}
\end{equation}
where $b_I\left(\bm{v}^+-\bm{v}^-\right)$ identifies the force exerted by the lower monolayer on the upper monolayer due to intermonolayer friction.
Finally, balance of linear momentum perpendicular to the bilayer leads to
\begin{equation}
\sum_{\pm}\left\{\bm{\sigma}^\pm \bm{:}\bm{k}-\bnabla\bcdot\bm{\sigma}_n^\pm \right\} =P.
\end{equation}
Seifert and Langer first introduced a linearized version of these equations around a planar state \citep{Seifert1993-ld}, which has been recently reviewed in the context of Onsager's principle \citep{Fournier2015-ov}. The stress tensors in equation \eqref{bilayerstress1} and equation \eqref{bilayerstress2} are similar to those found in \citep{Rahimi2012-sa}, using the Doyle-Ericksen formula of continuum mechanics. Our general and systematic derivation shows the ability of Onsager's formalism to derive complex models mixing different physics in a fully non-linear setting, which would otherwise be difficult to rationalize. For instance, although not unconceivable, it is difficult to postulate the constitutive relation for the in-plane stress in equation \eqref{bilayerstress1}. 

\subsection{The cell cortex: A viscous layer driven by active tension\label{cortexSec}}
\begin{figure}
	\begin{center}
		\includegraphics[width=5in]{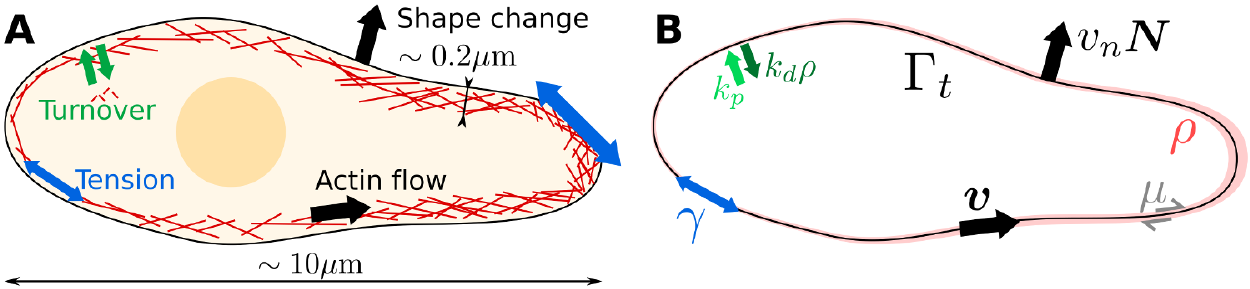}
		\caption{\label{cortex} In a simple model, the cortex is described as a surface $\Gamma_t$ with a space varying thickness $\rho$. Active tension and the turnover are taking cast into Onsager's formalism.}
	\end{center}
\end{figure}
The cell cortex is a layer of cross-linked actin filaments lying just beneath the plasma membrane of animal cells \citep{Bray1988-ak}. The thickness of this layer is of hundreds of nanometers, while the typical size of an animal cell is of tens of microns. Thus, this layer can be considered as a quasi two-dimensional material. In addition to actin, this network is crowded with polymerization regulators, cross-linkers, or myosin motors, which bind to actin filaments. By consuming ATP, these molecular motors pull on actin filaments and generate active tension. In turn, this active tension, if non-uniform, generates actin flows and drives shape changes \citep{Salbreux2012-fa}. Another important property of this actin network is that undergoes dynamic remodelling, with a continuous {turnover}   by polymerization and depolymerization of actin filaments and binding and unbinding of cross-linking proteins \citep{Howard2001-nl}. This process is characterized by a time-scale in the order of a few tens of seconds. At time-scales shorter than the turnover time, the cortex behaves as an elastic network. At longer time-scales, the dynamic remodelling of the cortex leads to a fluid-like viscous behaviour with active tension. 

Following previous works \citep{Turlier2014-wh,Prost2015-yl}, we consider an active gel model of the cortex as an isotropic viscous material with active tension confined to a surface and undergoing turnover, in which viscosity, active tension, and depolymerization depend on the thickness of the cortex. This model can describe phenomena at time-scales of a minute and longer, where elastic energy storage in the network becomes negligible, and does not account for the architecture of the network, e.g.~the orientation of the actin filaments, which may not be appropriate in some important examples such as during cytokinesis  \citep{Reymann2016-ti}. Furthermore, we assume that the viscous forces exerted by the cytosol and the external fluid medium are negligible. Using common estimates of cortex 2D viscosity ($\mu= 27\times 10^{-4}\SI{}{\pascal\second\meter}$ \citep{Bergert2015-my}), an estimate for the Saffman-Delbr\"uck length scale is $l_\text{DS}\approx\SI{3}{\meter}$. Thus, and given the size of cells, neglecting bulk viscosity is well-justified. \cite{Turlier2014-wh} and previous works \citep{Bergert2015-my,Saha2016-hg} were restricted to axisymmetric or to two-dimensional configurations, and derived the active gel equations from the stress tensor and force balance. Here, we develop a fully three-dimensional and geometrically non-linear version of this active gel model, and derive the governing equations using Onsager's formalism. 

Mathematically, we characterize the cortex as a fluid surface $\Gamma_t$, described here for the purpose of deriving the governing equations with a Lagrangian parametrization $\bm{\phi}$, with a space-varying thickness $\rho$, see figure \ref{cortex}. $\bm{\phi}$ and $\rho$ are our state variables. The process variable in this problem is the velocity field $\bm{V}$ of actin, with a tangential component $\bm{v}$, characterizing the flow of actin on $\Gamma_t$, and a normal component $v_n\bm{N}$ describing the change of shape of the actin cortex. The viscous rheology of the cortex 
is characterized by a dissipation potential, similar to that of lipid bilayers
\begin{equation}
\mathcal{D}[\bm{\phi},\rho;\bm{V}]=\int \mu \left[ |\bm{d}|^2 + (\text{tr}\bm{d})^2\right] \rho dS,
\end{equation}
where here $\mu$ is the bulk shear viscosity of the cortex. {This dissipation potential can be obtained by integrating over the thickness a three-dimensional shear dissipation potential $\int \mu |\bm{D}|^2dV$, with $\bm{D}$ the three-dimensional rate-of-deformation tensor, for an incompressible slab of gel with a plain stress assumption, i.e.~assuming that $\bm{D} = d_{ab}\bm{e}^a\otimes\bm{e}^b + D_{nn} \bm{n}\otimes\bm{n}$ and $D_{nn}=-\text{tr}\bm{d}$ \citep{Salbreux2009-nc,Turlier2014-wh}.} To introduce the active tension generated by the activity of myosin motors, we consider a power input of the form
\begin{equation}
\mathcal{P}[\bm{\phi},\rho;\bm{V}]=\int \xi(\rho) \rho\,\text{tr}\bm{d}~dS,
\end{equation}
where $\xi$ is a measure of myosin activity, which may depend on cortical density $\rho$, see discussion in section \ref{exCortex}. This leads to an active surface tension $\gamma = \xi(\rho) \rho$.  Since $\text {tr}\bm{d}$ measures the rate at which local area expands (positive $\text {tr}\bm{d}$) or contracts (negative $\text {tr}\bm{d}$), for a positive $\gamma$ the power input functional will drive the contraction of cortex area. As we neglect the elastic behaviour of the cortex, there is no free energy associated to the problem. Introducing a cell volume constraint, we obtain the Lagrangian
\begin{equation}
\mathcal{L}[\bm{\phi},\rho;\bm{V},P] = \mathcal{D}[\bm{\phi},\rho;\bm{V}] + \mathcal{P}[\bm{\phi},\rho;\bm{V}] - P Q_2[\bm{\phi};\bm{V}],
\end{equation}
and the dynamics follows from
\begin{equation}
\{\bm{V},P\} =\underset{\bm{U}}{\arg\min}~ \underset{S}{\arg\max}\,\mathcal{L}\left[\bm{\phi},\rho;\bm{U},S\right].
\end{equation}
From the Euler-Lagrange equations we identify the constitutive law 
\begin{equation}
\bm{\sigma} = \rho\left\{2\mu \bm{d} + \mu \text{tr}\bm{d} \,\bm{g} + \xi(\rho)\bm{g}\right\},
\end{equation}
and the statement of balance of linear momentum, this time in the absence of bending moments, 
\begin{equation}
\bnabla\bcdot\bm{\sigma}=\bm{0},\qquad \bm{\sigma}\bm{:}\bm{k} = P,
\end{equation}
with the last equation generalizing Laplace's law. To relate the rate of change of $\rho$ and $\bm{V}$, we consider balance of cortex material
\begin{equation}
D_t\rho + \rho \left( \bnabla\bcdot\bm{v}-v_nH \right) = k_p - k_d\rho,
\end{equation}
where the first term in the right hand side stands for actin polymerization, which, since polymerization nucleators are located at the plasma membrane, is assumed to occur at a constant rate $k_p$ independent of the thickness, and the second term stands for actin depolymerization, which is proportional to the local thickness, $k_d\rho$. The ratio $\rho_0=k_p/k_d$ determines the thickness at steady-state. By defining the characteristic turnover time as $\tau=1/k_d$, we can rewrite the previous equation as
\begin{equation}
D_t\rho + \rho \left( \bnabla\bcdot\bm{v}-v_nH \right) = (\rho_0-\rho)/\tau.
\end{equation}
\section{Discretization of the mechanics of fluid surfaces\label{computational}}

In this section, we introduce a general discretization framework for the simulation of fluid surfaces. First, we introduce a variational time-integrator based on Onsager's principle, which is unconditionally stable by construction. Then, we introduce the spatial discretization of the different fields on the time-evolving surface. We end with the derivation of the discrete equations for an inextensible fluid surface with bending elasticity as a reference example.

\subsection{Time discretization: Variational time-integrator based on Onsager's principle\label{timeint}}
To integrate in time the dynamics of continuum mechanical systems, a common approach is to first discretize in space, obtain a system of ordinary differential equations, which is then integrated in time with specialized algorithms. The fact that the dynamics in the models examined here emerge from a variational principle provides an alternative approach: to discretize in time the variational principle itself.  Time-integrators based on the discretization of a variational principle are usually referred to as {variational time-integrators}, and have been widely employed, for instance, for the discretization of Hamilton's principle in conservative systems including molecular dynamics \citep{Frenkel2001-sm} and elastodynamics \citep{Lew2004-st}, and in the context of dissipative systems \citep{Ortiz1999-fl,Peco2013-ws}. Variational time-integrators inherit qualitative properties of the associated time-continuous problem. For instance, in the case of time-integrators based on Hamilton's principle, Noether's theorem ensures that symmetries in the discrete action result in conserved currents as in the original continuous theory. Here, we propose a first order variational time-integrator for Onsager's principle that inherits that $\mathcal{F}$ is a Lyapunov functional of the dynamics, see equation \eqref{lyapunov1}. This feature provides nonlinear stability to the resulting discrete dynamics by construction.

We consider here a general statement of Onsager's variational principle, with a set of state variables $X$, a set of process variables $V$, obeying Onsager's principle in equation \eqref{OnsagerGeneral} and a process operator as in equation \eqref{procgen}. For simplicity, we neglect constraints in our discussion but they can be added by substituting the Rayleighian by the corresponding Lagrangian without changing the essence of the proposed variational integrator.
Let us consider a time discretization $\{t^1,\dots,t^N\}$ and let us start with a trivial process operator $\partial_t{X}=V$. We will consider here the simplest low order version of implicit variational time-integrator based on Onsager's principle, and leave the investigation of higher-order schemes to future work. We approximate $V^{n}=V(t^n)$ with a simple backward difference
\begin{equation}
V^{n+1} \approx \frac{X^{n+1} - X^{n}}{\Delta t^n},
\end{equation}
where $X^{n}=X(t^n)$ and $\Delta t^n = t^{n+1}-t^n$.
The dissipation potential and the power can now be approximated as
\begin{equation}
\begin{aligned}
\mathcal{D}(X;V) &\approx \mathcal{D}\left(X^n;\frac{X^{n+1}  - X^{n}}{\Delta t^n}\right),\\
\mathcal{P}(X;V) &\approx \mathcal{P}\left(X^n;\frac{X^{n+1}  - X^{n}}{\Delta t^n}\right).
\end{aligned}
\end{equation}
To discretize the Rayleighian, we also need to discretize the rate of change of the free energy. 
Rather than resorting to an expression like $\dot{\mathcal{F}}\approx D\mathcal{F}\bcdot\left(X^{n+1}-X^{n}\right)/\Delta t^n$, 
we consider
\begin{equation}
\label{meth_nonlin}
\dot{\mathcal{F}}(X,\partial_t{X}) \approx \frac{\mathcal{F}\left(X^{n+1}\right) - \mathcal{F}\left(X^{n}\right) }{\Delta t^n},
\end{equation}
or a similar higher-order finite difference.
This approach ensures that $\mathcal{F}$ is a Lyapunov functional of the dynamics, as we prove below, and retains the full non-linearity of $\mathcal{F}$ in the formulation. Using the previous expressions we define the discrete Rayleighian as
\begin{equation}
\label{rayleighian_dicrete_1}
\mathcal{R}^n\left(X^n;X^{n+1}\right) \equiv \frac{\mathcal{F}\left( X^{n+1} \right)}{\Delta t^n} + \mathcal{D}\left(X^n;\frac{X^{n+1}- X^{n}}{\Delta t^n}\right)\\+   \mathcal{P}\left(X^n;\frac{X^{n+1}  - X^{n}}{\Delta t^n}\right),
\end{equation}
where we have ignored the constant term $\mathcal{F}(X^n)/\Delta t^n$. Then, the incremental Onsager's principle is given by
\begin{equation}
\label{onsager_discrete_1}
X^{n+1} = \underset{X}{\arg\min}~\mathcal{R}^n\left(X^n;X\right).
\end{equation}
Thus, the dynamical problem arising from our variational time-discretization can be interpreted as an energy minimization problem for $\mathcal{F}$, which is usually a non-linear function of $X^{n+1}$, with the addition of a convex (and often quadratic) function of $X^{n+1}$, $\mathcal{D}$, subject to the external forces represented in $\mathcal{P}$. The weight of $\mathcal{F}$ relative to $\mathcal{D}$ is controlled by $\Delta t^n$, which can be decreased to ease the solvability of the problem by increasing the influence of the convex functional $\mathcal{D}$, or increased to allow the system to reach equilibrium faster. Let us now prove that, for a homogeneous problem ($\mathcal{P}(X;V)=0$), the free energy is a Lyapunov functional of the dynamics. We evaluate the Rayleighians
\begin{equation}
\begin{aligned}
\mathcal{R}^n\left(X^n;X^{n+1}\right) &= \frac{\mathcal{F}\left( X^{n+1} \right)}{\Delta t^n} + \mathcal{D}\left(X^n;\frac{X^{n+1}- X^{n}}{\Delta t^n}\right),\\
\mathcal{R}^n\left(X^n;X^n\right) &= \frac{\mathcal{F}\left( X^{n} \right)}{\Delta t^n} + \mathcal{D}\left(X^n;0\right) = \frac{\mathcal{F}\left( X^{n} \right)}{\Delta t^n},
\end{aligned}
\end{equation}
where we have used that $\mathcal{D}\left(X^n;0\right)=0$, as discussed in section \ref{inextmodel}. Since $X^{n+1}$ minimizes $\mathcal{R}^n$, it is clear that $\mathcal{R}^n\left(X^n;X^{n+1}\right)-\mathcal{R}^n\left(X^n;X^n\right) \leq 0$. Then,
\begin{equation}
\begin{aligned}
0&\geq\mathcal{R}^n\left(X^n;X^{n+1}\right)-\mathcal{R}^n\left(X^n;X^n\right) \\
&=  \frac{\mathcal{F}\left( X^{n+1} \right)-\mathcal{F}\left( X^{n} \right)}{\Delta t^n} + \mathcal{D}\left(X^n;\frac{X^{n+1}- X^{n}}{\Delta t^n}\right)\\
&\geq \frac{\mathcal{F}\left( X^{n+1} \right)-\mathcal{F}\left( X^{n} \right)}{\Delta t^n},
\end{aligned}
\end{equation}
where we have used that $\mathcal{D}\left(X^n;\frac{X^{n+1}- X^{n}}{\Delta t^n}\right)$ is positive. Therefore, we obtain
\begin{equation}
\mathcal{F}\left( X^{n+1} \right) \leq \mathcal{F}\left( X^{n} \right).
\end{equation}
which shows that {$\mathcal{F}$ is a Lyapunov functional of the discrete dynamics}. Thus, the time-step is not limited by stability, but rather by accuracy and solvability of the non-linear optimization problem in equation \eqref{onsager_discrete_1}, which becomes ``easier'' or ``more convex'' for small $\Delta t^n$. The ability to take stably large time-steps is particularly useful in stiff problems, such as those involving the Helfrich curvature energy.

When the process operator is not trivial, i.e.~$\partial_t{X}\neq V$, the approach above needs to be modified. For those cases, we can keep $V^{n+1}$ as the variable of the discrete Onsager's principle and discretize the process operator in different ways. As a first approach, we can consider a simple forward Euler approximation for the process operator
\begin{equation}
\label{discrete_process}
\partial_t{X} \approx \frac{X^{n+1}-X^n}{\Delta t^n} =\Pi \left(X^n\right)V^{n+1}\Longrightarrow X^{n+1} = X^{n} + \Delta t^n \Pi \left(X^n\right)V^{n+1}.
\end{equation}
We can then rewrite equation \eqref{meth_nonlin} as
\begin{equation}
\dot{\mathcal{F}} \approx \frac{\mathcal{F}\left(X^{n+1}\right) - \mathcal{F}(X^{n}) }{\Delta t^n} = \frac{\mathcal{F}\left( X^{n} + \Delta t^n \Pi \left(X^n\right)V^{n+1} \right) - \mathcal{F}(X^{n}) }{\Delta t^n}.
\end{equation}
This approximation still retains the non-linearity of $\mathcal{F}$ and is thus implicit in this sense. We can now define the Rayleghian as
\begin{equation}
\mathcal{R}^n(X^n;V^{n+1}) = \frac{\mathcal{F}\left( X^{n} + \Delta t^n \Pi (X^n)V^{n+1} \right)}{\Delta t^n} + \mathcal{D}\left(X^n;V^{n+1}\right) +  \mathcal{P}\left(X^n;V^{n+1}\right),
\end{equation}
and solve
\begin{equation}
\label{vartim1}
V^{n+1} = \underset{V}{\arg\min}~\mathcal{R}^n(X^n;V).
\end{equation}
Finally, we can recover $X^{n+1}$ from equation \eqref{discrete_process}. 
With this simple forward approximation for the process operator, however, the accuracy and stability of the integration can be very limited. As a better alternative, we consider a backward Euler approximation of the process operator, which involves solving
\begin{equation}
\label{discrete_process2}
X^{n+1} - X^{n} - \Delta t^n \Pi \left(X^{n+1}\right)V^{n+1}=0,
\end{equation}
together with the minimization of the Rayleighian
\begin{equation}
\begin{aligned}
\mathcal{R}^n(X^n,X^{n+1};V^{n+1}) = &~\frac{\mathcal{F}\left( X^{n} + \Delta t^n \Pi \left(X^{n+1}\right)V^{n+1} \right)}{\Delta t^n} + \mathcal{D}\left(X^n;V^{n+1}\right) \\
&+  \mathcal{P}\left(X^n;V^{n+1}\right).
\end{aligned}
\end{equation}
That is, one needs to solve the system
\begin{equation}
\label{vartim2}
\begin{aligned}
0&=X^{n+1} - X^{n} - \Delta t^n \Pi \left(X^{n+1}\right)V^{n+1},\\
V^{n+1} &= \underset{V}{\arg\min}~\mathcal{R}\left(X^n,X^{n+1};V\right),
\end{aligned}
\end{equation}
 for $X^{n+1}$ and $V^{n+1}$ simultaneously. It is easily shown that with any of these discretizations, $\mathcal{F}$ is also a Lyapunov function of the dynamics in the absence of power input, thus retaining the nonlinear stability of the time-discretization scheme.
  
\subsection{Spatial discretization\label{spacedis}}
\begin{figure}
	\begin{center}
		\includegraphics[width=4in]{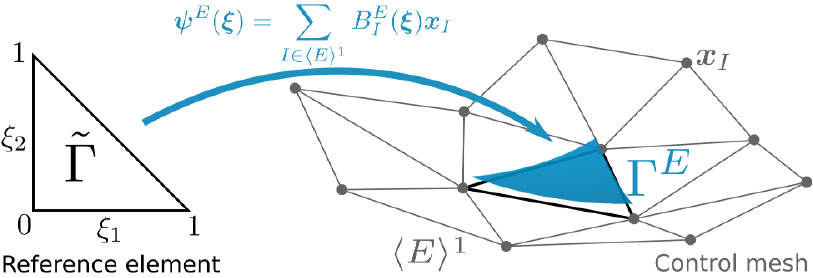}
		\caption{\label{trianchart} In subdivision surfaces, a control mesh is used to parametrize the surface $\Gamma$. For each triangle in the control mesh, $E$, the mapping equation \eqref{subdiv1}, depending on the control points of the first ring of neighbours to $E$, $\bm{x}_I$ with $I\in\langle E\rangle^1$, define the surface $\Gamma^E$ (blue).  The union of $\Gamma^E$ for each $E$ in the control mesh forms the $H^2$ surface $\Gamma$.}
	\end{center}
\end{figure}

In this section, we examine the spatial discretization of $\Gamma_t$ and the different fields defined on it. For simplicity, let us start by examining the numerical parametrization of a generic surface $\Gamma$. We first note that, since models for fluid surfaces usually involve the shape operator $\bm{k}$, this tensor needs to be square-integrable on $\Gamma$. For that reason, the parametrization of $\Gamma$ must be a square-integrable function with square-integrable first- and second-order derivatives; we call such a surface a $H^2$ surface. The problem of discretizing a $H^2$ surface may be addressed resorting to different numerical frameworks, such as higher-order B-splines as in isogeometric methods \citep{Piegl2012-db,Sauer2017-oc} or max-ent approximants \citep{Millan2011-lg}. Another versatile technique to discretize smooth surfaces based on meshes with arbitrary connectivity is {subdivision surfaces}. Here we focus on {Loop subdivision surfaces} based on triangular meshes \citep{Loop1987-of,Stam1999-fk,Biermann2000-bz,Cirak2000-tq, Cirak2001-bz, Cirak2011-wf,Torres-Sanchez2017-ay}.
To define the discretization of $\Gamma$ with subdivision surfaces, we consider a control mesh made of triangles $E=1,\dots,N_e$ whose edges join the set of control points with positions $\{\bm{x}_I\}_{I=1}^{N_n}$. For each triangle in the mesh, we define the parametrization $\bm{\psi}^E(\bm{\xi}) :\tilde{\Gamma}\rightarrow\mathbb{R}^3$, with $\tilde{\Gamma}$ the reference triangle (see figure \ref{trianchart}), by
\begin{equation}
\label{subdiv1}
\bm{\psi}^E(\bm{\xi}) = \sum_{I\in\langle E\rangle^1}  \bm{x}_{I} B_I^E(\bm{\xi}),
\end{equation}
where $B_I^E$ represents the {subdivision basis function} associated to node $I$ at element $E$ and $\langle E \rangle^1$ identifies the first ring of nodes surrounding the element, including the nodes forming the element and all first neighbours to them.  We denote by  $\Gamma^E=\bm{\psi}^E\left(\tilde{\Gamma}\right)$ the curved triangle obtained by the local parametrization in equation \eqref{subdiv1}. It can be shown that these curved triangles are disjoint (except at the edges) and that their union $\Gamma=\cup_{E=1}^{N_e} \Gamma^E$ defines a 
$C^2$-continuous surface almost everywhere, except at a finite number of points where it is $C^1$. These points coincide with the image of irregular nodes in the control mesh, which are those with a connectivity different from 6. There, the surface is continuous with continuous derivative but presents a discontinuity in the second derivative. 
Thus, from the perspective of differential geometry, the set $\cup_{E=1}^{N_e} \left(E,\tilde{\Gamma},\psi^E\right)$ defines an atlas of charts that parametrize the $H^2$ surface $\Gamma$. 

Now we consider ALE parametrizations of the form derived in section \ref{aleparam}. For the parametrization of the surface $\Gamma_{t_0}$, we write
\begin{equation}
\bm{\psi}_{0}^E(\bm{\xi}) = \sum_{I\in \langle E \rangle^1} \bm{x}_{0I} B_I^E(\bm{\xi}) .
\end{equation}
Thus, the control mesh in our scheme is given by the position of the control points $\left\{\bm{x}_{0I}\right\}_{I=1}^{N_n}$. We also define the fields
\begin{align}
h^E(\bm{\xi},t)&=\sum_{I\in \langle E \rangle^1} h_I(t) B_I^E(\bm{\xi}) ,\\
\bm{M}^E(\bm{\xi})&=\sum_{I\in \langle E \rangle^1} \bm{M}_I B_I^E(\bm{\xi}).\label{directors}
\end{align}
The parametrization of the deformed surface $\Gamma_t$ then reads
\begin{equation}
\label{discrete_param}
\bm{\psi}^E(\bm{\xi},t) = \sum_{I\in\langle E \rangle^1}  \bm{x}_{0I} B_I^E(\bm{\xi}) + \left(\sum_{J\in\langle E \rangle^1} h_J(t) B_J^E(\bm{\xi}) \right)\left(\sum_{K\in\langle E \rangle^1} \bm{M}_K B_K^E(\bm{\xi}) \right).
\end{equation}
We note that, given that $\bm{\psi}_0$, $h$ and $\bm{M}$ are in $H^2$, $\bm{\psi}$ is also in $H^2$. 
We also note that, if we had used the normal to the reference surface $\bm{N}_0$ instead of $\bm{M}$, because the calculation of $\bm{N}_0$ already involves first order derivatives of $\bm{\psi}_0$, we would need $\Gamma_{t_0}$ be $C^2$ everywhere, which cannot be achieved with subdivision surfaces. This is the reason why we choose the field of directors as in equation (\ref{directors}), where $\bm{M}_I$ can be chosen to approximate the true field of normals, for instance in a least-squares sense.

We can consider other kinds of basis functions. In particular, we consider the set of linear basis functions $N_I(\bm{\xi})$ with $I\in\langle E \rangle^0$, the zeroth-ring of nodes of the element, defined by
\begin{equation}
\label{linearelements}
 N_{E_1} = (1-\xi_1,1-\xi_2)\quad N_{E_2} = (\xi_1,0), \quad N_{E_3} = (0,\xi_2),
\end{equation}
where $E_1,E_2$ and $E_3$ denote the labels of the three nodes forming the element $E$. 
We can then discretize fields on $\Gamma_t$ with $N_I$ if they only need to be in $H^1$ (that is, square-integrable functions with a square-integrable derivative). For instance, a density field, which only appears in the free energy and dissipation potentials through its value and first order derivatives, can be discretized as 
\begin{equation}
\rho=\sum_{I\in \langle E \rangle^0}  \rho_IN_I^{E}.
\end{equation}
For notational simplicity in this and following equations, we write $N_I^E$ where we should write $N_I^{E}\circ\left(\bm{\psi}^{E}\right)^{-1}$ since $N_I^E$ take values in the parametric domain $\tilde{\Gamma}$ whereas $\rho$ is a field on $\Gamma_t$.
Following the same arguments, we could be tempted to discretize the components of $\bm{v}$ as
\begin{equation}
v^{a}=\sum_{I\in \langle E \rangle^0} v^a_{I}  N_I^{E},
\end{equation}
since the Rayleighian also depends on $\bm{v}$ through its value and its first-order derivatives only.
This, however, requires that a basis $\bm{e}_a$, continuous and with first square-integrable derivatives, is defined everywhere on the surface so that
$\bm{v} = v^a\bm{e}_a$ is continuous and with square-integrable derivatives. 
However, one cannot define such a basis for a closed surface as a consequence of the hairy ball theorem (for instance, polar coordinates in the sphere present singularities at the poles). 
Using the canonical basis of the parametrization, we could try to discretize
$
\bm{v}= v^{a} \partial_a \bm{\psi}^{E},
$
but $\partial_a \bm{\psi}^{E}$ is discontinuous across elements due to the jump in the definition of local coordinates.  A possible solution to this problem is to increase the number of degrees of freedom used to describe $\bm{v}$ and discretize the three components of $\bm{v}$ in the global basis of Euclidean space
\begin{equation}
\bm{v} = v^\alpha \bm{i}_\alpha.
\end{equation}
Being the basis vectors $\bm{i}_\alpha$ constant, we could discretize
\begin{equation}
v^{\alpha}=\sum_{I\in \langle E \rangle^0}  v_{I}^\alpha  N_I^{E} .
\end{equation}
However, $\bm{v}$ being tangent to $\Gamma_t$, its three components $v^1,v^2$ and $v^3$ are not independent and one would need to introduce the additional constraint
$\bm{v}^E\bcdot\bm{N}=0$ such as in \citep{Fries2018-lv, Reuther2018-ky}. 
A more convenient option is to recall the Hodge decomposition of $\bm{v}$ in equation \eqref{velhodge} and discretize the scalar fields $\alpha$ and $\beta$. We note that $\alpha$ and $\beta$ need  to be in $H^2$  for $\bm{d}$ to be well-defined, and for this reason we use subdivision basis functions to discretize them
\begin{align}
\label{disAlpha}
\alpha&=\sum_{I\in \langle E \rangle^1}\alpha_I B_I^{E},\\
\label{disBeta}
\beta&=\sum_{I\in \langle E \rangle^1} \beta_I  B_I^{E}.
\end{align}

Apart from $h$ and the vector potentials $\alpha$ and $\beta$, in some models we need to discretize Lagrange multiplier fields, such as the surface tension $\gamma$. Since $\gamma$ acts as a Lagrange multiplier, the space of basis functions for $\gamma$ needs to be chosen with care to ensure that the discretization satisfies the discrete inf-sup condition \citep{Brezzi2012-ol}.
\begin{figure}
	\begin{center}
		\includegraphics[width=3.5in]{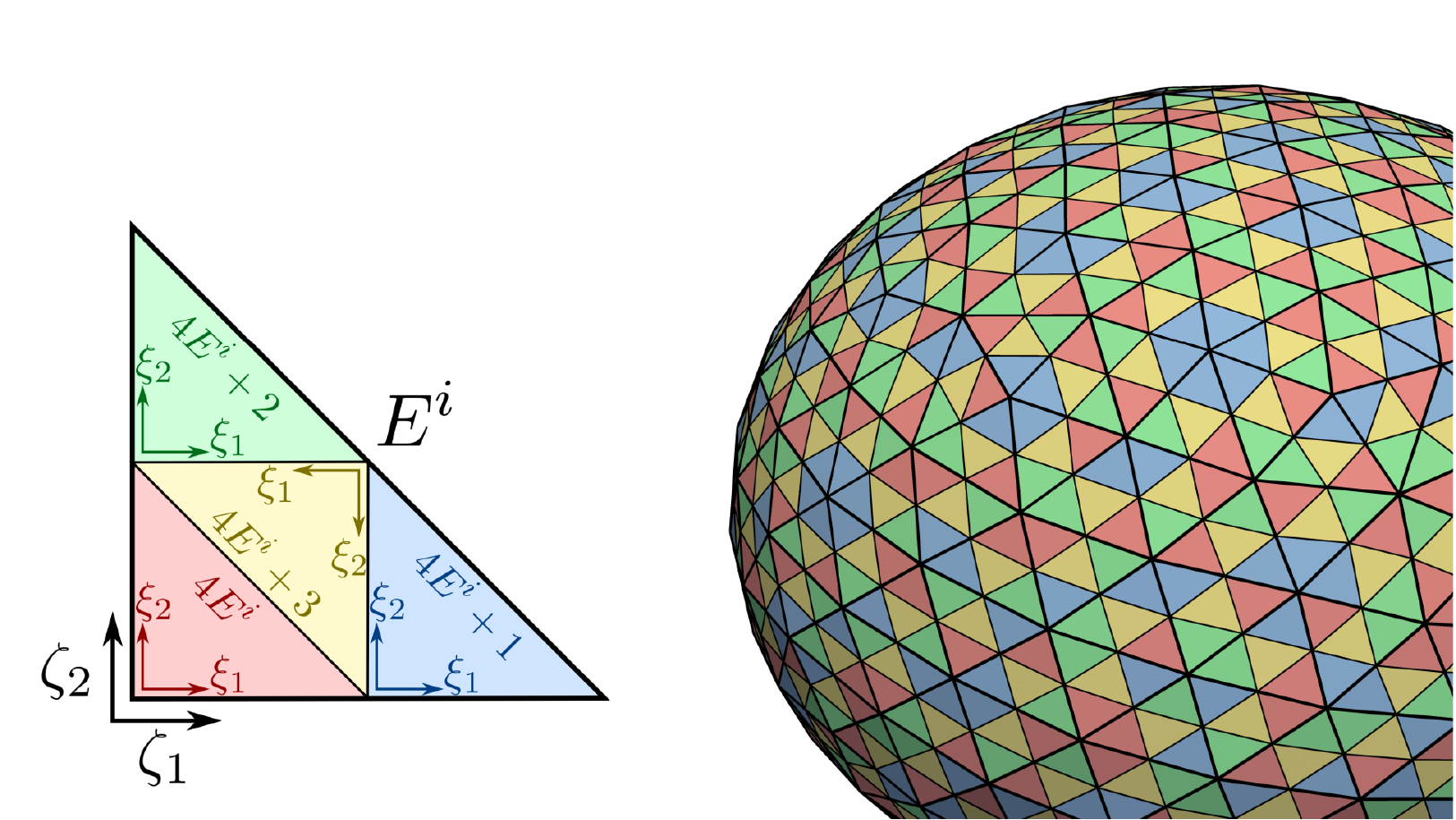}
		\caption{\label{dualmesh} To discretize the surface tension, a Lagrange multiplier field enforcing local inextensibility, we consider a macro-element approach to build compatible finite element spaces for velocities and surface tensions. Each element in the coarser mesh is subdivided into 4 finer elements. The subdivision mapping can be defined locally in parametric space without resorting to the positions of the nodes in the mesh. Both parametric spaces, corresponding to the coarser and finer meshes, cover the surface $\Gamma$ with triangles.}
	\end{center}
\end{figure}
Similarly to previous works in isogeometric analysis \citep{Dortdivanlioglu2018-js}, we consider a macro-element approach where Lagrange multipliers are approximated using a coarsened mesh. We consider a coarse mesh of triangles $E^c\in 1,\dots,N_e^c$, and subdivide each triangle $E^c$ into four triangles $(4E^c,4E^c+1,4E^c+2,4E^c+3)$; the nodes of these meshes are $N_n^c$ and $N_n^f$ respectively. This subdivision can be regarded as a map between the parametric domains of two atlases $\cup_{E=1}^{N_e^c}\left(E,\tilde{\Gamma}\right)$ and $\cup_{E=1}^{N_e^f}\left(E,\tilde{\Gamma}\right)$, with $N_e^f=4N_e^c$, defined by
\begin{equation}
\left(E^f,\bm{\xi}\right) = O\left(E^c,\bm{\zeta}\right)=\left\{\begin{array}{ll}
\left(4E^c,2\bm{\zeta}\right)&\text{if } \zeta_1+\zeta_2 < \frac{1}{2},\\
\left(4E^c+1,2\bm{\zeta}-\left(1,0\right)\right) &\text{if } \zeta_1>\frac{1}{2},\\
\left(4E^c+2,2\bm{\zeta}-\left(0,1\right)\right)&\text{if } \zeta_2>\frac{1}{2},\\
\left(4E^c+3,\left(1,1\right)-2\bm{\zeta}\right)&\text{otherwise},
\end{array}\right.
\end{equation}
see figure \ref{dualmesh}. We note that the function $O$ does not depend on the positions of the nodes of the mesh. 
The finer atlas $\cup_{E^f=1}^{N_e^f}\left(E^f,\tilde{\Gamma}\right)$ is then used to discretize the geometry of the surface and the vector potentials, following Eqs.~\eqref{discrete_param}, \eqref{disAlpha} and \eqref{disBeta}. 
On the other hand, we discretize Lagrange multiplier fields with linear elements defined by equation \eqref{linearelements} in the coarser atlas
\begin{equation}
\gamma = \sum_{I\in \langle E^c \rangle^0} \gamma_IN_I^{E^c}.
\end{equation}

\subsection{Finite element formulation of Onsager's principle\label{sec:finel}}

Here, we show the application of our methodology, based on the variational time-integrator described in section \ref{timeint} and on the space discretization described in section \ref{spacedis}, to the model of an inextensible viscous fluid surface with bending elasticity (section \ref{inextmodel}). We define the following vectors of nodal coefficients
\begin{equation}
\mathsf{h} = \begin{pmatrix}
h_1\\
\vdots\\
h_{N_n^f}\\
\end{pmatrix},\qquad 
\mathsf{a} = \begin{pmatrix}
\alpha_1\\
\vdots\\
\alpha_{N_n^f}\\
\end{pmatrix},\qquad 
\mathsf{b} = \begin{pmatrix}
\beta_1\\
\vdots\\
\beta_{N_n^f}\\
\end{pmatrix},\qquad
\mathsf{s} = \begin{pmatrix}
\mathsf{\gamma}_1\\
\vdots\\
\mathsf{\gamma}_{N_n^c}\\
\end{pmatrix},
\end{equation}
containing the degrees of freedom describing the offset, the irrotational and solenoidal vector potentials, and the surface tension. The discrete Lagrangian, now a function, can then be written as 
\begin{equation}
\label{disLag}
\begin{aligned}
\mathcal{L}^n(\mathsf{h},\mathsf{a},\mathsf{b},\mathsf{s}) =&~ \frac{1}{\Delta t^n} \mathcal{F}_H(\mathsf{h}) + \frac{1}{2} 
\begin{pmatrix}
\left(\mathsf{h}-\mathsf{h}^n\right)^T&
\mathsf{a}^T &
\mathsf{b}^T
\end{pmatrix}
\begin{pmatrix}
\mathsf{D}_{\mathsf{h}\mathsf{h}} & \mathsf{D}_{\mathsf{h}\mathsf{a}} & \mathsf{D}_{\mathsf{h}\mathsf{b}}\\
\mathsf{D}_{\mathsf{h}\mathsf{a}}^T & \mathsf{D}_{\mathsf{a}\mathsf{a}} & \mathsf{D}_{\mathsf{a}\mathsf{b}}\\
\mathsf{D}_{\mathsf{h}\mathsf{b}}^T & \mathsf{D}_{\mathsf{a}\mathsf{b}}^T & \mathsf{D}_{\mathsf{b}\mathsf{b}}\\
\end{pmatrix}
\begin{pmatrix}
\mathsf{h}-\mathsf{h}^n\\
\mathsf{a}\\
\mathsf{b}\\
\end{pmatrix}\\
&+ \mathsf{s}^T\begin{pmatrix}\mathsf{Q}_\mathsf{h}& \mathsf{Q}_\mathsf{a}\end{pmatrix}
\begin{pmatrix}
\frac{\mathsf{h}-\mathsf{h}^n}{\Delta t^n}\\
\mathsf{a}
\end{pmatrix} -\frac{1}{\Delta t^n}P\left[\Omega(\mathsf{h})-\Omega^n\right],
\end{aligned}
\end{equation}
where the explicit form for the different terms can be found in appendix \ref{appDisc}.
The discrete version of Onsager's variational principle then leads to the saddle-point problem
\begin{equation}
\{\textsf{h}^{n+1},\textsf{a}^{n+1},\textsf{b}^{n+1},\textsf{s}^{n+1},P^{n+1}\} =  \underset{\textsf{l},\textsf{e},\textsf{f}}{\arg\min}~\underset{\textsf{t},S}{\arg\max}\,\mathcal{L}\left(\mathsf{l},\mathsf{e},\mathsf{f},\mathsf{t},S\right),
\end{equation}
the stationarity conditions of which form a non-linear algebraic system of equations, which we solve using Newton's method.

\subsection{Restricting rigid body motion in simulations}

The simulation of fluid surfaces lacking of interaction with the surrounding viscous fluid requires of restricting rigid body motions of the interface since these do not dissipate energy or affect the free energy of the system. To restrict these motions, we impose three translational constraints 
\begin{equation}
\int_{\Gamma_t} \partial_t h \bm{M} dS = \bm{0}.
\end{equation}
and three rotational constraints
\begin{equation}
\int_{\Gamma_t} \bm{x}\times \bm{V} dS = \bm{0},
\end{equation}
using six additional Lagrange multipliers.

\subsection{Mass conservation: Stabilized finite element formulation}

We now address the discretization of mass conservation, which is required in the Seifert-Langer model of lipid bilayers as well as for the simulation of the cell cortex. Since we consider an ALE parametrization, we need to discretize the ALE version of equation \eqref{masscons}. We consider an implicit backward Euler scheme in time for this advection-reaction equation. For its space discretization, we consider a stream-upwind Petrov Garlerkin (SUPG) method \citep{Donea2003-oa}, which treats the convective term by adding controlled numerical diffusion in a consistent manner. The equations of conservation of mass and balance of linear momentum  are solved monolithically using Newton's method. See appendix \ref{dissmasscons} for details.

\section{Representative simulations of fluid surfaces\label{examples}}

In this section, we revisit the models developed in section \ref{models}. We use the numerical framework described in the previous section to simulate these models under different conditions, which exemplify the mechanical behaviour of lipid bilayers and the cell cortex.

\subsection{Lipid bilayers:  An inextensible viscous layer with bending energy}
\begin{figure}
	\begin{center}
		\includegraphics[width=3.5in]{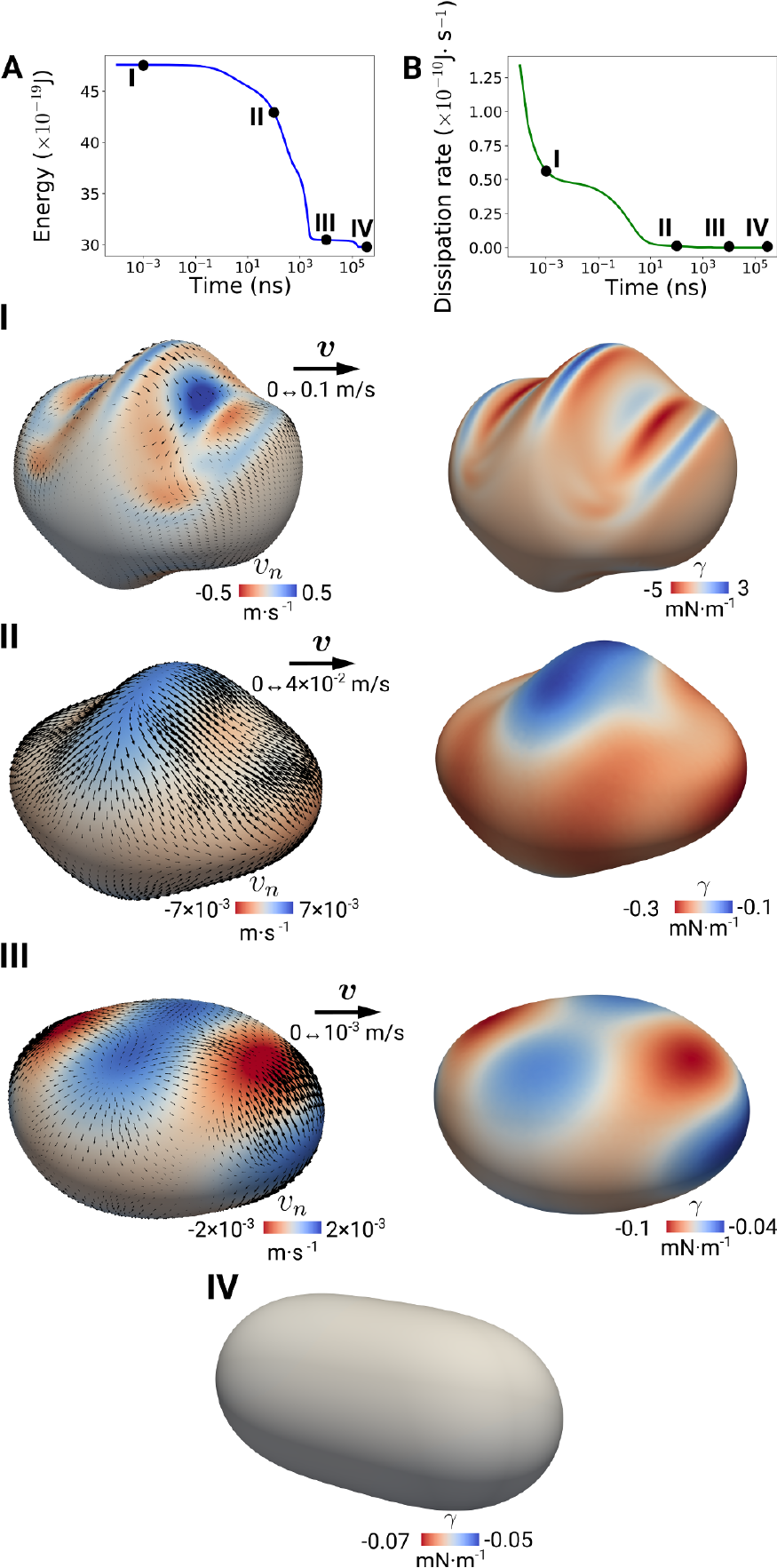}
		\caption{\label{relaxation1} Relaxation dynamics of an inextensible viscous layer with bending elasticity. (A) Helfrich energy as a function of time. (B) Shear dissipation as a function of time. Snapshots I-IV represent different stages of the dynamics. In the left panel, we plot the normal (colormap) and tangential (arrows) components of the velocity. In the right panel we plot the Lagrange parameter $\gamma$, representing the contribution to surface tension of the inextensibility constraint. }
	\end{center}
\end{figure}
\begin{figure}
	\begin{center}
		\includegraphics[width=5.25in]{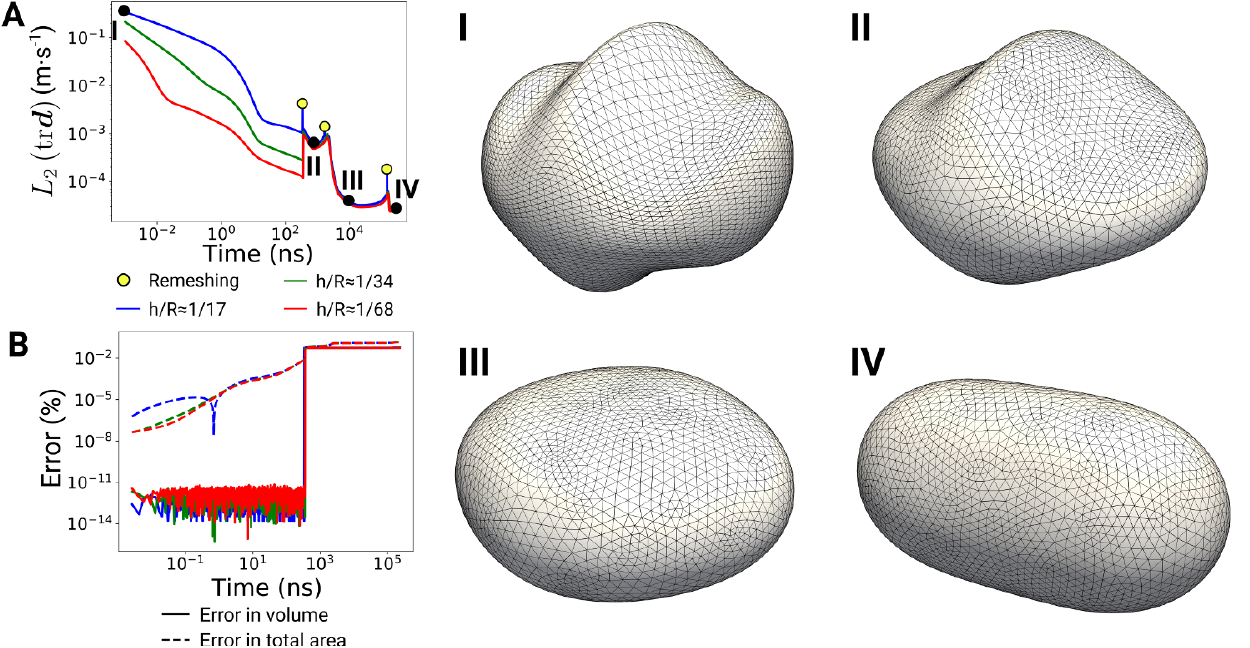}
		\caption{\label{relaxation2} (A) $L_2$ norm of $\text{tr}\bm{d}$ for different refinement levels; blue $h/R\approx1/17$, green $h/R\approx1/34$, red $h/R\approx1/68$. (B) Error committed in the conservation of volume and total mass (solid and dashed curves, respectively). In (I-IV) we plot the meshes used for the coarser subdivision level.}
	\end{center}
\end{figure}

\textbf{Example 1: Relaxation dynamics from a non-equilibrium non-axisymmetric shape.}
We first simulate the behaviour of an inextensible viscous layer with curvature elasticity as a first approach to model the elasto-hydrodynamics of lipid bilayers. To test the performance of the numerical methods described in the previous section, we first examine the relaxation of an out-of-equilibrium (and non-axisymmetric) shape given by 
\begin{equation}
\bm{x}(\varphi,\theta) = \begin{pmatrix}
R\sin\theta\cos\varphi\\
 R\sin\theta\sin\varphi\\
  \lambda_1 R\cos\theta  (1+\lambda_2\cos(2\pi\sin\theta\sin\varphi))
\end{pmatrix}
\end{equation}
with $R=\SI{100}{\micro\meter}$, $\lambda_1 = 0.7$ and $\lambda_2=0.3$ (see figure \ref{relaxation1}). Using common estimates for the model parameters \citep{Dimova2006-tu,Rahimi2012-sa}, we choose $\kappa=10^{-19}\SI{}{\joule}$, $\mu =  10^{-9}\SI{}{\joule\second\meter}^{-2}$. As expected for a dissipative system in the absence of external inputs, the free energy $\mathcal{F}$ decreases monotonically with time (figure \ref{relaxation1}A) by dissipating energy (figure \ref{relaxation1}B).  Note that, because of the semi-logarithmic scale, it is difficult to appreciate in Figs.~\ref{relaxation1}A and B that the negative of the rate of change of free energy is equal to the rate of dissipation.
 The initial shape given in figure \ref{relaxation1}I relaxes through different non-equilibrium states, figure \ref{relaxation1}II and figure \ref{relaxation1}III, until reaching the final equilibrium shape, figure \ref{relaxation1}IV. In the left panel of Figs.~\ref{relaxation1}I-IV we show the velocity field, which has been split for visualization purposes into its normal (colormap) and tangential (arrows) components. In the right panel of these figures, we show the Lagrange multiplier field $\gamma$ representing the contribution to surface tension of the inextensibility constraint, which shows a smooth behaviour, suggesting that the macro-element approach described in section \ref{spacedis} satisfies the discrete inf-sup condition; a more detailed study of this specific will be presented elsewhere. 
 In figure \ref{relaxation2}A, we compute the $L_2$ norm of $\text{tr}\bm{d}$ for three different levels of refinement, marked in blue (with average triangle side  $h/R\approx 1/17$, $N_e=8128$ and $N_n=4066$), green  ($h/R\approx 1/34$, $N_e=32512$ and $N_n=16258$) and red ($h/R\approx1/68$, $N_e=130048$ and $N_n=65026$) as a function of time, which measures the error committed in the enforcement of inextensibility.  {Initially, we observe that the error converges linearly in a log-log scale as the mesh is refined.} Even though we use an ALE method to reduce mesh distortion, the dramatic shape changes during the relaxation dynamics require four full remeshing operations, which are marked with yellow circles in Fig. \ref{relaxation2}A. The resulting meshes are shown in panel of figure \ref{relaxation2}I-IV and in Movie 1 for the coarser refinement level. To remesh, we follow a three-step procedure. First, we update the reference surface $\Gamma_{t_0}$ to $\Gamma_t$ using least-squares. This reference surface serves then  as a seed for the remeshing algorithm implemented in the VMTK library \citep{Antiga2008-in}, which assigns an element area following
\begin{equation}
A = \frac{1}{1/A_0+c_s |\bm{k}|^2},
\end{equation}
where $A_0$ is a reference element area for a planar patch, and $c_s$ specifies the sensitivity to curvature (in these simulations, $c_s=0.1$). 
 Finally, another least squares fit is performed to parametrize the new surface to the initial one $\Gamma_t$, which finally sets the new $\Gamma_{t_0}$. We note that the first least-squares fit is only performed to give a seed to the meshing algorithm; the essential least-squares fit, using the initial shape as a seed to fit the geometry of a parametrization based on the new mesh, is performed after remeshing.  We observe that remeshing increases the error associated to local inextensibility noticeably, but this error remains small. Thus example illustrates the benefit of the ALE method to reduce the frequency of remeshing events. We finally note that the relative error in total area and volume conservation is smaller than 0.1\% over the whole dynamics, see figure \ref{relaxation2}B. The error in volume conservation is very small ($<10^{-11}\%$) until the first remeshing step, where the error presents a jump. This illustrates the success of our non-linear method to impose volume conservation, see section \ref{sec:finel}. On the other hand, it shows the lack of explicit control on volume (and area) conservation during remeshing, which could be incorporated into the least-squares procedure underlying remeshing. Errors in area conservation are smoother in time and larger in magnitude, since it is imposed weakly in terms of local area conservation based on the discretization of $\text{tr}\bm{d}$ and the Lagrange multiplier $\gamma$.

\begin{figure}
	\begin{center}
		\includegraphics[width=3.75in]{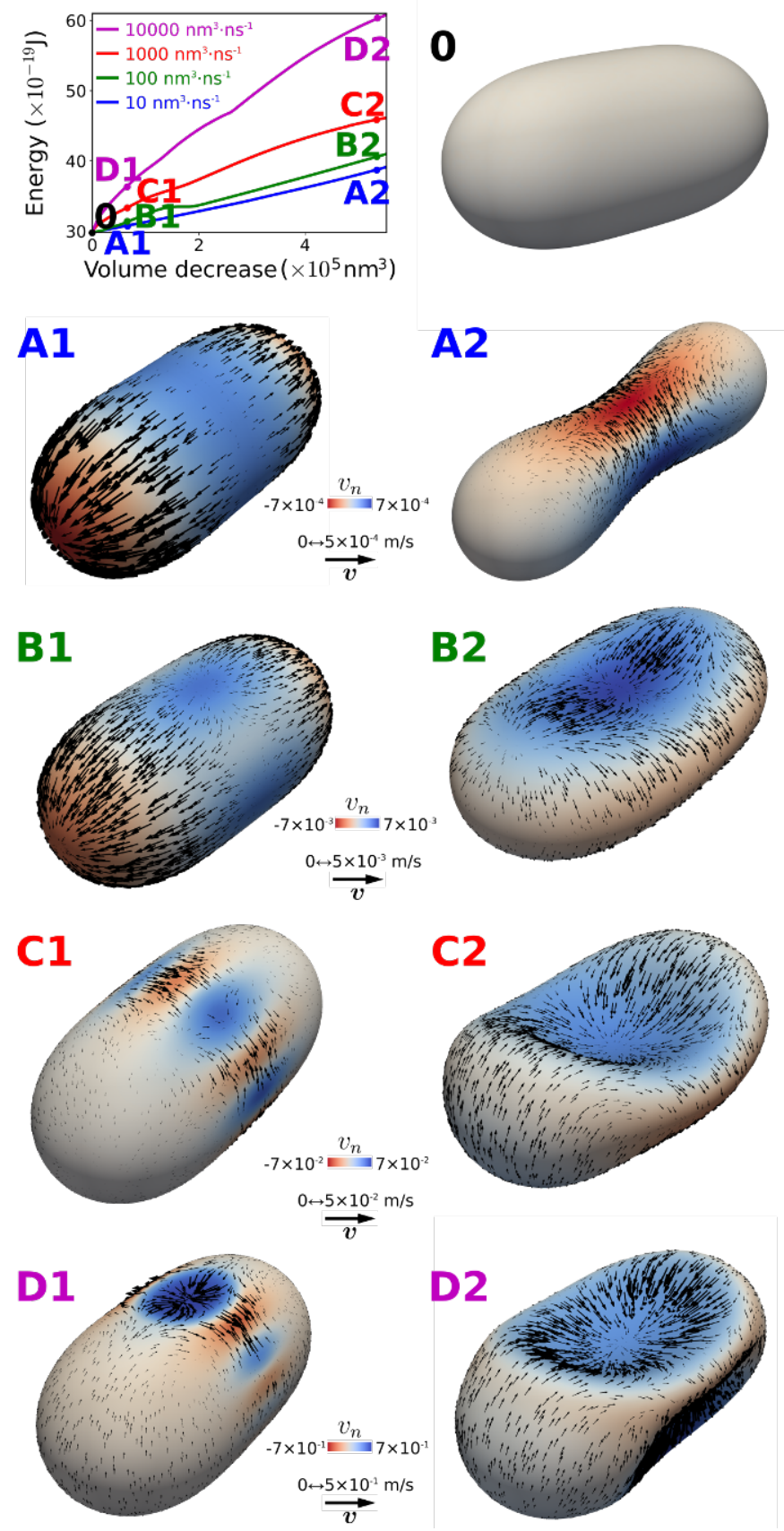}
		\caption{\label{osm_shock} Bending energy as a function of volume decreases during the deflation of a vesicle at different deflation rates; blue $\SI{10}{nm}^3\SI{}{\nano\second}^{-1}$, green $\SI{100}{nm}^3\SI{}{\nano\second}^{-1}$, red $\SI{1000}{nm}^3\SI{}{\nano\second}^{-1}$, magenta $\SI{10000}{nm}^3\SI{}{\nano\second}^{-1}$. Snapshots show the vesicle shape and normal and tangential velocities (colormap and arrows respectively) for the different deflation rates.}
	\end{center}
\end{figure}

\textbf{Example 2: Dynamics following hyper-osmotic shocks.}
As a second example, we examine the effect of osmotic shocks in vesicles. Cells and vesicles are often exposed to changes in the inner and outer chemical composition, which create flows of water through the semipermeable lipid membrane, increasing or decreasing their enclosed volume, and generating shape changes \citep{Staykova2013-fu,Kosmalska2015-ow}. Here, we simulate the effect of a hyper-osmotic shock by decreasing the enclosed volume at different deflation rates.  We start with the equilibrium shape of the previous example using the finest mesh (figure \ref{osm_shock}-0), and apply a deflation rate of $\SI{10}{\nano\meter}^3 \SI{}{\nano\second}^{-1}$. In a plot comparing the elastic energy stored during  deflation and the total volume decrease (blue curve in figure \ref{osm_shock}), we observe a linear dependence. In fact, at this rather small deflation rate, we observe that the shape of the vesicle (figure \ref{osm_shock}A1-A2) follows a sequence of prolate shapes for the given area and volume that are equivalent to those found at equilibrium \citep{Feng2006-gi}. We observe, however, a small fluctuation of normal and tangential velocities in the equator of the vesicle, which are a signature of a non-equilibrium symmetry-breaking process.
These deviations from axisymmetry become more noticeable at higher deflation rates. For instance, for a deflation rate of $\SI{100}{\nano\meter}^3\SI{}{\nano\second}^{-1}$, we observe that the shape starts to deviate from quasi-equilibrium path and velocity variations disturbing axisymmetry are very pronounced (see figure \ref{osm_shock}-B1), leading to a very different shape as compared to the  equilibrium one for the same volume decrease (see figure \ref{osm_shock}-B2). In agreement with this, we observe that the energy stored during this faster deflation is now higher (green curve).  The viscous dissipation of the lipid membrane becomes increasingly dominant as  deflation rate increases (see figure \ref{osm_shock}-C and D respectively). Similarly, we observe that the final shape gets further away from the equilibrium shape, by storing much more elastic energy for a given amount of volume decrease (red and magenta curves). Mechanically, the viscous dissipative forces can be interpreted as a dynamical confinement for the elastic membrane, causing it to transiently buckle and break symmetry.

\subsection{Lipid bilayers: Seifter-Langer model}

In this section we examine the response of the Seifert-Langer model to monolayer density imbalances, which may arise from chemical perturbations. Membranes in cells and organelles are often exposed to changes in their local lipid density. For instance, proteins and other membrane inclusions, such as polymers, insert in the membrane and locally change the lipid packing \citep{Shibata2009-zt,Tsafrir2003-rl}. Chemical signals, such as pH disturbances \citep{Khalifat2008-pw,Fournier2009-zs}, can also alter lipid packing.  Furthermore, changes in the local density can occur asymmetrically, affecting only one of the two monolayers. Local density perturbations lead to transient dynamics, where lipid flows and shape changes are tightly coupled and dictated by the interplay between stretching, bending, shear and intermonolayer friction. Thus, these processes constitute an excellent example of application of our theoretical and computational framework. Furthermore, these processes have been previously examined under the assumption of axisymmetry \citep{Rahimi2012-sa}, which can be used as a reference to verify our numerical procedure.

Following \cite{Rahimi2012-sa}, we examine deflated spheroidal prolate vesicles, initially at equilibrium, to which we apply a density disturbance. To prepare the initial state, we start with a sphere of radius $R$ and, fixing its volume $V$, we increase its surface area $S$ to obtain a given reduced volume $v$, which is defined as the ratio between $V$ and the volume of a sphere with surface area $S$,
$v = \frac{3 \sqrt{4\pi} V}{S^{3/2}}$.
For a sphere $v=1$ and $v<1$ otherwise. During the area increase, we solve the shape that minimizes the Helfrich energy. Once the prolate shape has been obtained, we initialize the lipid densities on each monolayer close to their equilibrium state for the given shape, i.e. satisfying
$\rho^\pm = \rho_0(1\mp dH)$. 
To perturb the initial density profiles, we add a localized perturbation $\delta\rho^\pm = \delta\breve{\rho}^\pm(1\mp dH)$,
where $\delta\breve{\rho}^\pm = \delta\breve{\rho}^\pm_m\, f(\theta,\phi)$
is the perturbation of the densities at the neutral surfaces of each monolayer, $\delta\breve{\rho}^\pm_m$ is the maximum value of the perturbation at the outer and inner monolayers respectively, and $f(\theta,\phi)$ is a function with values from 0 to 1 of the angles $(\theta,\phi)$ of a set of spherical coordinates adapted to the prolate shape. Following  \cite{Dimova2006-tu,Rahimi2012-sa}, we choose $\kappa=10^{-19}\SI{}{\joule}$, $k_S=5\times10^{-2}\SI{}{\joule\meter}^{-2}$, $b_I = 10^{9}\SI{}{\joule\second\meter}^{-4}$, $\mu = 5 \times 10^{-10}\SI{}{\joule\second\meter}^{-2}$, and the dilatational viscosity $\lambda = 0$ (this parameter seems to play a minor role in the dynamics).

\begin{figure}
	\begin{center}
		\includegraphics[width=5.25in]{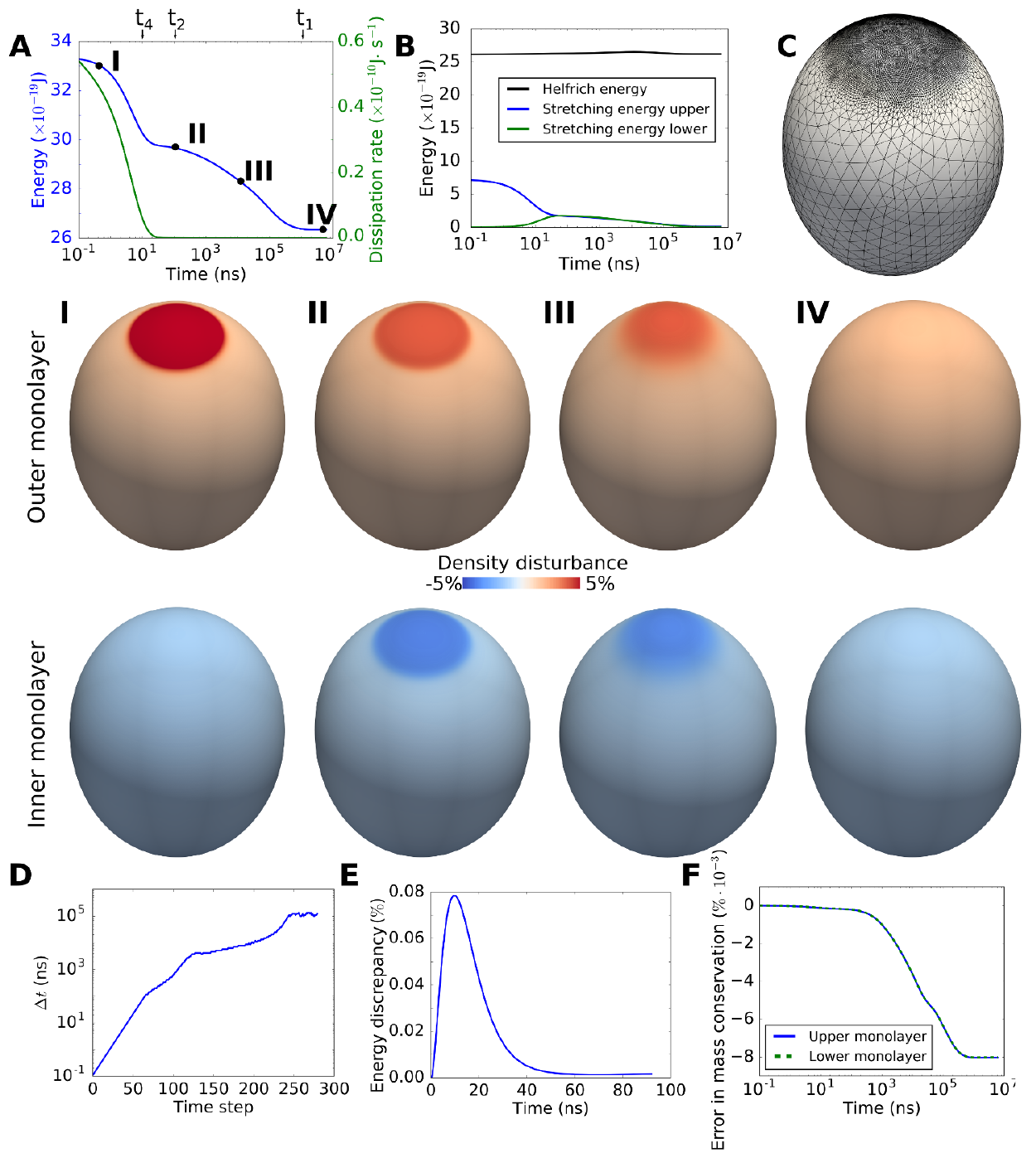}
		\caption{\label{densities} Relaxation dynamics of a density perturbation on the outer monolayer of a small vesicle of $R=\SI{200}{\nano\meter}$ with $\delta\breve\rho_m^+=5\%$. (A) Energy (blue) and dissipation (green) along the time-evolution of the system. Note that the $x-$axis is in log-scale to enhance the different time-scales in the problem. The different scales of the system $t_1$, $t_2$ and $_4$ (see main text) are depicted for comparison. (B) Time-evolution of the different energies of the problem. (I-IV) show snapshots of the shape and the densities of outer and inner monolayers at different stages of the dynamics. (C) Mesh used for the simulations with a much higher resolution at the pole where the density disturbance is imposed. (D) Time-evolution of the time-step. (E) Energy discrepancy when comparing our time-adaptive simulations with one with fixed and very small time-step for the first 100 ns of dynamics. (F) Time-evolution of the relative error in total mass conservation.}
	\end{center}
\end{figure}

\textbf{Example 1: Relaxation dynamics of a density disturbance in an axisymmetric vesicle of 200 nm.} To compare with \citep{Rahimi2012-sa}, we start by examining a small vesicle ($R=200$ nm) with a reduced volume $v=0.99$, to which we apply a disturbance of  $5\%$ in the outer monolayer, $\delta\breve\rho_m^+/\rho_0=5\%$, with a distribution $f(\theta) = \tanh\left((w-\theta)/\pi\right)$,
where $\displaystyle w={\pi}/{10}$ controls the width of the disturbance. We show some snapshots of the dynamics along with the time-evolution of the dissipation and the main energy contributions, see figure \ref{densities}. Again, we observe  that the total energy $\mathcal{F}$ decays with time (figure \ref{densities}A), as expected. Furthermore, from figure \ref{densities}B we observe that the largest energetic component is $\mathcal{F}_\text{H}$, the Helfrich energy. 
However, it does not play a significant role in this problem since its variation is very small. Instead, we observe that the relaxation of the stretching energy in the upper monolayer, which transiently increases that of the lower monolayer, is the main driver of the dynamics (see figure \ref{densities}B). In snapshot III, we can observe how the local density asymmetry results in a small but noticeable shape change, whose signature can be seen in the curvature energy. Note that, given the versatility of subdivision surfaces to deal with meshes of arbitrary connectivity, we have used a surface mesh with a much higher resolution at the pole where the density disturbance is imposed, see figure \ref{densities}C.


These dynamics can be rationalized introducing several time-scales for this model following \cite{Rahimi2012-sa}. Gradients of the average density relax with a time-scale given by $t_4=\mu/k_S$, as they are driven by stretching energy and dragged by shear dissipation. This time-scale  is size-independent, and usually very fast, $t_4\approx\SI{10}{\nano\second}$ for our choice of model parameters. Gradients of density differences between monolayers are also penalized by the stretching energy. However, at fixed shape, these gradients relax by intermonolayer slippage. Indeed, density differences have been shown to diffuse with a diffusivity $D=k_S/b_I$ \citep{Evans1994-yg}, which results in a time-scale 
$t_1=\bar{S}/D=\bar{S} b_I/k_S$, where $\bar{S}$ is the area of the density disturbance. 
However, density differences can also relax by curving the membrane, not mobilizing intermonolayer slippage, with a time-scale given by $t_2=\sqrt{\bar{S}} \mu/(k_S d)$. 
For the $200$ nm vesicle, we find that $t_1\approx0.15\SI{1}{\milli\second}$ and $t_2\approx \SI{1}{\micro\second}$. All these time-scales are apparent in figure \ref{densities}A and highlight the dramatic gap between time-scales in this model, which need to be resolved by the simulations. To address this challenge, we adapt the time-step as shown in figure  \ref{densities}D, with time-steps spanning six orders of magnitude, from $\SI{0.1}{\nano\second}$ to $\SI{0.1}{\milli\second}$. 
To adapt the time-step we follow the following prescription: if Newton's method is solved less than $N_S$ steps, with $N_S$ given initially (usually a number between 4 and 6), we increase $\Delta t^{n+1}=f\Delta t^{n}$ with $f$ a scaling factor greater than $1$. If, however, Newton's method does not converge in $N_S$ steps, we reduce $\Delta t^{n+1}$ as $\Delta t^{n+1}=\Delta t^{n}/f$. This adaptive time-stepping algorithm allows us to perform the simulation in less than 300 time-steps, whereas a fixed time-step algorithm with the required initial resolution would need 10 million of time-steps. To show that the dynamics is not affected by the adaptive time-stepping, we plot the difference in the total energy between a simulation with a fixed and very small time-step  ($\Delta t = \SI{0.1}{\nano\second}$) and the simulation with the adaptive time-steping for the first 100 ns of dynamics, which shows a difference smaller than $0.1\%$ (figure \ref{densities}E). Another important aspect of the numerical method is the global conservation of mass and volume. Conservation of the total mass depends on the local mass conservation imposed weakly through the process operator, whereas conservation of volume is imposed as a non-linear constraint at every time-step. We show the time-evolution of the relative error in total mass for the outer and inner monolayers in figure \ref{densities}F, where we observe errors smaller than $10^{-2}\%$. We find errors in enclosed volume conservation smaller than $10^{-4}\%$.

\begin{figure}
	\begin{center}
		\includegraphics[width=4.25in]{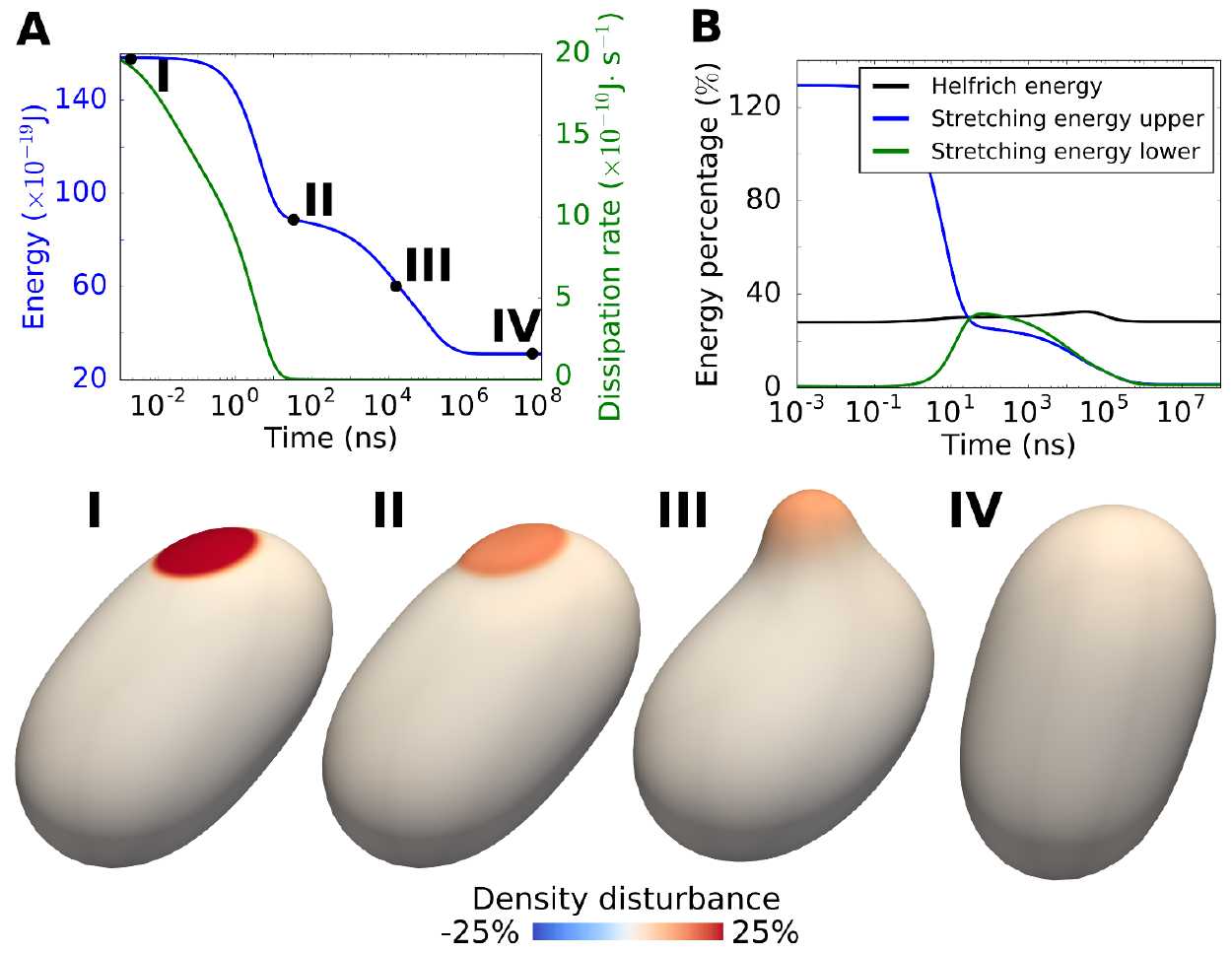}
		\caption{\label{dynamics_25}Relaxation dynamics of a density perturbation on the outer monolayer of a small vesicle of $R=\SI{200}{\nano\meter}$ with $\delta\breve\rho_m^+=25\%$ for a non-axisymmetric case. (A) Energy (blue) and dissipation (green) along the time-evolution of the system. (B) Time-evolution of the different energies of the problem. (I-IV) show snapshots of the shape and the density of outer monolayer at different stages of the dynamics.}
	\end{center}
\end{figure}

	\begin{figure}
		\begin{center}
			\includegraphics[width=4in]{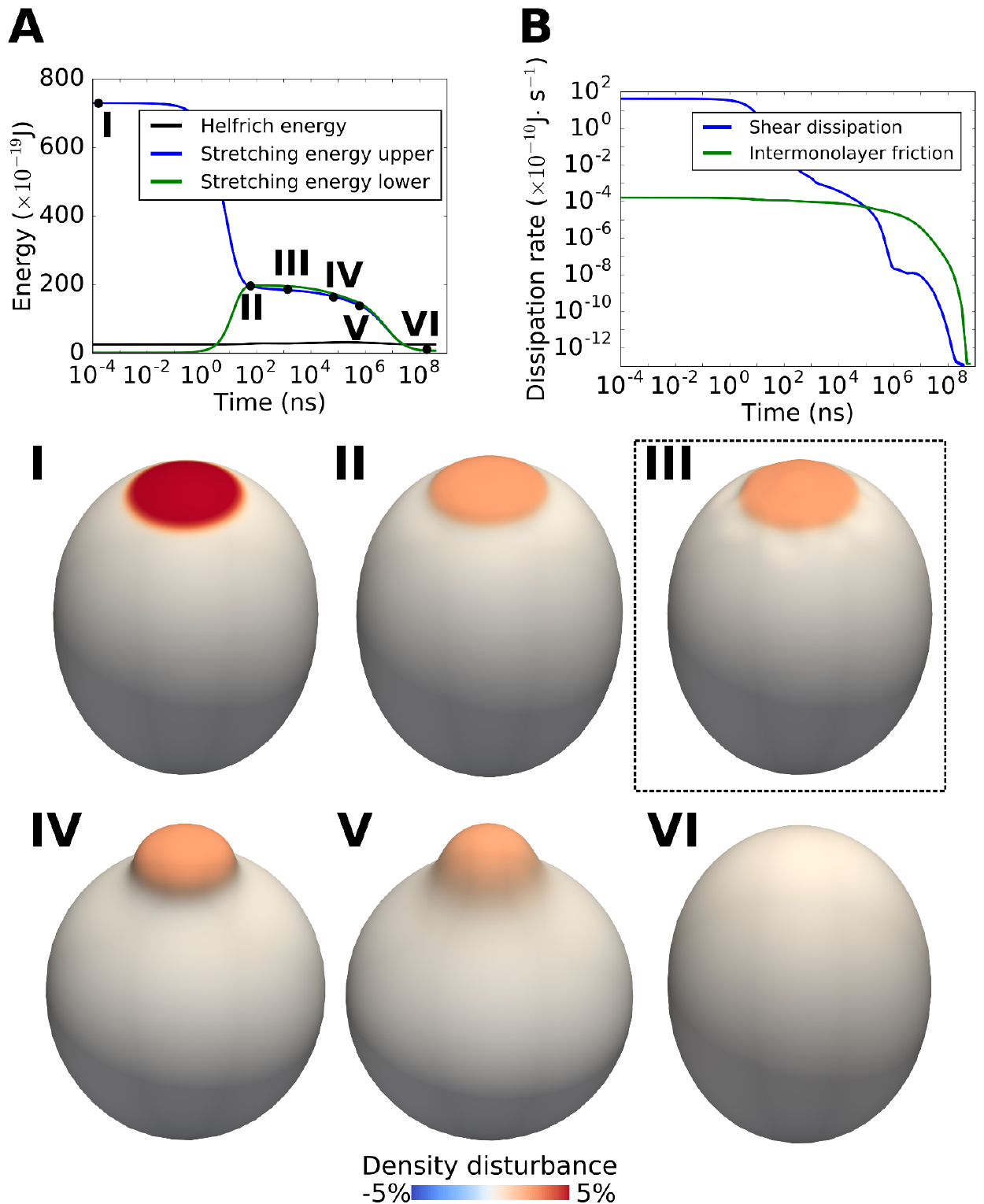}
			\caption{\label{dynamics_large}Relaxation dynamics of a density perturbation on the outer monolayer of a small vesicle of $R=\SI{2}{\micro\meter}$ with $\delta\breve\rho_m^+=5\%$. (A) Time-evolution of the different energies of the problem. (B) Time-evolution of the different sources of dissipation of the problem. (I-IV) show snapshots of the shape and the density of outer monolayer at different stages of the dynamics. }
		\end{center}
	\end{figure}
	\begin{figure}
		\begin{center}
			\includegraphics[width=3.25in]{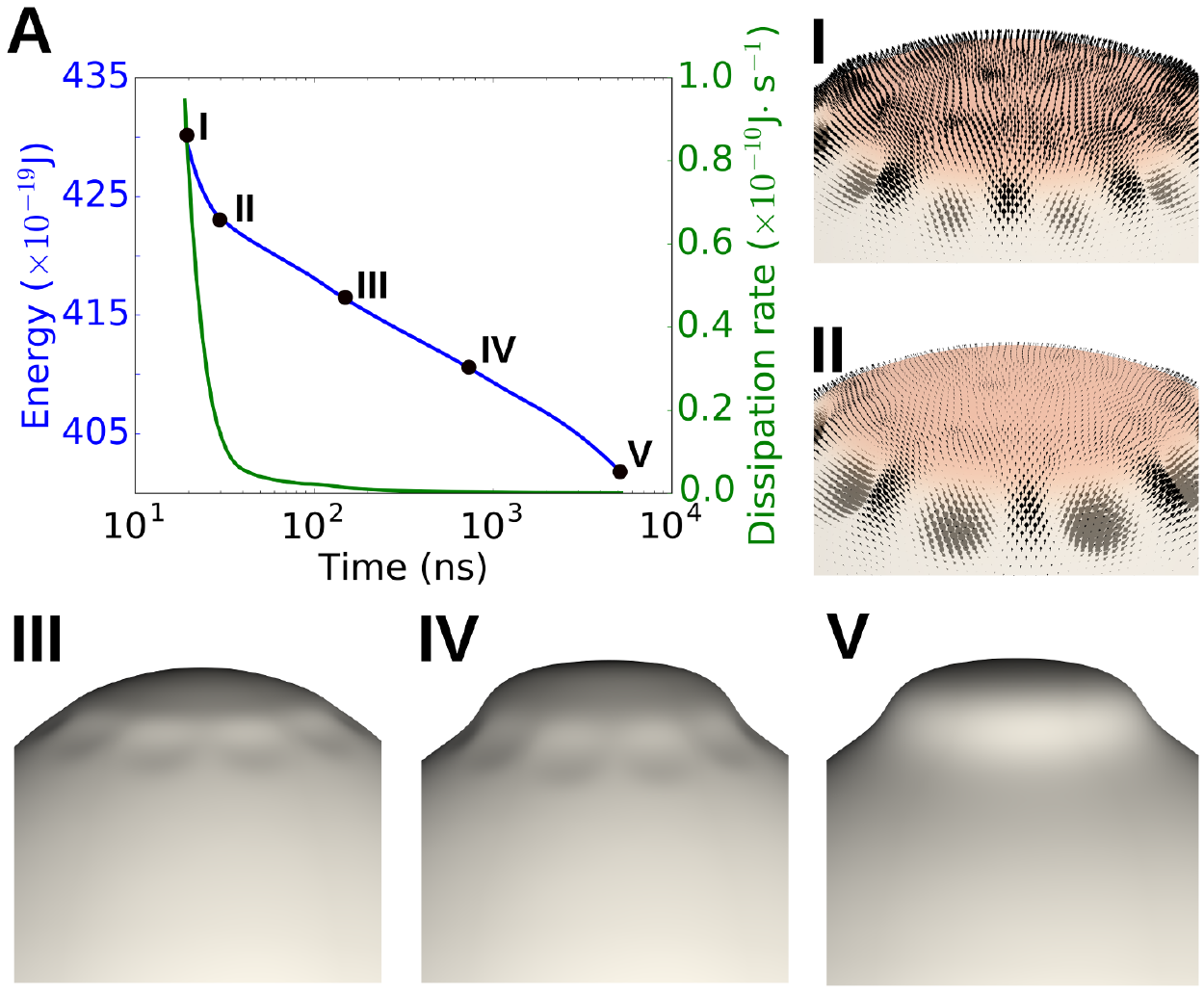}
			\caption{\label{dynamics_large_pattern} Zoom of A in the region in which the pattern forms. (I) and (II) show the velocity field with arrows, which lead to the pattern formation. After the pattern has formed, the bulge continues growing (III) and (IV). Finally, once the bulge grows large enough, the wrinkles associated to the pattern smoothly dissappear.}
		\end{center}
	\end{figure}
\textbf{Example 2: Relaxation dynamics of a density disturbance in a non-axisymmetric vesicle of 200 nm.} To further show the versatility of the numerical method, we examine the dynamics of a non-axisymmetric system, in which the density disturbance is larger, $\delta\rho_m^+=25\%$, and not aligned with the symmetry axis of the prolate initial vesicle  (see figure \ref{dynamics_25} and Movie 2). We observe a similar dynamics, now with a larger bulge due to the larger density difference, and with an initial stretching energy 4 times larger than the bending energy.

\textbf{Example 3: Relaxation dynamics of a density disturbance in an initially axisymmetric vesicle of 2 micron.} Finally, we analyze a vesicle of $R=\SI {2}{\micro\meter}$ with $\delta\breve\rho_m^+/\rho_0=5\%$. For this size, the stretching energy becomes even  more dominant than for the $R=\SI {200}{\nano\meter}$ vesicle. Indeed, the relative influence between the different energetic components is highly size-dependent. Given two vesicles, say 1 and 2, related by a geometric scaling factor $X$, we have that $\mathcal{F}_\text{H}(2)= \mathcal{F}_\text{H}(1)$ (the Helfrich energy is size independent), whereas $\mathcal{F}_\text{S}(2)= X^2\mathcal{F}_\text{S}(1)$. In agreement with this, the dynamics for $R=\SI {2}{\micro\meter}$ show the formation of a large bulge that affects the shape of the whole vesicle and with a stretching energy 20-fold larger than the Helfrich energy (see figure \ref{dynamics_large}). The time-scales associated to this problem are $t_1\approx\SI{15}{\milli\second}$ and $t_2\approx \SI{10}{\micro\second}$, with $t_4=\SI{20}{\nano\second}$ as before. In agreement with these time-scales, we observe again the first energy decrease in a scale comparable with $t_4$, and a total duration of the relaxation dynamics of $\SI{10}{\milli\second}$, similar to $t_1$. In figure \ref{dynamics_large}B we plot the different dissipation contributions, shear viscosity and intermonolayer friction, in a log-log plot. This plot shows that, during the initial equilibration of the total density and during the bulge formation, shear dissipation dominates. However, at later stages, density differences relax due to intermonolayer slippage. In this time-adaptive simulation, the smallest and largest time-steps differ by 7 orders of magnitude.

Interestingly, in the initial stages of the bulge formation (figure \ref{dynamics_large}III), we observe that a pattern resembling buckling forms at the edge of the bulge, presumably caused by a transient and local compression in a large enough region compared to the F\"oppl-von K\'arm\'an length-scale $\displaystyle l_{FvK}=\sqrt{\kappa/\sigma}\approx\SI {5}{\nano\meter}$, where surface tension is dominated by stretching energy $\sigma=k_S\left(\left(\rho^\pm/\rho_0\right)^2-1\right)\approx10^{-2}$ J$\bcdot$m$^{-2}$  for $\rho^\pm=1.05\rho_0$.  This kind of transient buckling deformation is a three-dimensional phenomenon that could not develop in the axisymmetric simulations by \cite{Rahimi2012-sa}. To examine this phenomenon further, we zoom figure \ref{dynamics_large}III in the region where the pattern forms, see figure \ref{dynamics_large_pattern}. The formation of the pattern does not result from an increase of the total energy, which suggest that it is not caused by a numerical instability of our method. In Figs.~\ref{dynamics_large_pattern}I and II, we show the velocity field of the outer monolayer near the bulge at two different instants during the pattern formation. After the pattern has formed, Figs.~\ref{dynamics_large_pattern}III and IV, the amplitude of the bulge continues to increase, and the oscillatory deformation pattern progressively disappears, see figure \ref{dynamics_large_pattern}V. 
The rest of the dynamics is similar to that obtained by \cite{Rahimi2012-sa}, which suggest that the pattern forms due to an initial buckling instability that does not affect the final fate of the dynamics. To further test the stability of our scheme, we used a finer mesh and found the same dynamics; fluctuations develop with the same length scale, which suggest that it is a physical outcome of the model rather than an instability of the method. Our model lacks the dissipative forces induced by the bulk medium, which may modify this buckling-induced transient pattern formation. Indeed, the size of the disturbance is close to the Saffman-Delbr\"uck length, $l_\text{SD}=\SI{5}{\micro\meter}$, and therefore bulk dissipation could start playing a role \citep{Saffman1975-xx,Arroyo2009-xz}

\subsection{The cell cortex: A viscous layer driven by active tension\label{exCortex}}

The elementary model of the actomyosin cortex introduced in section \ref{cortexSec} exhibits a non-trivial phenomenology and reproduces to a large extent the mechanics of cells in different processes, such as during cytokinesis \citep{Turlier2014-wh} or in rheological assays \citep{Torres-Sanchez2017-ay}. Here, we focus on the ability of this model to describe adhesion-independent cell migration. 
\begin{figure}
\begin{center}
\includegraphics[width=5in]{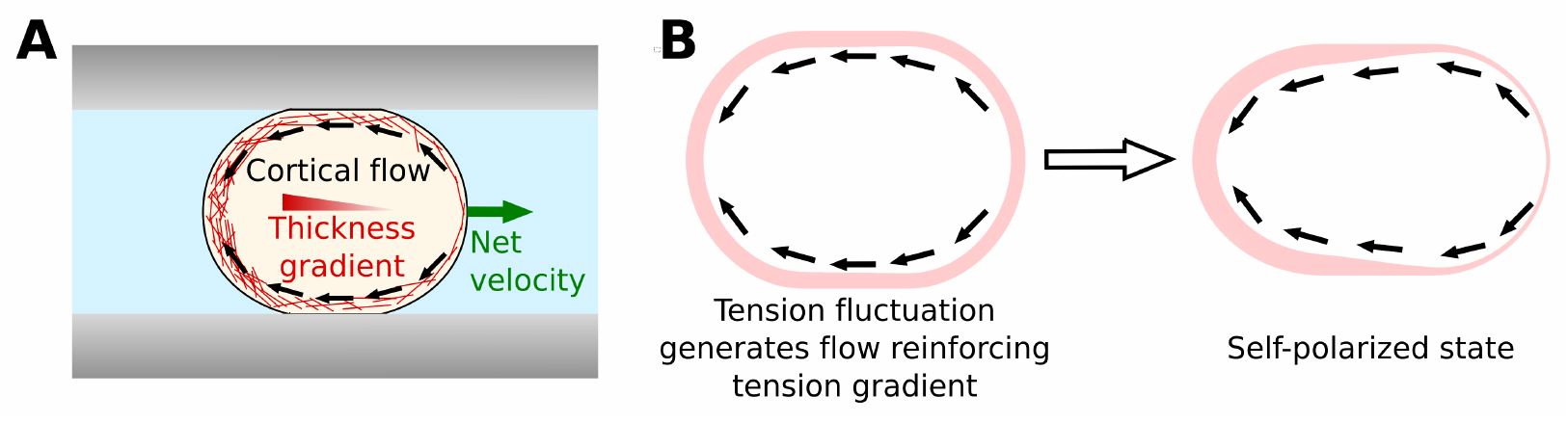}
\caption{\label{schememig} (A) In adhesion-independent migration, confined cells develop a self-sustained cortical flow. By friction with the surroundings, here friction with the confining plates, the cell migrates in a direction opposite to the gradient of tension. (B) An initial myosin activity fluctuation or density disturbance can trigger a cortical flow, which in turn reinforces the gradient in tension. This leads to a self-polarized state in which a steady state flow is achieved.}
\end{center}
\end{figure}
In this kind of migration \citep{Bergert2015-my,Ruprecht2015-hz}, cells develop a persistent cortical flow from the front to the rear of the cell that propel the cell forward by unspecific friction under confinement, see figure \ref{schememig}A. This friction is independent of specific adhesion molecules.  Adhesion-independent locomotion plays a major role in three-dimensional cell migration through the extracellular matrix or in confined environments \citep{Poincloux2011-rf,Liu2015-bc}. 

Adhesion-independent migration raises several questions. First, what is the mechanism by which cells acquire such a polarized state? Second, how can this flow be made persistent to allow for a self-sustained motion? And, how does the tight interplay between interfacial flows on the cortex and cell shape changes manifest itself in this process? Several models based on the theory of active gels have been developed over the past decade to try to answer these questions \citep{Hawkins2011-pz,Tjhung2012-mu,Callan-Jones2013-wb}. In these models, myosin-mediated contraction of the cell cortex is identified as the main driver of the self-polarization. In particular, a spatial fluctuation in myosin activity can lead to a tension gradient in the cortex. This tension gradient triggers cortical flows, which further reinforce the gradient of tension, see figure \ref{schememig}B. This mechanism works against actin turnover, which tries to homogenize the system. Thus, adhesion-independent migration depends on the competition between myosin activity and actin turnover. Most of previous models rely on simple one-dimensional or fixed-shape assumptions that cannot address the effect of shape in locomotion. Recently, \citet{Callan-Jones2016-tp} studied the shape transformations that cells suffer during migration, but this work is restricted to small deformations around a sphere. Here, we present, to our best knowledge, the first numerical results of a fully three-dimensional and nonlinear model connecting cortical flows and cell shape dynamics during locomotion. This work opens the door to a more systematic study of adhesion-independent migration in the future.

A critical ingredient controlling the formation of a self-polarized cortical flow is myosin activity, which is described by the function $\xi(\rho)$ in our model. If this function is constant, as assumed in previous works \citep{Turlier2014-wh}, then the active tension is proportional to cortical thickness,  $\gamma=\xi_0\rho$. In this case, the positive feedback illustrated in figure \ref{schememig}B leads to an instability with unbounded actin accumulation at the rear of the cell.
Recently, \citet{Chugh2017-lj} found that active tension does not depend linearly on cortical density in general. They found that in mitotic cells tension depends non-monotonically on cortical thickness, which they identified as a proxy for filament length. They proposed a conceptual model according to which active tension would be modulated by network architecture. Along the lines of this work, here we model the a dependence of specific contractility on cortical thickness as
\begin{equation}
\xi(\rho)=\xi_0\left[1-\frac{1}{3} \left(\frac{\rho}{\omega\rho_0}\right)^2\right],
\end{equation}
where $\xi_0$ measures a basal myosin activity and $\omega$ characterizes its dependence with cortex thickness.
This leads to an active tension
\begin{equation}
\gamma(\rho) = \xi(\rho)\rho = \xi_0\left[\rho-\frac{\omega\rho_0}{3} \left(\frac{\rho}{\omega\rho_0}\right)^3\right],
\end{equation}
which has a maximum at $\rho=\omega\rho_0$; at equilibrium $\gamma_0=\gamma(\rho_0)=\xi\rho_0(1-1/3\omega^2)$. We note that the second term in the  active tension looks very similar to the osmotic contribution introduced by \cite{Callan-Jones2013-wb}  to stabilize the dynamics of polarization.

Following the experimental work by \cite{Ruprecht2015-hz}, we examine the migration of cells confined between two plates. To represent this confinement mathematically, we introduce a free energy contribution of the form
\begin{equation}
\mathcal{F}_\text{c} = \int_{\Gamma_t} U(z) dS,
\end{equation}
where $z$ is a coordinate perpendicular to the plates, and $U(z)$  is a repulsive potential modelling contact with the plates and given by
\begin{equation}
U(z) = \left\{
\begin{array}{ll}
\displaystyle 0 & \displaystyle\text{if } |z| < h/2,\\
\displaystyle\frac{K_c}{3}\left(\frac{|z|-h/2}{\delta_c}\right)^3 &\displaystyle \text{if } |z| \geq h/2.
\end{array}
\right.
\end{equation}
with $K_c$ and $\delta_c$ characterizing the strength and the width of the repulsive interaction respectively. 

\begin{figure}
\begin{center}
\includegraphics[width=5in]{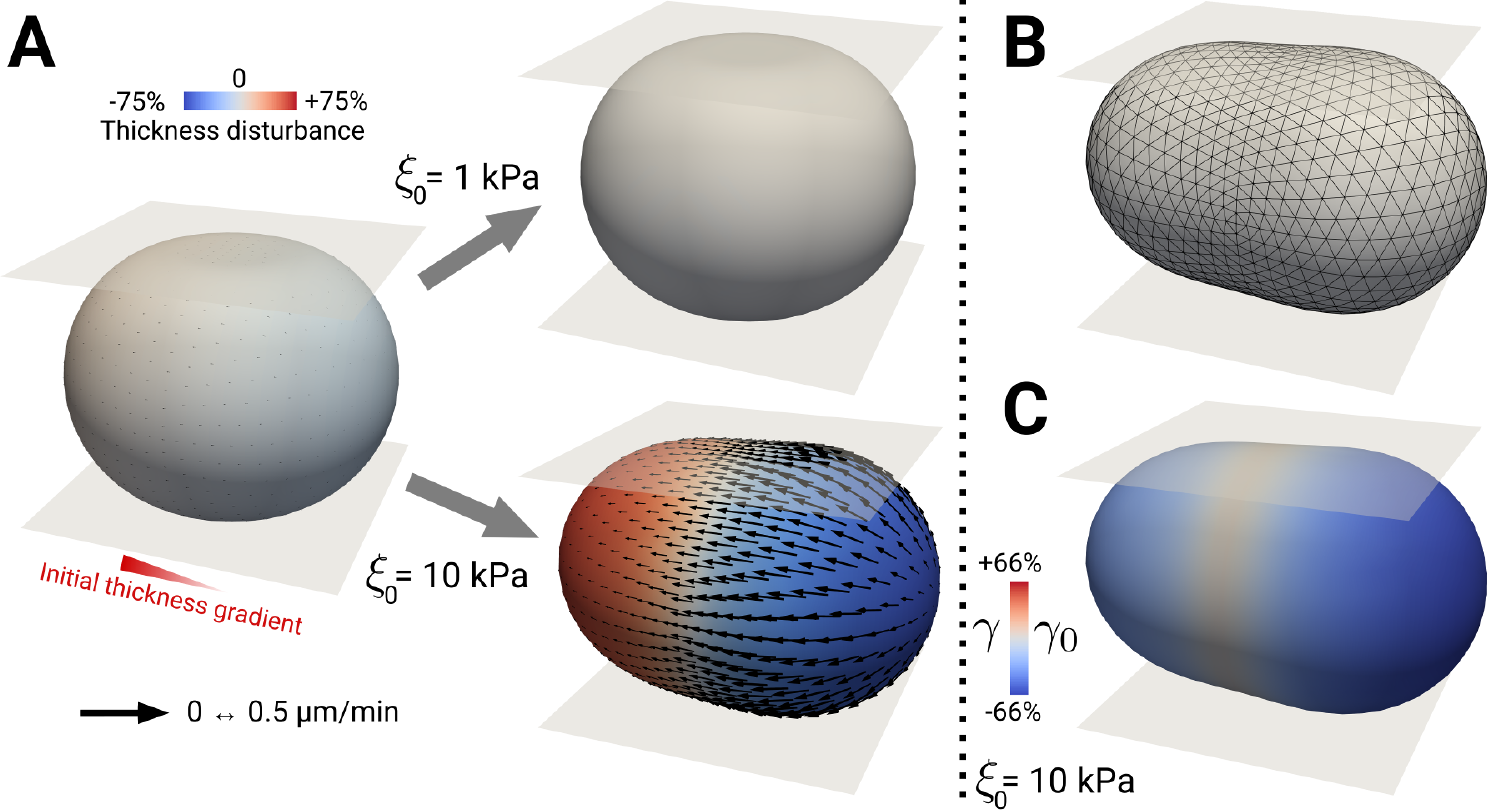}
\caption{\label{selfpolar} (A) An initial thickness gradient (left) can be homogenized due to turnover for low tension ($\xi_0=\SI{1}{\kilo\pascal}$, right top) or can lead to a sustained self-polarized steady-state for higher tension ($\xi_0=\SI{10}{\kilo\pascal}$, right bottom). Thickness is depicted with a colormap, whereas velocity is shown with arrows. (B) The ALE mesh is able to cope with this kind of directed flow without remeshing, which would continuously distort any Lagrangian mesh and require frequent remeshing operations. (C) Active tension profile for a self-polarized cell ($\xi_0=\SI{10}{\kilo\pascal}$). Since tension is a non-monotonic function of actin density, it has a maximum between the front and the rear of the cell. }
\end{center}
\end{figure}
\begin{figure}
	\begin{center}
		\includegraphics[width=5.in]{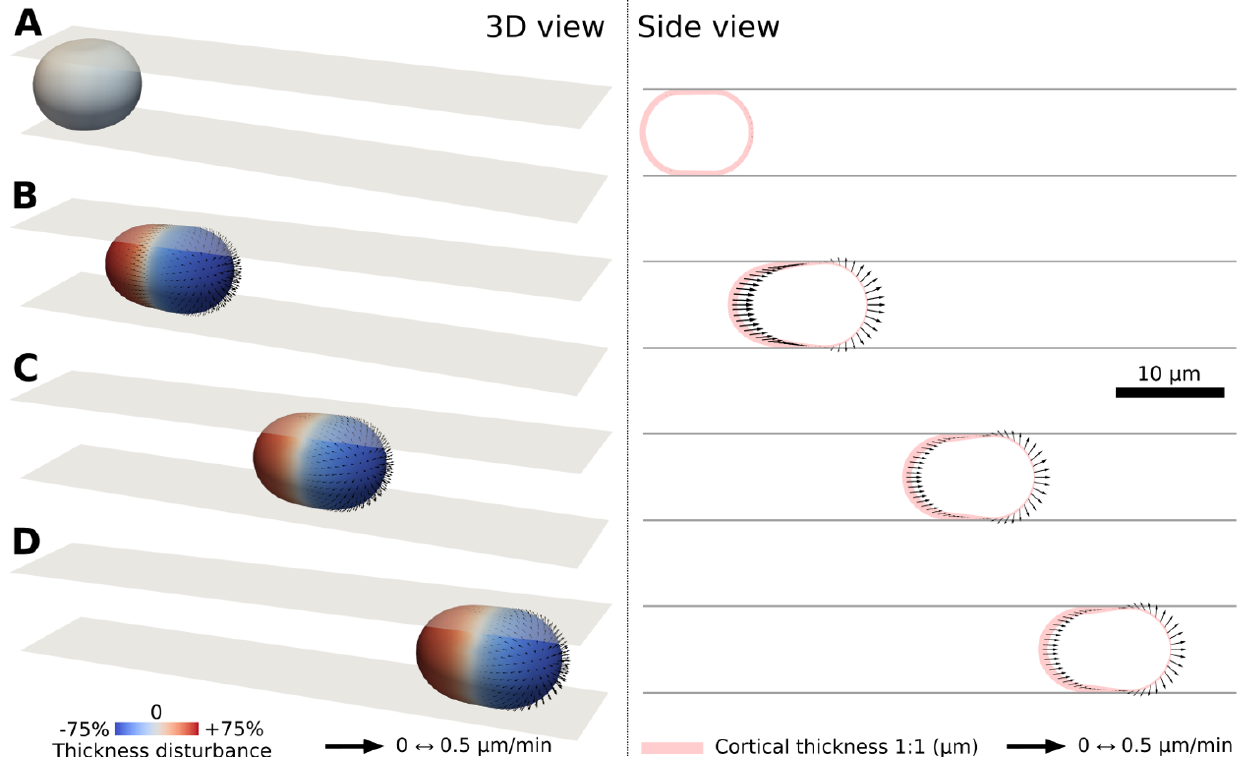}
		\caption{\label{crawler} With friction with the plates, the cell is able to crawl.  On the left, we show a 3D viewpoint of cell locomotion, with thickness shown in colormap, and velocity shown with arrows. On the right, we show a side view of the motion, with cortical thickness depicted in light red in 1:1 scale. }
	\end{center}
\end{figure}
We now perform simulations of the model on a model cell of average radius $R=\SI{5}{\micro\meter}$. Material parameters are obtained from literature $\rho_0=\SI{500}{\nano\meter}$ \citep{Clark2013-jm}, $\mu=10\SI{}{\kilo\pascal\second}$ \citep{Bergert2015-my}, $\tau=10\SI{}{\second}$ \citep{Fritzsche2013-bd}, $\xi_0=1-10\SI{}{\kilo\pascal}$ \citep{Chugh2017-lj}. We also choose $\omega =2\sqrt{3}/3$. We first compress the cell between the plates with $h=\SI{4}{\micro\meter}$ and let the system relax. To drive the cortex out of the equilibrium state at constant density $\rho=\rho_0$, we perturb the system with a gradient in density of $1\%$ in the $x$ direction simulating a possible fluctuation of myosin activity within the cortex, see figure \ref{selfpolar}A left. We first simulate the system with $\xi_0=\SI{1}{\kilo\pascal}$. For this small degree of contractility, the tension difference generated by the initial thickness perturbation is not high enough to overcome cortical turnover, and the system quickly relaxes to a situation of homogeneous cortical density, see figure \ref{selfpolar}A top right. For a higher value of myosin activity, $\xi_0=\SI{10}{\kilo\pascal}$, the cortical flow generated by the activity gradient overcomes turnover, and the cell becomes self-polarized with a sustained cortical flow, see figure \ref{selfpolar} bottom right. Together with the flow, the cell experiences a shape change during the transient dynamics towards the steady self-polarized state. Our ALE method is able to sustain such shape changes without remeshing, see figure \ref{selfpolar}B. More remarkably, we observe that the mesh is not affected by the constant flow of actin from the front to the rear of the cell. In a Lagrangian framework, such steady state would continuously distort the mesh, and very frequent remeshing would be required. Finally, we observe that, since tension is not a monotonic function of cortical thickness, it exhibits a maximum between the front and the rear of the cell, see figure \ref{selfpolar}C. This self-polarized state, however, cannot lead to cell migration by itself unless we introduce a mechanical interaction with the confining plates. To represent unspecific friction, we introduce the dissipation potential
\begin{equation}
\mathcal{D}_c = \int_{\Gamma_t} \frac{\eta_c U'(z)}{2} \left[V_x^2+V_y^2\right] dS,
\end{equation}
where $\eta_c$ measures friction with the plates and $U'(z)$ identifies the pressure exerted by the cell on the plates. This pressure is equal to the internal pressure of the cell $P$, which is essentially determined by the cell radius of curvature and its surface tension and is $P\approx \SI{0.3}{\kilo\pascal}$ in our simulations. Resorting to experimental measurements of the product of $\eta_c P=1-10^4\SI{}{\kilo\pascal\second\meter}^{-1}$ on somewhat larger cells \citep{Bergert2015-my}, we choose $\eta_c = \SI{600}{\second\meter}^{-1}$. We note, however, that our results are largely independent of friction because we do not consider a hydrodynamical resistive force in the relatively unconfined situation of cell motion between parallel plates.  We repeat the previous simulation at $\xi_0=\SI{10}{\kilo\pascal}$, and observe how the self-polarization of the cell now leads to cell migration, in a direction opposite to the cortical flow, see figure \ref{crawler} and Movie 3.  We note that, aside from a small disturbance of cortical velocity due to friction with the plates, the velocity field of actin in the steady state is the sum of a constant center of mass velocity plus a velocity profile similar to the one in figure \ref{selfpolar}A bottom right (data not shown). For these simulations and to deal with cell migration, we consider the following ALE parametrization
\begin{equation}
\bm{\psi}(\xi_1,\xi_2;t) = \bm{\psi}_0(\xi_1,\xi_2) + h(t)\bm{M}(\xi_1,\xi_2)+ \bm{R}(t), 
\end{equation}
where we impose zero net displacement due to the offset, $\int_{\Gamma_t} h(t)\bm{M} dS = \bm{0}$, and incorporate a rigid body translation $\bm{R}(t)$ as an unknown.

In conclusion, our theoretical and computational framework allows us to formulate and simulate thin and curved active gels with high generality. We have illustrated that this approach can be used to examine systematically adhesion-independent cell migration under confinement. Remarkably, our ALE formulation allows us to deal with the shape changes that the cell experiences during self-polarization and confinement, and with the steady cortical flows that are established.

\section{Summary, discussion and future work\label{conclusions}}

We have introduced a novel theoretical and computational framework to model and simulate  fluid surfaces. Fluid surfaces are a common motif in cell and tissue biology. Thanks to increasingly quantitative biophysical experiments, there is a growing need for accurate theoretical predictions. Yet, modelling these systems requires overcoming significant theoretical and computational challenges, which we have addressed in this work. First, based on time-evolving parametrizations, we have rigorously extended the notion of ALE methods to fluid surfaces. We have also used Onsager's formalism, a general variational framework for the dissipative dynamics of soft-matter systems, to derive thermodynamically consistent models of fluid surfaces coupling multiple physics in a fully geometrically non-linear manner. From a numerical perspective, we have proposed a new framework for the simulation of fluid surfaces based on a variational and nonlinearly stable time-integrator rooted in Onsager's variational formalism, allowing us to bridge time-scales over 7 decades,  and on a combination of subdivision and linear finite elements.

We have applied the previous theoretical and numerical methods to derive the governing equations and simulate the dynamics of canonical models of fluid surfaces with unprecedented generality (in three dimensions, for general shapes, and accounting for full geometric nonlinearity). We have first studied the dynamics of lipid bilayers in a number of interesting assays, including membrane relaxation, deflation due to osmotic shocks or perturbations due to density disturbances.  Our framework opens new possibilities in the study of shape pattern formation under dynamical changes in lateral strain or osmotic conditions in supported membranes \citep{Staykova2013-fu} beyond axisymmetry, relevant to cell membrane mechano-adaptation \citep{Kosmalska2015-ow}. Our method could also be useful to understand the effective rheology of a bilayer populated by transmembrane proteins, limiting inter-monolayer slippage in a heterogeneous manner, which could explain the unexpected and highly viscous behaviour of complex biomembranes \citep{Campillo2013-pr}, or coupled to additional fields describing the concentration of membrane proteins to understand the dynamics of curvature sensing and generation (see \citep{Arroyo2018-vd,Baumgart2011-de} and references therein).
While interfacial hydrodynamics are dominant at length-scales smaller than the Saffman-Delbr\"uck length, the bulk hydrodynamics may be a relevant ingredient in processes involving larger scales. Including the bulk hydrodynamics is straightforward conceptually, but requires specialized computational methods, such as immersed boundary methods \citep{Liu2006-op}.

We have also applied our methodology to model and simulate the cell cortex. Our model is based on a viscous isotropic fluid layer, which is able to reproduce a number of rheological experiments and could be employed to infer material parameters in conjunction with experiments \citep{Torres-Sanchez2017-ay}. Here, we have shown that our model is capable of reproducing adhesion-independent cell migration. Our simulations show how our ALE method can deal with the shape transformations that cells experience during migration and at the same time it can withstand  steady flows from the front to the rear of the cell during migration. While our model for the cortex can reproduce a number of cellular behaviours, it is insufficient to reproduce phenomena where the transient elastic behaviour of the cortex becomes important, e.g.~during laser ablation \citep{Saha2016-hg}, or situations in which the orientational order of actin filaments becomes relevant \citep{Reymann2016-ti}. This would require introducing tensorial fields on the surface \citep{Nestler2018-el}. Furthermore, a more detailed mechano-chemical model of activity, the explicit treatment of the cytosol, and models capable of spontaneously producing polarization would provide a more complete understanding of the mechanics of the cortex. These and other extensions of the active gel model presented here are enabled by the theoretical and computational tools introduced here.

\section*{Acknowledgements}
We acknowledge the support of the European Research Council (CoG-681434), the European Commission (project H2020-FETPROACT-01-2016-731957), the Spanish Ministry of Economy and Competitiveness/FEDER (DPI2015-71789-R to MA and BES-2012-05489 to ATS), and the Generalitat de Catalunya (SGR-1471,  ICREA Academia€ award to MA). We also thank Nikhil Walani, Sohan Kale and Daniel Santos-Oliv\'an for useful discussions.

\appendix
\section{Relation between Lagrangian and ALE velocities\label{LagALEvel}}
\begin{align}
\bm{V} &= \partial_t \bm{\phi}\circ \bm\phi^{-1} \nonumber
= \partial_t \left(\bm{\psi}\circ\bm{\psi}^{-1}\circ\bm{\phi}\right)\circ \bm\phi^{-1} \nonumber
=  \partial_t \left(\bm{\psi}\circ\bm{\theta}\right)\circ \bm\phi^{-1}\nonumber\\
&=\partial_t\bm{\psi}\circ\bm{\theta}\circ \bm\phi^{-1} + \left[\left(D\bm{\psi}\right)\circ\bm{\theta}\circ \bm\phi^{-1} \right] \left[\partial_t\bm{\theta}\circ \bm\phi^{-1}\right]\nonumber\\
&=\partial_t\bm{\psi}\circ\bm\psi^{-1} + \left[\left(D\bm{\psi}\right)\circ\bm\psi^{-1}\right] \left[\bar{\bm{c}}\circ \bm\phi^{-1}\right]\\
&=\bm{W}  + \left[\left(D\bm{\psi}\right)\circ\bm\psi^{-1}\right] \left[\tilde{\bm{c}}\circ \bm\theta\circ \bm\phi^{-1}\right]\nonumber
=\bm{W}  + \left[\left(D\bm{\psi}\right)\circ\bm\psi^{-1}\right] \left[\tilde{\bm{c}}\circ \bm\psi^{-1}\right]\nonumber\\
&=\bm{W}  + \left[\left(D\bm{\psi}\right)\tilde{\bm{c}}\right]\circ\bm\psi^{-1}\nonumber
=\bm{W}  + \bm{\psi}_{*} \tilde{\bm{c}}  \nonumber
=\bm{W} + \bm{c}.\nonumber
\end{align}
\section{Relation between Lagrangian and ALE time-derivatives\label{LagALEder}}
\begin{equation}
\begin{aligned}
D_t f &= \partial_t \bar{f} \circ \bm{\phi}^{-1} = \partial_t \left(\tilde{f} \circ \bm{\theta}\right) \circ \bm{\phi}^{-1} 
\\
&= \partial_t \tilde{f} \circ \bm{\theta}\circ \bm{\phi}^{-1} + \left[D\tilde{f} \circ \bm{\theta}\circ \bm{\phi}^{-1}\right]  \left[\partial_t\bm{\theta}\circ  \bm{\phi}^{-1}\right]\\
&= L_{\bm{W}} f + \left[D\tilde{f} \circ \bm{\psi}^{-1}\right]  \left[\bar{\bm{c}}\circ  \bm{\phi}^{-1}\right]
= L_{\bm{W}} f + \left[D\tilde{f} \circ \bm{\psi}^{-1}\right]  \left[\tilde{\bm{c}}\circ \bm\theta\circ \bm\phi^{-1}\right]
\\
&= L_{\bm{W}} f + \left[D\tilde{f} \circ \bm{\psi}^{-1}\right]  \left[\tilde{\bm{c}}\circ  \bm{\psi}^{-1}\right]
= L_{\bm{W}} f + \left[D\tilde{f} \tilde{\bm{c}}\right]  \circ \bm{\psi}^{-1}= L_{\bm{W}} f + \bnabla f \bcdot \bm{c}.\\
\end{aligned}
\end{equation}
Here we identify $\left[D\tilde{f} \tilde{\bm{c}}\right]  \circ \bm{\psi}^{-1}$ as the pull-back of $\bnabla f \bcdot \bm{c}$, where $\bnabla f$ is the surface gradient of $f$. 

\section{Rate-of-deformation tensor in terms of velocities \label{RateofDef}}
To obtain the form $\bm{d}$ in terms of $\bm{V}$, let us consider the components of $\bar{\bm{g}}$, which coincide with those of $\bm{g}$ in the convected basis by $\bm{\phi}$ given by the tangent vectors $\bm{e}_a= \partial_a\bm{\phi}, ~a=1,2$
\begin{equation}
\left[\bar{\bm{g}}\right]_{ab}(\bm{\xi}) = g_{ab}\circ \bm{\phi}_t(\bm{\xi}) = \partial_a\bm{\phi}_t \bcdot\partial_b\bm{\phi}_t.
\end{equation}
Then, we have
\begin{align}
\left[L_{\bm{V}} \bm{g}\right]_{ab}  &= \partial_t \bar{g}_{ab} \circ\bm{\phi}^{-1}\nonumber\\
&=\left[\partial_t\partial_a\bm{\phi}\circ\bm{\phi}_t^{-1}\bcdot\bm{e}_b+ \bm{e}_a\bcdot\partial_t\partial_b\bm{\phi}\circ\bm{\phi}_t^{-1}\right] \\
&= \left[\partial_a\left[\left(\bm{v}+ v_n\bm{N}\right)\circ\bm{\phi}_t\right]\circ\bm{\phi}_t^{-1}\bcdot \bm{e}_b + \bm{e}_a \bcdot \partial_a\left[\left(\bm{v}+ v_n\bm{N}\right)\circ\bm{\phi}_t\right]\circ\bm{\phi}_t^{-1}\right]\nonumber\\
&= \bnabla_av_b + \bnabla_bv_a-2v_nk_{ab},\nonumber
\end{align}
where we have used the conmutativity of partial derivatives, the definition of covariant derivative 
\begin{equation}
\bnabla_av_b=\partial_a\left(\bm{v}\circ\bm{\phi}_t\right)\circ\bm{\phi}_t^{-1}\bcdot\bm{e}_b,
\end{equation}
the orthogonality of $\bm{N}$ to the tangent space of $\Gamma_t$ $\bm{e}_a \bcdot \bm{N}=0$, and the definition of the second fundamental form
\begin{equation}
k_{ab} =- \partial_b\left( \bm{N}\circ\bm{\phi}_t\right)\circ\bm{\phi}_t^{-1}\bcdot\bm{e}_a.
\end{equation} 

\section{Weak form of an inextensible monolayer with bending rigidity\label{weakFormInext}}
To derive the weak form of the problem, we rewrite the material time derivative of the free energy (equation \eqref{matderhelf}) as 
\begin{equation}
\begin{aligned}
D_t\mathcal{F}_H[\bm{\phi};\bm{V}] =&~ \int_{\Gamma_t}\kappa \left\{-H\Delta v_n+ \left(\frac{1}{2}H^2-k_{ab}k^{ab} \right)Hv_n  + \bnabla_a\left(\frac{1}{2} \kappa H^2 v^a\right)\right\}dS\\
=&~ \int_{\Gamma_t}\kappa \left\{-\Delta H + \left(\frac{1}{2}H^2-k_{ab}k^{ab} \right)H  \right\}v_n dS\\
&+ \int_{\Gamma_t}\left\{\bnabla_a\left[H\left(k^{ab}-\frac{1}{2} g^{ab} H\right)\right] - \bnabla_a H k^{ab}\right\}v_bdS\\
=&~ \int_{\Gamma_t}\kappa \left\{\left[-\bnabla_a\left[H\left(\frac{1}{2} g^{ab} H-k^{ab}\right)\right] - \bnabla_a H k^{ab}\right] \bm{e}^b\right. \\ 
&\left.+\left[\Delta H - H \left(\frac{1}{2}H g^{ab}-k^{ab} \right)k_{ab}  \right] \bm{N}\right\} \bcdot \bm{V} dS,
\end{aligned}
\end{equation}
where we have used that $\bnabla_a H g^{ab}= \bnabla_a k^{ab}$, that
\begin{equation}
\begin{aligned}
\bnabla_a\left(\frac{1}{2}  H^2 v^a\right) =&~ \frac{1}{2}  H^2 \bnabla_a v^a + v^a  H \bnabla_a H  =  \frac{1}{2}  H^2 \bnabla_a v^a + v_b  H \bnabla_a k^{ab}\\
=&~\frac{1}{2}  H^2 \bnabla_a v^a + \bnabla_a\left( H k^{ab} v_b\right) -  H k^{ab}\bnabla_a v_b - v_b \bnabla_a H k^{ab}\\
=&~H\left(\frac{1}{2} g^{ab} H-k^{ab} \right)\bnabla_a v_b-v_b \bnabla_a H k^{ab} + \bnabla_a\left( H k^{ab} v_b\right) \\
=&~-\bnabla_a\left[H\left(\frac{1}{2} g^{ab} H-k^{ab}\right)\right] v_b -v_b \bnabla_a H k^{ab}+ \bnabla_a\left( H k^{ab} v_b\right) \\
&+ \bnabla_a\left( H\left(\frac{1}{2} g^{ab} H-k^{ab}\right) v_b\right),\\
\end{aligned}
\end{equation}
and taken into account that the last two terms are null Lagrangians. Thus, variations of the velocity field around the solution $\bm{V}$ of the form $\bm{V}+\bm{U}$ lead to
\begin{equation}
\begin{aligned}
\delta_{\bm{U}}\left\{D_t\mathcal{F}_H[\bm{\phi};\bm{V}]\right\}= \int_{\Gamma_t}&\kappa \left\{\left[-\bnabla_a\left[H\left(\frac{1}{2} g^{ab} H-k^{ab}\right)\right] - \bnabla_a H k^{ab}\right] \bm{e}^b\right.\\
&\left. + \left[\Delta H - H \left(\frac{1}{2}H g^{ab}-k^{ab} \right)k_{ab}  \right] \bm{N}\right\} \bcdot \bm{U} dS.
\end{aligned}
\end{equation}
Equivalently, taking variations of the dissipation potential (equation \eqref{sheardiss}), we get
\begin{equation}
\begin{aligned}
\delta_{\bm{U}}\mathcal{D}_S\left[\bm{\phi};\bm{V}\right] &= \int_{\Gamma_t}2\mu d^{ab} \left\{\bnabla_a u_b - u_n k_{ab}\right\} dS \\
&= -\int_{\Gamma_t}2\mu \bnabla_ad^{ab}  u_b dS - \int_{\Gamma_t} 2\mu d^{ab} u_n k_{ab} dS\\
&=-\int_{\Gamma_t}2\mu \left\{\bnabla_ad^{ab} \bm{e}^b + \mu d^{ab} k_{ab} \bm{N}\right\} \bcdot \bm{U} dS,
\end{aligned}
\end{equation}
where, again, we have set to zero null Lagrangians. Variations of the inextensibility constraint result in
\begin{equation}
\delta_{\bm{U}}\int_{\Gamma_t} \gamma \text{tr}\bm{d} dS= \int_{\Gamma_t} \gamma (\bnabla_a u^a- u_n H) dS = -\int_{\Gamma_t} \left(\bnabla_a \gamma \bm{e}^a+\gamma g^{ab} k_{ab} \bm{N}\right) \bcdot\bm{U} dS.
\end{equation}
Finally, the last two terms have trivial variations
\begin{equation}
\delta_{\bm{U}}\int_{\Gamma_t}P\bm{N}\bcdot\bm{V} dS =  \int_{\Gamma_t}P\bm{U}\bcdot\bm{N} dS,
\end{equation}
and
\begin{equation}
\delta_{\bm{U}}\mathcal{P}[\bm{\phi};\bm{V}] =  -\int_{\Gamma_t}\bm{F}\bcdot\bm{U} dS.
\end{equation}
Collecting all these variations, we have the following statement of stationarity 
\begin{equation}
\begin{aligned}
0 = \delta_{\bm{U}}\mathcal{L} &= \int_{\Gamma_t}\left\{-\bm{F}+ \left[-\bnabla_a\left[H\left(\frac{1}{2} g^{ab} H+k^{ab}\right) + 2\mu d^{ab} + \gamma g^{ab}\right] - \bnabla_a H k^{ab} \right] \bm{e}_b \right.\\
&\,~\,\left.+ \left[\bnabla_a \bnabla_b H g^{ab}  - \left[H \left(\frac{1}{2}H g^{ab}-k^{ab} \right) + 2\mu d^{ab} +\gamma g^{ab}\right]k_{ab}  + P\right] \bm{N} \right\} \bcdot \bm{U} dS  \\
&= -\int_{\Gamma_t} \left\{\bnabla_a\bm{\Sigma}^a+\bm{B}\right\}\bcdot\bm{U}dS,
\end{aligned}
\end{equation}
which should hold for all admissible variations $\bm{U}$, where 
\begin{equation}
\bnabla_a \bm{\Sigma}^a  = \left(\bnabla_a \sigma^{ab} + k\indices{_a^b}  \sigma^a\right)\bm{e}_b + \left(-\sigma^{ab}k_{ab} + k\indices{_a^b} \bnabla_b \sigma^a\right) \bm{N},
\end{equation}
from where one can identify
\begin{equation}
\bm{\Sigma}^a = \sigma^{ab} \bm{e}_b + \sigma_n^a\bm{N},
\end{equation}
with
\begin{equation}
\sigma^{ab} = \kappa H\left(\frac{1}{2}Hg^{ab}-k^{ab}\right)+2\mu d^{ab}+\gamma g^{ab},
\end{equation}
and
\begin{equation}
\sigma_n^a = \kappa g^{ab} \bnabla_bH.
\end{equation}
Finally, 
\begin{equation}
\bm{B} =  \bm{F}+P\bm{N}.
\end{equation}

\section{Weak form of the three-dimensional non-linear Seifert-Langer model\label{weakFormSL}}
In this case, we focus on the stretching energy, dilatation dissipation and intermonolayer friction, since the rest of terms were already derived for an inextensible monolayer (see \ref{weakFormInext}). The rate of change of the stretching energy is
\begin{equation}
\begin{aligned}
\label{weakFormSL1}
D_t & \mathcal{F}_S   \left[\bm{\chi};\rho^\pm;v_n,\bm{v}^\pm\right] = \int_{\Gamma_t} k_S\left\{\left[ \rho^\pm(1\pm dH)-1\right] \times
\vphantom{\frac{1}{2}}\right.
\left[\left(1\pm dH\right)\left(\overbrace{-\bnabla\bcdot\left(\rho^\pm \bm{v}^\pm\right)}^1 +\overbrace{\rho^\pm v_n H}^2\right)\right. \\
&\overbrace{\pm d \rho^\pm \left(\Delta v_n+k_{ab} k^{ab} v_n\right)}^3
\left.\underbrace{-\frac{1}{2}\left(\rho^\pm(1\pm dH) - 1\right)v_nH}_{4}\right]
\left.+\underbrace{\frac{1}{2}\bnabla\bcdot\left( \left[ \rho^\pm(1\pm dH)-1\right]^2 \bm{v}^\pm\right)}_{5} \right\}dS.
\end{aligned}
\end{equation}
Accounting for multiplicative factors, term 1 leads to
\begin{multline}
-\bnabla\bcdot\left(\rho^\pm \bm{v}^\pm\right) \left[ \rho^\pm(1\pm dH)-1\right] (1\pm dH)\\ =  -\bnabla\rho^\pm  (1\pm dH) \left[ \rho^\pm(1\pm dH)-1\right] \bcdot \bm{v}^\pm-\rho^\pm \bnabla\bcdot \bm{v}^\pm\left[ \rho^\pm(1\pm dH)-1\right],
\end{multline}
and term 5 to
\begin{equation}
\begin{aligned}
\frac{1}{2}\bnabla\bcdot\left( \left[ \rho^\pm(1\pm dH)-1\right]^2 \bm{v}^\pm\right) =&~ (1\pm dH) \left[\rho^{\pm} (1\pm dH) -1\right] \bnabla \rho^{\pm} \bcdot\bm{v}^\pm \\
&\pm \rho^\pm d  \left[\rho^{\pm} (1\pm dH) -1\right] \bnabla H\bcdot\bm{v}^\pm\\
&+ \frac{1}{2}\left[\rho^{\pm} (1\pm dH) -1\right] ^2 \bnabla\bcdot\bm{v}^\pm,
\end{aligned}
\end{equation}
Summing them, we note that their first terms cancel out with each other. Rearranging the last terms, we get
\begin{multline}
\label{e5}
\frac{1}{2}\left[1-\left[\rho^{\pm} (1\pm dH)\right]^2 \right]\bnabla\bcdot\bm{v}^\pm\pm \rho^\pm d \left[\rho^{\pm} (1\pm dH) -1\right] \bnabla H \bcdot\bm{v}^\pm \\
=\frac{1}{2}\left[1-\left[\rho^{\pm} (1\pm dH)\right]^2 \right]\bnabla\bcdot\bm{v}^\pm \mp H  \bnabla\left\{ \rho^\pm d \left[\rho^{\pm} (1\pm dH) -1\right] \right\} \bcdot\bm{v}^\pm \\
\mp \rho^\pm d \left[\rho^{\pm} (1\pm dH) -1\right] H \bnabla\bcdot \bm{v}^\pm,
\end{multline}
plus null Lagrangians, which we neglect for the sake of simplicity since we are dealing with a closed surface.
Let us define the stress tensors 
\begin{equation}
\left(\sigma^\pm_{S}\right)^{ab} =  \left(\sigma^\pm_{S_1}\right)^{ab}+ \left(\sigma^\pm_{S_2}\right)^{ab},
\end{equation}
\begin{equation}
\left(\sigma^\pm_{S_1}\right)^{ab}=\frac{1}{2}\left[1-\left[\rho^{\pm} (1\pm dH)\right]^2 \right] g^{ab} ,
\end{equation}
\begin{equation}
\left(\sigma^\pm_{S_2}\right)^{ab}= \mp \rho^\pm d \left[\rho^{\pm} (1\pm dH) -1\right] k^{ab},
\end{equation}
and the normal stress vector
\begin{equation}
\begin{aligned}
\left(\sigma^\pm_{Sn}\right)^a&= \pm \bnabla_b\left\{g^{ab} \rho^\pm d \left[\rho^{\pm} (1\pm dH) -1\right] \right\} \\
&= \pm g^{ab}\bnabla_b \rho^{\pm} d \left[\rho^{\pm} (1\pm dH) -1\right] \pm g^{ab} \rho^\pm d \bnabla_b \rho^{\pm}  (1\pm dH) + g^{ab} (\rho^\pm d)^2 \bnabla_b H\\
&=g^{ab}  d \left[2\rho^\pm (1\pm dH)-1 \right] \bnabla_b \rho^{\pm}  + g^{ab} (\rho^\pm d)^2 \bnabla_b H.
\end{aligned}
\end{equation}
Then, equation \eqref{e5} can be rewritten as
\begin{equation}
\bm{\sigma}^\pm_S\bm{:}\bnabla\bm{v}^\pm-\bm{k}\,\bm{\sigma}_{Sn}\,\bm{v}^\pm.
\end{equation}
Terms 2 plus 4 lead to
\begin{equation}
\left\{\left[ \rho^\pm(1\pm dH)-1\right]  \rho^\pm\left(1\pm dH\right) - \frac{1}{2}\left[ \rho^\pm(1\pm dH)-1\right] ^2\right\}  v_n H = -  \bm{\sigma}_{S_1}^\pm\bm{:}\bm{k} v_n.
\end{equation}
Term 3, neglecting null Lagrangians,
\begin{multline}
\pm d \rho^\pm \left(\Delta v_n+k_{ab} k^{ab} v_n\right)\left[ \rho^\pm(1\pm dH)-1\right] \\
=\left\{ \pm\Delta \left[ \rho^\pm d\left[ \rho^\pm(1\pm dH)-1\right]\right] \pm d \rho^\pm  \left[ \rho^\pm(1\pm dH)-1\right]  k^{ab} k_{ab}\right\}v_n\\
= \left\{\bnabla\bcdot \bm{\sigma}_{Sn}^\pm \mp  \bm{\sigma}_{S_2}^\pm\bm{:}\bm{k}\right\} v_n.
\end{multline}
Altogether, we can write the rate of change of the free energy as
\begin{equation}
\begin{aligned}
D_t\mathcal{F}_S\left[\bm{\chi};\rho^\pm;v_n,\bm{v}^\pm\right] =\int_{\Gamma_t}&\left\{ -\left[\bnabla \bcdot \bm{\sigma}_S^\pm + \bm{k} \bm{\sigma}_{Sn}^\pm\right]\bcdot \bm{v}^\pm\right.\\
& \left.+ \left[ \bnabla\bcdot \bm{\sigma}_{Sn}^\pm - \bm{\sigma}_{Sn}^\pm\bm{:}\bm{k}\right] v_n\right\}dS.
 \end{aligned}
\end{equation}
Thus,
\begin{equation}
\delta_{\bm{u}^\pm} \left\{D_t\mathcal{F}_S\left[\bm{\chi};\rho^\pm;v_n,\bm{v}^\pm\right]\right\} =-\int_{\Gamma_t} \left[\bnabla \bcdot \bm{\sigma}_S^\pm + \bm{k} \bm{\sigma}_{Sn}^\pm\right]\bcdot \bm{u}^\pm dS,
\end{equation}
and 
\begin{equation}
\delta_{u_n} \left\{D_t\mathcal{F}_S\left[\bm{\chi};\rho^\pm;v_n,\bm{v}^\pm\right]\right\} =-\int_{\Gamma_t} \left[ \bnabla\bcdot \bm{\sigma}_{Sn}^\pm - \bm{\sigma}_{Sn}^\pm\bm{:}\bm{k}\right] u_n dS.
\end{equation}
From variations of the dilatation dissipation potential, we get
\begin{equation}
\begin{aligned}
\delta_{\bm{u}^\pm}\mathcal{D}_D\left[\bm{\chi};\rho^\pm;v_n,\bm{v}^\pm\right] &= -\int_{\Gamma_t} \lambda \bnabla\bcdot\left(\text{tr}{\bm{d}}^\pm\bm{g}\right)  \bm{u}^\pm dS,
\end{aligned}
\end{equation}
and
\begin{equation}
\begin{aligned}
\delta_{u_n}\mathcal{D}_D\left[\bm{\chi};\rho^\pm;v_n,\bm{v}^\pm\right] &= -\int_{\Gamma_t} \lambda \bnabla\left(\text{tr}\bm{d}^\pm\bm{g}\right) \bm{:}\bm{k} u_ndS.
\end{aligned}
\end{equation}
Finally, variations of the intermonolayer friction dissipation potential lead to
\begin{equation}
\begin{aligned}
\delta_{\bm{u}^\pm}\mathcal{D}_I\left[\bm{\chi};\rho^\pm;\bm{v}^\pm\right] &= \pm\int_{\Gamma_t} b_I  \left(\bm{v}^+-\bm{v}^{-}\right)\bcdot \bm{u}^\pm dS,
\end{aligned}
\end{equation}

\section{Discrete free energy and dissipation potentials for an inextensible viscous monolayer with bending energy\label{appDisc}}
We have defined the discrete Helfrich energy,
\begin{equation}
\label{discrete_helfrich}
\mathcal{F}_H\left(\mathsf{h}\right) =  \sum_{E=1}^{N_e^f} \int_{\tilde{\Gamma}} \frac{\kappa}{2}H\left(\mathsf{h}\right)^2 J\left(\mathsf{h}\right) d\bm{\xi},
\end{equation}
where we have split integration on $\Gamma_t$ as a sum of integration on the curved triangles $\Gamma_t^E$, which are evaluated at the parametric domains $\tilde{\Gamma}$. Functions $H\left(\mathsf{h}\right)$ and $J\left(\mathsf{h}\right)$ describe the mean curvature and the surface  Jacobian in terms of the discretized parametrization; these can be computed by plugging the form of $\bm{\psi}$ (equation \eqref{discrete_param}) in terms of $\mathsf{h}$ in the expressions for the curvature and metric in the natural or convected basis of the parametrization.
We have also defined the matrices representing  dissipation and the inextensibility constraint
\begin{align}
\left[D_{\mathsf{h}\mathsf{h}}\right]_{IJ} &= \mu \sum_{E=1}^{N_e^f} \int_{\tilde{\Gamma}}  \left(\frac{\bm{M}(\mathsf{h})\bcdot\bm{N}(\mathsf{h})}{\Delta t^n}\right)^2 |\bm{k}(\mathsf{h})|^2 B_I^E B_J^E J^n d\bm{\xi},\\
\left[D_{\mathsf{h}\mathsf{a}}\right]_{IJ} &= -\mu  \sum_{E=1}^{N_e^f} \int_{\tilde{\Gamma}}  \frac{\bm{M}(\mathsf{h})\bcdot\bm{N}(\mathsf{h})}{\Delta t^n} B_I^E  \bm{k}(\mathsf{h})\bm{:}\bnabla\bnabla B_J^E J^nd\bm{\xi},\\
\left[D_{\mathsf{h}\mathsf{b}}\right]_{IJ} &=  -\mu  \sum_{E=1}^{N_e^f} \int_{\tilde{\Gamma}}  \frac{\bm{M}(\mathsf{h})\bcdot\bm{N}(\mathsf{h})}{\Delta t^n} B_I^E  \bm{k}(\mathsf{h})\bm{:}\bnabla\left(\bnabla \times B_J^E\right) J^nd\bm{\xi},\\
\left[D_{\mathsf{a}\mathsf{a}}\right]_{IJ} &= \mu \sum_{E=1}^{N_e^f} \int_{\tilde{\Gamma}} \bnabla\bnabla B_I^E \bm{:} \bnabla\bnabla B_J^E   J^nd\bm{\xi},\\
\left[D_{\mathsf{a}\mathsf{b}}\right]_{IJ} 
&= \mu  \sum_{E=1}^{N_e^f}  \int_{\tilde{\Gamma}} \bnabla\bnabla B_I^E \bm{:} \bnabla\left(\bnabla \times B_J^E\right) J^nd\bm{\xi},\\
\left[D_{\mathsf{b}\mathsf{b}}\right]_{IJ} &= \mu  \sum_{E=1}^{N_e^f}  \int_{\tilde{\Gamma}} \left(\bnabla\left(\bnabla\times B_I^E\right)\right)^S \bm{:} \left(\bnabla\left(\bnabla \times B_J^E\right)\right)^S  J^n d\bm{\xi}, 
\\
\left[Q_{\mathsf{h}}\right]_{IJ} &= - \sum_{E=1}^{N_e^f}  \int_{\tilde{\Gamma}} \left(N_I^E\circ O^{-1}\right) (\bm{M}\bcdot\bm{N}) H(\mathsf{h}) B_J^EJ^nd\bm{\xi},\\
\left[Q_{\mathsf{a}}\right]_{IJ} &= \sum_{E=1}^{N_e^f}  \int_{\tilde{\Gamma}} \left(N_I^E\circ O^{-1}\right) \Delta B_J^E J^nd\bm{\xi},
\end{align}
where, in the last two equations, the functions $N_I^E$, interpolating the surface tension $\gamma$ are composed with $O^{-1}$ and evaluated at the macroelement.
We note that $\bnabla$ is the covariant derivative, calculated from partial derivatives in parametric space and using Christoffel symbols \citep{Do_Carmo1992-bx}.
We also note that we have also discretized the rate of change of volume$\dot{\Omega}$ as $(\Omega-\Omega^n)/\Delta t^n$ instead of discretizing equation \eqref{volumeconstraint} directly, similarly to our discretization of the energy release rate. This leads to a discrete dynamics that keeps a constant volume by construction, up to numerical error, regardless of the value of $\Delta t^n$.  To exercise this formulation, we compute $V(\mathsf{h})$ using Gauss theorem on the surface
\begin{equation}
\Omega\left(\mathsf{h}\right) = \frac{1}{3} \int_{\tilde{\Gamma}} \bm{\psi}\left(\mathsf{h}\right)\bcdot\bm{N}\left(\mathsf{h}\right) J^n d\bm{\xi}.
\end{equation}
We finally note that we use Gauss quadrature in the reference element $\tilde{\Gamma}$, although other integration schemes specially suited for subdivision surfaces have been recently proposed \citep{Juttler2016-yi}. 

\section{Discretization of mass conservation\label{dissmasscons}}
We consider an implicit Euler scheme to discretize in time the process operator in the transport problem as in equation \eqref{discrete_process2}, which leads to
\begin{equation}
\frac{\rho^{n+1}-\rho^n}{\Delta t^n} +
\rho^{n+1}\text{tr}\bm{d}+\bm{c}\bcdot \bnabla\rho^{n+1} = 0.
\end{equation}
In this case, $\bm{d}$ and $\bm{c}$ depend on $(\mathsf{h},\mathsf{a},\mathsf{b})$, but we do not write it for simplicity. This is a reaction-advection problem in $\rho^{n+1}$ and its discretization with finite elements has to be carefully considered, since Garlerkin methods cannot deal with large convective terms. Discretizing we obtain
\begin{equation}
\sum_I\rho_I^{n+1} \left[N_I^E\left(1+\Delta t^n\,\text{tr}\bm{d}\right)+\Delta t^n\,\bm{c}\bcdot \bnabla N_I^E \right] = \rho^n.
\end{equation}
To deal with the convective term appropriately, we use the test functions
\begin{equation}
w_J=N_J^E+\gamma_s\Delta t^n\,\bm{c}\bcdot \bnabla N_J^E,
\end{equation}
following a Petrov-Garlerkin method in which the weight functions do not coincide with the basis functions used in the approximation of the solution $\rho{n+1}$. This method is called {stream-upwind Petrov Garlerkin} (SUPG) \citep{Donea2003-oa}, which is able to treat the convective term of the transport problem by adding numerical diffusion controlled by the SUPG parameter $\gamma_s$. Then, the weak form is
\begin{multline}
\sum_I \rho_I^{n+1} \sum_{E=1}^{N_e^f}\int_{\tilde{\Gamma}} \left(N_J^E+\gamma_s\Delta t^n\,\bm{c}\bcdot \bnabla N_J^E\right) \left[N_I^E\left(1+\Delta t^n\,\text{tr}\bm{d}\right)+\Delta t^n\,\bm{c}\bcdot \bnabla N_I^E \right] Jd\bm{\xi}
\\=\sum_{E=1}^{N_e^f}\int_{\tilde{\Gamma}}  \left(N_J^E+\gamma_s\Delta t^n\,\bm{c}\bcdot \bnabla N_J^E\right)\rho^nJd\bm{\xi},
\end{multline}
where here $J$ is also a function of $\mathsf{h}$. This equation can also be written as a linear system
\begin{equation}
\hat{\mathsf{M}} \mathsf{r}^{n+1} = \hat{\mathsf{s}},
\end{equation}
with
\begin{equation}
\hat{\mathsf{M}}_{IJ} = \sum_{E=1}^{N_e^f}\int_{\tilde{\Gamma}} \left(N_J^E+\gamma_s\Delta t^n\,\bm{c}\bcdot \bnabla N_J^E\right) 
\left[N_I\left(1+\Delta t^n\,\text{tr}\bm{d}\right)+\Delta t^n\,\bm{c}\bcdot \bnabla N_I^E \right]Jd\bm{\xi},
\end{equation}
and
\begin{equation}
\hat{\mathsf{s}}_{J} = \sum_{E=1}^{N_e^f}\int_{\tilde{\Gamma}}  \left(N_J^E+\gamma_s\Delta t^n\,\bm{c}\bcdot \bnabla N_J^E\right)\rho^nJd\bm{\xi}.
\end{equation}
We note that $\hat{\mathsf{M}}$ is not symmetric and $\hat{\mathsf{M}}$ and $\hat{\mathsf{s}}$ depend non-linearly on $(\mathsf{h},\mathsf{a},\mathsf{b})$ through $\bm{d}$, $\bm{c}$ and $J$. The coupled system of finite element equations involving balance of linear momentum and mass transport, corresponding to the spatial discretization of equation \eqref{vartim2}, are solved simultaneously using a Newton-Raphson method.

\bibliographystyle{jfm}
\bibliography{ref}

\end{document}